\documentclass{article}
\usepackage{graphicx}
\usepackage{amsmath}
\usepackage{amssymb}
\usepackage{epsfig}
\include{header}               % newcommand statements
\begin{document}
\pagenumbering{roman}
\title{
%\Huge {\bf (DRAFT)}\\
\Huge {\bf Letter of Intent for KASKA}\\
\Large{ High Accuracy Neutrino Oscillation Measurements \\ 
with $\bar{\nu}_e$s from \\
{\it Kas}hiwazaki-{\it Ka}riwa Nuclear Power Station.}\\
{\ }\\
{\ }\\
}
\begin{figure}[t]
\includegraphics[width=0.2\textwidth] {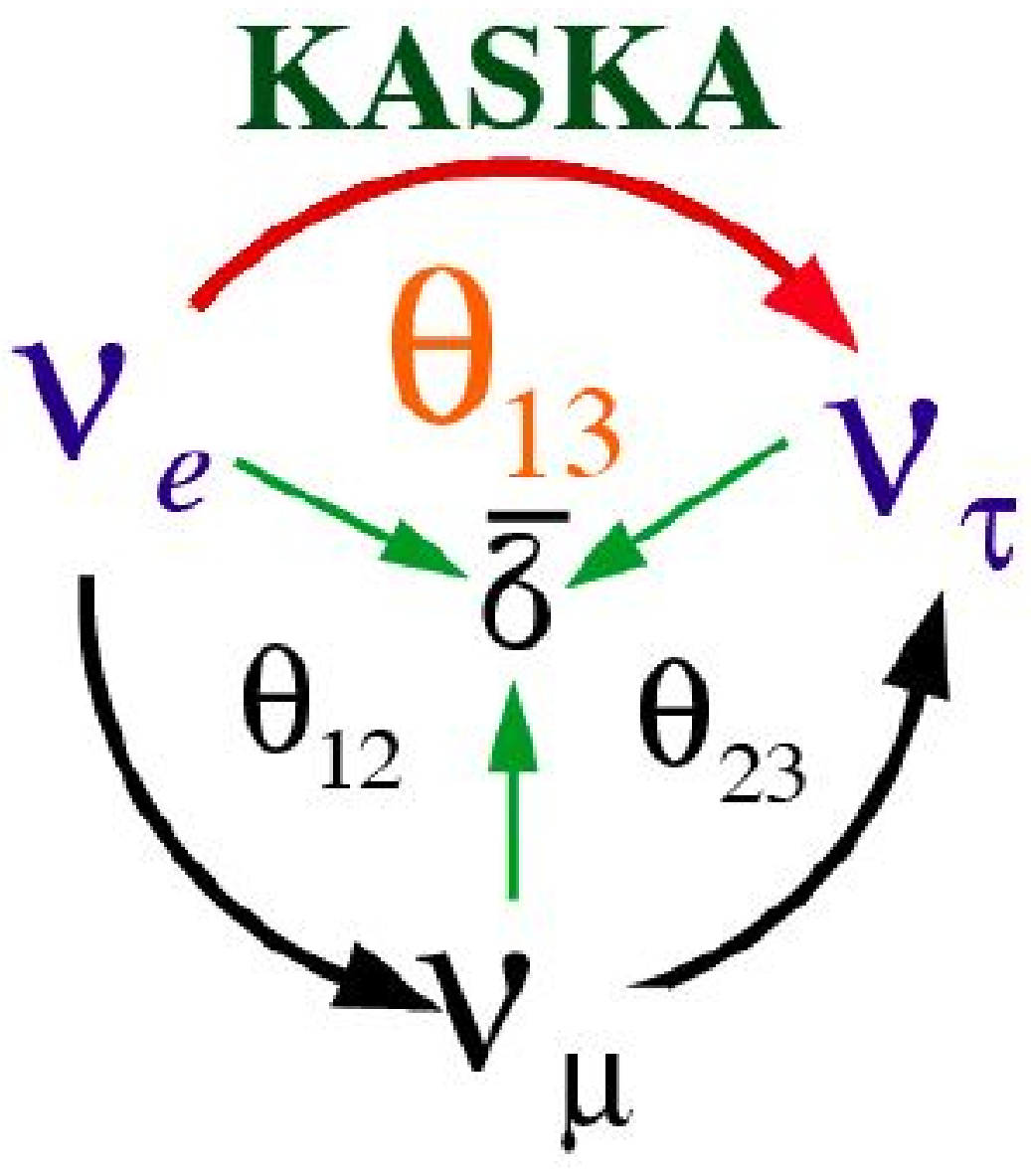}  
\end{figure}
%\vskip 1cm
\author{
M. Aoki$^{\ref{Niigata}}$, \and
$\it{K. Akiyama}$$^{\ref{MUE}}$$^{,a}$, \and
Y. Fukuda$^{\ref{MUE}}$,  \and
$\it{A. Fukui}$$^{\ref{Kobe}}$$^{,b}$,  \and
Y. Funaki$^{\ref{TIT}}$,  \and
H. Furuta$^{\ref{TIT}}$, \and
T. Hara$^{\ref{Kobe}}$, \and
T. Haruna$^{\ref{TMU}}$, \and
N. Ishihara$^{\ref{KEK}}$, \and
$\it{T. Iwabuchi}$$^{\ref{Niigata}}$$^{,c}$, \and
M. Katsumata$^{\ref{Niigata}}$, \and
T. Kawasaki$^{\ref{Niigata}}$, \and
M. Kuze$^{\ref{TIT}}$, \and
J. Maeda$^{\ref{TIT}}$, \and
T. Matsubara$^{\ref{TIT}}$, \and
$\it{T. Matsumoto}$$^{\ref{TMU}}$$^{,d}$, \and
H. Minakata$^{\ref{TMU}}$, \and
H. Miyata$^{\ref{Niigata}}$, \and
Y. Nagasaka$^{\ref{HIT}}$, \and
$\it{T. Nakagawa}$$^{\ref{TMU}}$ $^{,e}$, \and
N. Nakajima$^{\ref{Niigata}}$, \and
H. Nakano$^{\ref{Niigata}}$, \and
K. Nitta$^{\ref{TIT}}$, \and
M. Nomachi$^{\ref{Osaka}}$, \and
K. Sakai$^{\ref{Niigata}}$, \and
Y. Sakamoto$^{\ref{TGU}}$, \and
$\it{K. Sakuma}$$^{\ref{TMU}}$$^{,f}$, \and
M. Sasaki$^{\ref{Niigata}}$, \and
F. Suekane$^{\ref{Tohoku}}$, \and
H. Sugiyama$^{\ref{KEK}}$, \and
T. Sumiyoshi$^{\ref{TMU}}$, \and
H. Tabata$^{\ref{Tohoku}}$, \and
N. Tamura$^{\ref{Niigata}}$, \and
M. Tanimoto$^{\ref{Niigata}}$, \and
Y. Tsuchiya$^{\ref{Tohoku}}$, \and
$\it{R. Watanabe}$$^{\ref{Niigata}}$\and 
 and \and
O. Yasuda$^{\ref{TMU}}$
}
%\end{center}
%\begin{itemize}
%\item[*] Editors
%\end{itemize}
\maketitle

The people with italic name are not current KASKA members, but have contributed to this LoI.

\begin{enumerate}
\item Hiroshima Institute of Technology  \label{HIT}
\item High Energy Accelerator Research Organization (KEK)   \label{KEK}
\item Kobe University                         \label{Kobe}
\item Miyagi University of Education  \label{MUE}
\item Niigata University                   \label{Niigata}
\item Osaka University                    \label{Osaka}
\item Tohoku University                  \label{Tohoku}
\item Tohoku Gakuin University  \label{TGU}
\item Tokyo Institute of Technology  \label{TIT}
\item Tokyo Metropolitan University  \label{TMU}

\end{enumerate}

Present Address: 

$^a$ Tohoku University \\              %  Akiyama

$^b$ Nagoya University  \\                 % Fukui

$^c$ Fujitsu Integrated Micro Technology \\  % Iwabuchi

$^d$ KEK   \\                                    % Matsumoto

$^e$ Hewlett-Packerd Japan, Ltd  \\           % Nakagawa

$^f$  Ministry of Economy, Trade and Industry  \\  % Sakuma

\include{abstract}
\tableofcontents
\pagenumbering{arabic}
\setcounter{page}{1}
% Main Part 

\section{Introduction}
\label{sec:introduction}

\subsection{Physics Background and $\theta_{13}$}
Since neutrinos have peculiar properties such as very small masses, very
much different mixing patterns from quark sector and  possibility of being
Majorana particle, study of neutrinos is very important to understand
the nature of the elementary particles and to construct its unified
theory in the future. 
Because the neutrinos have finite masses and mixing, they
perform neutrino oscillation if their masses are not identical.  
For two flavor case, the probability that a particular flavor 
neutrino $\nu_x$ with energy $E_\nu$ to remain as $\nu_x$ after 
traveling a distance $\it{L}$ becomes,
\begin{gather}
P_{\nu_x \rightarrow \nu_x} =1-\sin^22\theta \sin^2\Phi, \\ 
\Phi=\frac{\Delta m^2 L}{4E_{\nu}} \notag
\end{gather}
due to the neutrino oscillation, where $\Delta m^2$ is the difference
of the squared masses between the two mass eigen states; $ m^2_2-m^2_1$, and
$\theta$ is the mixing angle between mass eigenstate and flavor eigenstate. 
Through neutrino oscillation, it is possible to access to a
very small mass range where direct measurements are difficult to reach
and to measure the mixing angle. 

For 3 flavor case, the flavor eigenstates; $\nu_e, \nu_{\mu}$ and
$\nu_{\tau}$, and mass eigenstates; $\nu_1, \nu_2$ and $\nu_3$, are
mixed by $3\times3$ Maki-Nakagawa-Sakata (MNS) matrix~\cite{MNS}.

\begin{equation}
\left(\begin{array}{c}
\nu_e \\ \nu_{\mu} \\ \nu_{\tau} 
\end{array}\right)
=
\left(\begin{array}{ccc}
U_{e1} & U_{e2} &U_{e3} \\ U_{\mu 1} & U_{\mu 2} & U_{\mu 3} \\
U_{\tau 1} & U_{\tau 2} & U_{\tau 3}
\end{array}\right)
\left(\begin{array}{c}
\nu_1 \\ \nu_2 \\ \nu_3 
\end{array}\right)
=U_{MNS}
\left(\begin{array}{c}
\nu_1 \\ \nu_2 \\ \nu_3 
\end{array}\right)
\end{equation}

In this case, three flavor oscillation formula becomes,
%\begin{equation}
\begin{eqnarray}
&{\ }& \left\{
\begin{array}{c}
 P(\nu_{\alpha} \rightarrow \nu_{\beta})\\
 P(\bar{\nu}_{\alpha} \rightarrow \bar{\nu}_{\beta})
\end{array}\right\} \nonumber\\
&=&\delta_{\alpha \beta}
-4\sum_{i>j}\Re(U_{\alpha i}^* U_{\beta i}U_{\alpha j}U_{\beta
 j}^*)\sin^2\Phi_{ij}
\pm 2 \sum_{i>j}\Im(U_{\alpha i}^* U_{\beta i}U_{\alpha j}U_{\beta j}^*)\sin2\Phi_{ij}\nonumber\\
\label{eq:3nu_oscillation}
\end{eqnarray}
%\end{equation}
where $\alpha, \beta$ are indices of flavors and $i, j$ are indices
of masses with $m_i$ being the mass of $\nu_i$.
$\Phi_{ij}=\frac{\Delta m_{ij}^2L}{4E_{\nu}}$ represent oscillation
phase caused by the difference of mass square, $\Delta
m_{ij}^2=m_{j}^2-m_{i}^2$.\\

The MNS matrix can be expressed by 4 parameters;
\begin{equation}
\begin{split}
U_{MNS}=
\left(\begin{array}{ccc}
1 & 0&0\\ 0 & c_{23} & s_{23} \\ 0 & -s_{23} & c_{23}
\end{array}\right)
\left(\begin{array}{ccc}
c_{13}& 0&s_{13}e^{i\delta_l}\\ 0 & 1 & 0 \\ -s_{13}e^{-i\delta_l} & 0 & c_{13}
\end{array}\right)
\left(\begin{array}{ccc}
c_{12}& s_{12}&0\\ -s_{12} & c_{12} & 0 \\ 0 & 0 & 1
\end{array}\right) \\
=
\left(\begin{array}{ccc}
c_{13}c_{12}& c_{13}s_{12}&s_{13}e^{i\delta_l}\\
-c_{23}s_{12}-s_{13}s_{23}c_{12}e^{-i\delta_l} &
c_{23}c_{12}-s_{13}s_{23}s_{12}e^{-i\delta_l} & c_{13}c_{23} \\
s_{23}s_{12}-s_{13}c_{23}c_{12}e^{-i\delta_l} &
-c_{13}s_{23}-s_{13}c_{23}s_{12}e^{-i\delta_l}& c_{13}c_{23}
\label{eq:MNS_parameter}
\end{array}\right) \\
\end{split}
\end{equation}
where $c_{ij}$ and $s_{ij}$ express $\cos\theta_{ij}$ and
$\sin\theta_{ij}$, respectively.\\

\begin{figure}[htbp]
\begin{center}
\includegraphics[width=0.7\textwidth] {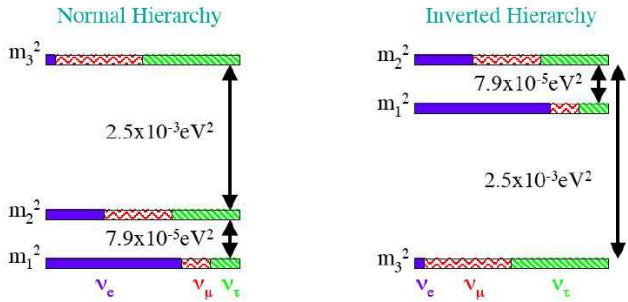}  
\caption{The relation of neutrino masses and flavor components of mass eigenstates. 
There are two possibilities in the mass relations according to the sign of $\Delta m_{23}^2$. 
The case of $\Delta m_{23}^2 > 0$ ($\Delta m_{23}^2<0$) is called 'normal hierarchy' ('inverted hierarchy'). 
The $\nu_e$ component of $\nu_3$ is less than 4\%. 
Other mixings are relatively large. }
\label{fig:mass_pattern}
\end{center}
\end{figure}

Experimental studies of the neutrino oscillations have been
progressing very rapidly in these days. 
The first firm evidence of the neutrino oscillation was discovered in the disappearance of muon type
atmospheric neutrinos by SuperKamiokande (SK) group in 1998~\cite{SK:1998}.
SK group has been taking more data and recently observed spectrum distortion~\cite{SK:2004}, which is consistent with oscillation pattern. 
K2K group confirmed the result~\cite{K2K:2004} by observing the disappearance of
$\nu_{\mu}$ produced by the KEK-PS accelerator, after traveling 250~km. 
These phenomena are naturally explained by the $\nu_{\mu} \rightarrow \nu_{\tau}$ oscillation since there is no
significant enhancement in $\nu_e$ spectrum.  
The measured oscillation parameters are 
$\Delta m^2_{23} \approx 2.4 \times 10^{-3} eV^2$ and $\sin^22\theta_{23}
\approx 1$.  
As for the electron type neutrinos there have been indications of the
neutrino oscillation in the deficit of the solar neutrinos for a long
time~\cite{Solar}. 
A transformation of the solar $\nu_e$ to other type
neutrinos was identified by combining SNO data and SK data in
2002~\cite{SNO,Fukuda:2001nj}. 
Finally, the KamLAND group observed a large deficit
in reactor $\bar{\nu}_e$ in 2002~\cite{KamLAND:2002} and then spectral
distortion in 2005~\cite{KamLAND:2004}. 
Combined with solar neutrino data, the oscillation parameters are measured as 
$\Delta m^2_{12} \approx7.9 \times 10^{-5} eV^2$ and $\tan^2\theta_{12} \approx 0.40$.
With these measurements, measurable 2 mass combinations and 2 mixing
angles out of 3 are already measured. 
The sign of $\Delta m^2_{12}$ is determined to be $>0$ by MSW matter effect of solar neutrinos. 
However, the sign of $\Delta m^2_{23}$ is not determined yet. 
Finite $\theta_{13}$ and $\delta_l$ have not been measured yet. 
The most stringent upper limit of $\theta_{13}$ was measured by CHOOZ group using reactor
$\bar{\nu}_e$s to be $\sin^22\theta_{13}<0.15$ at $\Delta m^2=2.5
\times 10^{-3}eV^2$~\cite{CHOOZ}. 
As for $\delta_l$ parameter, there is no information by now. 
By combining measured parameters, the
magnitudes of MNS matrix elements are  determined to be roughly,
\begin{equation}
|U_{MNS}|\approx
\left(\begin{array}{ccc}
0.8 &  0.5  &<0.2 \\ 
0.4 & 0.6   & 0.7 \\ 
0.4 & 0.6   & 0.7
\end{array}\right)
\end{equation}
and the relation of masses and flavor components of mass eigenstates are shown in Fig.~\ref{fig:mass_pattern}.
The elements of the MNS matrix are very much different from those
of the Cabbibo-Kobayashi-Maskawa Matrix, which is nearly diagonal.
The $U_{e3}$ element (and thus $\sin \theta_{13}$) is small compared
to other elements. 
This peculiar feature has to be explained by the unified theory of elementary particles and the smallness of the
$\theta_{13}$ may play a key role when constructing the unified theory.

\begin{figure}[htbp]
\begin{center}
\includegraphics[width=0.5\textwidth] {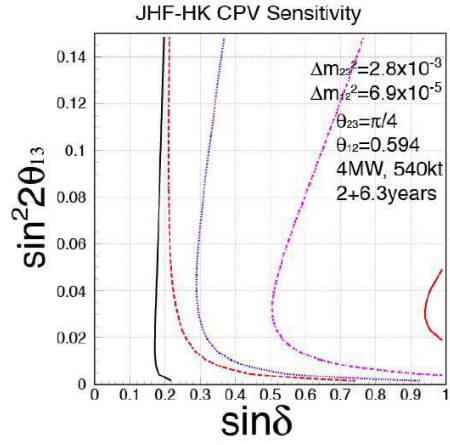}  
\caption{ Expected 3$\sigma$ discovery regions of $\sin\delta_l$ as a function of $\sin^22\theta_{13}$ after 2($\nu_{\mu}$) and 6.3 ($\bar{\nu}_{\mu}$) year of exposure in phase II of the JHF-Kamioka project (J-PARC). 
The (black) solid curve is the case of no background and only statistical error of the signal, (red) dashed one is 
2\% error for the background subtraction, and (blue) dotted, (magenta) dot-dashed and (red) solid curves are the cases for systematic errors of both background subtraction and signal detection of 2\%, 5\% and 10\%, respectively~\cite{TKobayashi:2003}. }
\label{fig:th13_delta}
\end{center}
\end{figure}

As for the $\delta_l$, the $e^{i\delta_l}$ term is multiplied by $\sin
\theta_{13}$ in the Eq.(\ref{eq:MNS_parameter}).  
Thus the effect of CP violation phenomena is proportional to $\sin\theta_{13}$.
For example, the detectability of finite $\delta_l$ from asymmetry of
$P(\nu_{\mu} \rightarrow \nu_e )$ and $P(\bar{\nu}_{\mu} \rightarrow
\bar{\nu}_e )$ in near-future accelerator based neutrino experiments
depends on the size of $\theta_{13}$ as shown in
Fig.~\ref{fig:th13_delta}.
Therefore it is important to know the magnitude of $\sin \theta_{13}$ to
evaluate the effect of leptonic CP violation, which may explain the
matter-antimatter asymmetry of the current universe.

KamLAND experiment measured a deficit of reactor neutrino at average baseline 180km. 
For three generations, what KamLAND observed is expressed as,
\begin{equation}
P(\bar{\nu}_e \rightarrow \bar{\nu}_e)
=\cos^4\theta_{13}(1-\sin^22\theta_{12}\sin^2\Phi_{12})
\end{equation}
There is $\sim 5 \%$ of ambiguity introduced by the $\theta_{13}$ term, when determining
$\sin^2\theta_{12}$ from KamLAND oscillation.  
Therefore precise measurement
$\theta_{13}$ helps to obtain more accurate $\theta_{12}$ from the
KamLAND data.

Thus there are number of reasons why measurement of $\theta_{13}$ is
one of the most urgent topics of the neutrino physics.\\

%==============================================
\subsection{Reactor Neutrino Oscillation}
In reactors, huge amount of low energy $\bar{\nu}_e$'s are produced in
the $\beta$-decays of the fission product nuclei. A typical reactor
with thermal power of 3GW isotropically produces the Avogadro number
of $\bar{\nu}_e$'s in 15 minutes. The average $\bar{\nu}_e$ energy
detected through $\bar{\nu}_e +p \rightarrow e^++n$ interaction is
around 4~MeV. 
These properties of  huge number and low energy of reactor $\bar{\nu}_e$ are suitable to use to
measure neutrino oscillation. 

In 3 generations, the $\bar{\nu}_e \rightarrow \bar{\nu}_e$
oscillation formula is calculated from the Eq.(\ref{eq:3nu_oscillation}) and (\ref{eq:MNS_parameter}) 
as ,
\begin{equation}
P( \bar{\nu}_e \rightarrow \bar{\nu}_e ) = 
1-4c_{13}^2(c_{13}^2s_{12}^2c_{12}^2\sin^2\Phi_{21}
                                           +s_{13}^2c_{12}^2\sin^2\Phi_{31}
                                           +s_{13}^2s_{12}^2\sin^2\Phi_{32})
                                           \label{eq:reactor_nu_oscillation}.\\
\end{equation}
Because $|\Phi_{31}|=|\Phi_{12}+\Phi_{23}|\approx|\Phi_{23}| $,
there are essentially two types of oscillations; one is driven by
$\Delta m_{13}^2$ at around 2km and the other one is driven by $\Delta
m_{12}^2$ at around 50km.

\begin{figure}[htbp]
\begin{center}
\includegraphics[width=0.8\textwidth] {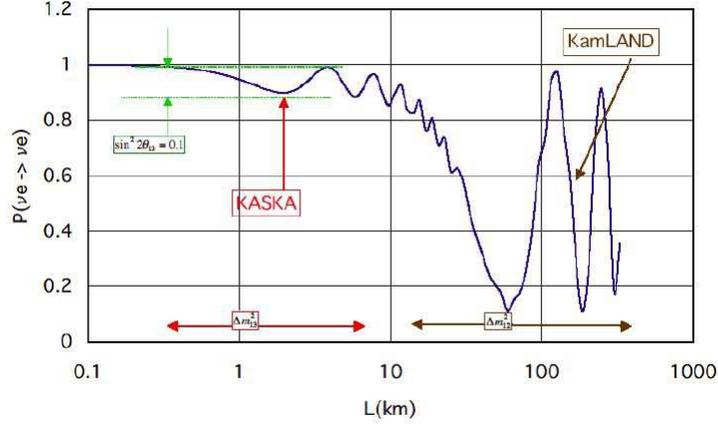}  
\caption{ $\bar{\nu}_e$ oscillation with 3 generations. }
\label{fig:3nue_oscillation}
\end{center}
\end{figure}

Fig.~\ref{fig:3nue_oscillation} shows the survival probability of 4~MeV
$\bar{\nu}_e$ in case $\sin^22\theta_{13}=0.1$.

\begin{figure}[htbp]
\begin{center}
\includegraphics[width=0.6\textwidth] {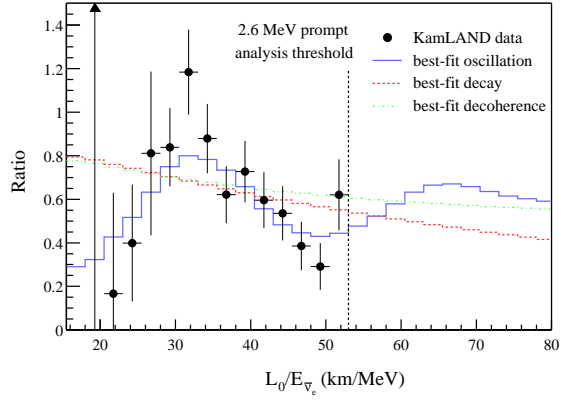}  
\caption{ Oscillation pattern of the reactor neutrinos measured by KamLAND at an average baseline 180km.~\cite{KamLAND:2004} }
\label{fig:KL_oscillation}
\end{center}
\end{figure}

KamLAND, which measured $\sin^22\theta_{12}$ and $\Delta m_{12}^2$ at
around $L\sim180km$, has demonstrated power of reactor measurement of neutrino oscillation. 
Fig.~\ref{fig:KL_oscillation} shows the oscillation pattern of the reactor neutrino measured by the KamLAND
experiment. 
KamLAND detected an energy-dependent deficit of the reactor neutrino.  
A clear oscillatory pattern is observed in Fig.~\ref{fig:KL_oscillation}.\\

\begin{figure}[htbp]
\begin{center}
\includegraphics[width=0.8\textwidth] {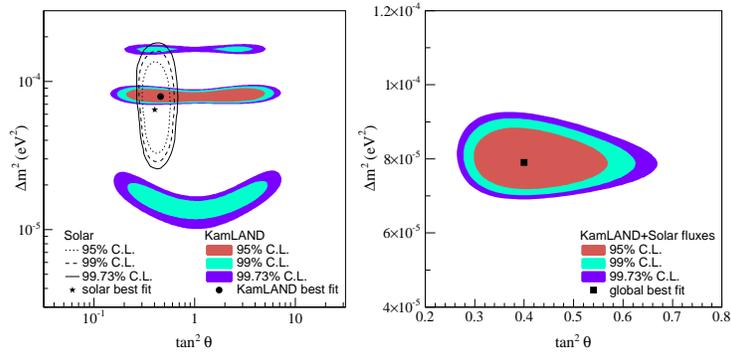}  
\caption{ Oscillation parameters of electron type neutrinos. 
The KamLAND results (Left) and after combination with solar neutrino experiments.~\cite{KamLAND:2004}  }
\label{fig:KL_cont}
\end{center}
\end{figure}

Fig.~\ref{fig:KL_cont} shows the oscillation parameters measured by
KamLAND only (left plot) and KamLAND+ Solar analysis (right plot). 
From these measurement, $\tan^2\theta \sim 0.4, \Delta m^2 \sim 8\times10^{-5}eV^2 $ are obtained.\\

\begin{figure}[htbp]
\begin{center}
\includegraphics[width=0.45\textwidth] {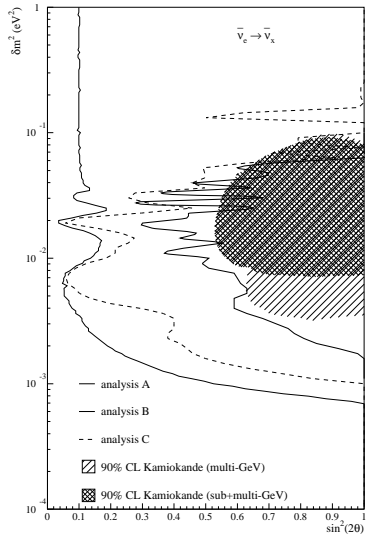}  
\caption{CHOOZ $\sin^22\theta_{13}$ limits~\cite{CHOOZ}.}
\label{fig:CHOOZ_result}
\end{center}
\end{figure}

KASKA aims to measure the $\sin^22\theta_{13}$ at around
$L\sim$1.6~km; 1/100th baseline of KamLAND.  At the distance, the Eq.(\ref{eq:reactor_nu_oscillation}) reduces to,
\begin{equation}
P( \bar{\nu}_e \rightarrow \bar{\nu}_e ) =
1-\sin^22\theta_{13}\sin^2\Phi_{13}+O(10^{-3})
\label{eq:reactor_oscillation}
\end{equation}
Therefore almost pure $\sin^22\theta_{13}$ can be measured.
The current best limit on $\sin^22\theta_{13}$ is obtained by CHOOZ
reactor experiment in France.  
The CHOOZ experiment used two powerful
reactors whose total thermal energy is 8.5~GW$_{th}$.  
The detector is placed at 1km from the reactors and the result of the measurements are shown in
Fig.~\ref{fig:CHOOZ_result}.
They did not confirm the positive result and showed the upper limit of
$\sin^22\theta_{13}<0.15$ for $\Delta m_{13}^2=2.5 \times 10^{-3}eV^2$.

\begin{figure}[htbp]
\begin{center}
\includegraphics[width=0.5\textwidth] {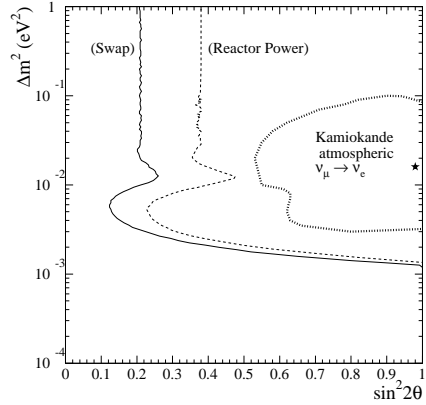}  
\caption{ Result of PaloVerde experiment.~\cite{PaloVerde}  }
\label{fig:PaloVerde}
\end{center}
\end{figure}

The PaloVerde experiment also searched for the deficit of reactor
neutrino with average baseline 0.8km using PaloVerde reactor complex
in Arizona, USA, whose total thermal energy is 11.63~GW$_{th}$.  
They did not observe the deficit of the reactor neutrino and the limit of
the oscillation measurement is shown in Fig.~\ref{fig:PaloVerde}.

In view of the currently known $\Delta m_{23}^2$, the baselines of
those experiments were somewhat shorter than most probable oscillation
maximum of 1.8~km. The measurement errors were rather large due to
uncertainties of reactor neutrino flux as well as detector associated
absolute errors and low statistics.

The KASKA experiment is going to measure this oscillation using the
$\bar{\nu}_e$'s from world most powerful Kashiwazaki-Kariwa nuclear
power station with optimized baseline. 
Four identical detectors, of which two are placed near the reactors and other two are placed at around
1.6~km from the reactors, are used. 
By comparing the data taken at near and far distances,  most of the systematic uncertainties are cancelled out
and precise measurement becomes possible. 
This far-near cancellation scheme for the $\theta_{13}$ measurement was first proposed by Krasnoyarsk group~\cite{Kr2Det}.  
Fig.~\ref{fig:sensitivity} shows the expected sensitivity for
$\sigma_{\text{rel}}=1\%$ and 0.5\% with 3 years of data.
It is possible to obtain 10 times better sensitivity than current CHOOZ limit.
 %

%==============================================
\subsection{Why reactor measurement?}

Near-future long baseline accelerator experiments (LBL) also have
sensitivity to $\sin^22\theta_{13}$~\cite{T2K, MINOS}.  
In such experiments, high energy $\nu_{\mu}$'s are produced by accelerator and appearance
of $\nu_e$'s is searched for after traveling a few hundred km. 
The probability of $\nu_{\mu}\rightarrow\nu_e$ transition at the oscillation maximum
is expressed by the following equation.  
\begin{eqnarray}
P(\nu_{\mu} \rightarrow \nu_e)&\approx&
f^2\sin^22\theta_{13}\sin^2\theta_{23}\nonumber\\ 
&-&
\left|\frac{\Delta m^2_{12}}{\Delta m^2_{23}}\right|
fg\cos \theta_{13}
\sin 2\theta_{12}\sin 2\theta_{23}\sin 2\theta_{13}\sin\delta_l 
\nonumber\\
&+&\left|\frac{\Delta m^2_{12}}{\Delta m^2_{23}}\right|^2
g^2\cos^2\theta_{23}sin^2 2\theta_{12},
\label{eq:LBL_appearance}
\end{eqnarray}
where
\begin{eqnarray}
f&\equiv&
\frac{\cos(AL/2)}{1\mp AL/\pi}
\nonumber\\
g&\equiv&
\frac{\sin(AL/2)}{AL/\pi}
\nonumber\\
A&\equiv&\sqrt{2}G_FN_e,
\nonumber
\end{eqnarray}
and $AL/\pi\simeq 5\times10^{-2}$ for $L$=300 km.
The sign in the definition of $f$ is $-$ $(+)$ for
the normal (inverted) hierarchy.
This measurement depends on many parameters which are not necessarily well known. 
Since
$\sin^22\theta_{12}$ and $\Delta m_{12}^2$ have turned out not so
small by KamLAND and solar neutrino experiments, the 2nd term in
Eq.(\ref{eq:LBL_appearance}) cannot be ignored any more (this very
reason enables the measurement of $\delta_l$ in future LBL
experiments).  
Because $\sin \delta_l$ is totally unknown, the 2nd
term becomes ambiguity when determining $\sin^22\theta_{13}$ from the
$\nu_e$ appearance probability. 
Furthermore, there is an ambiguity of
$\theta_{23}$ degeneracy, which means there are two possible
$\sin^2\theta_{23}$ corresponding to measured $\sin^22\theta_{23}$ value,
if it is not unity as,

\begin{equation}
\sin^2\theta_{23}=\frac{1+\sqrt{1-\sin^22\theta_{23}}}{2}
                                ~\mbox{\rm or}~
\frac{1- \sqrt{1-\sin^22\theta_{23}}}{2}
\end{equation}

\begin{figure}[htbp]
\begin{center}
 \includegraphics[width=0.7\textwidth] {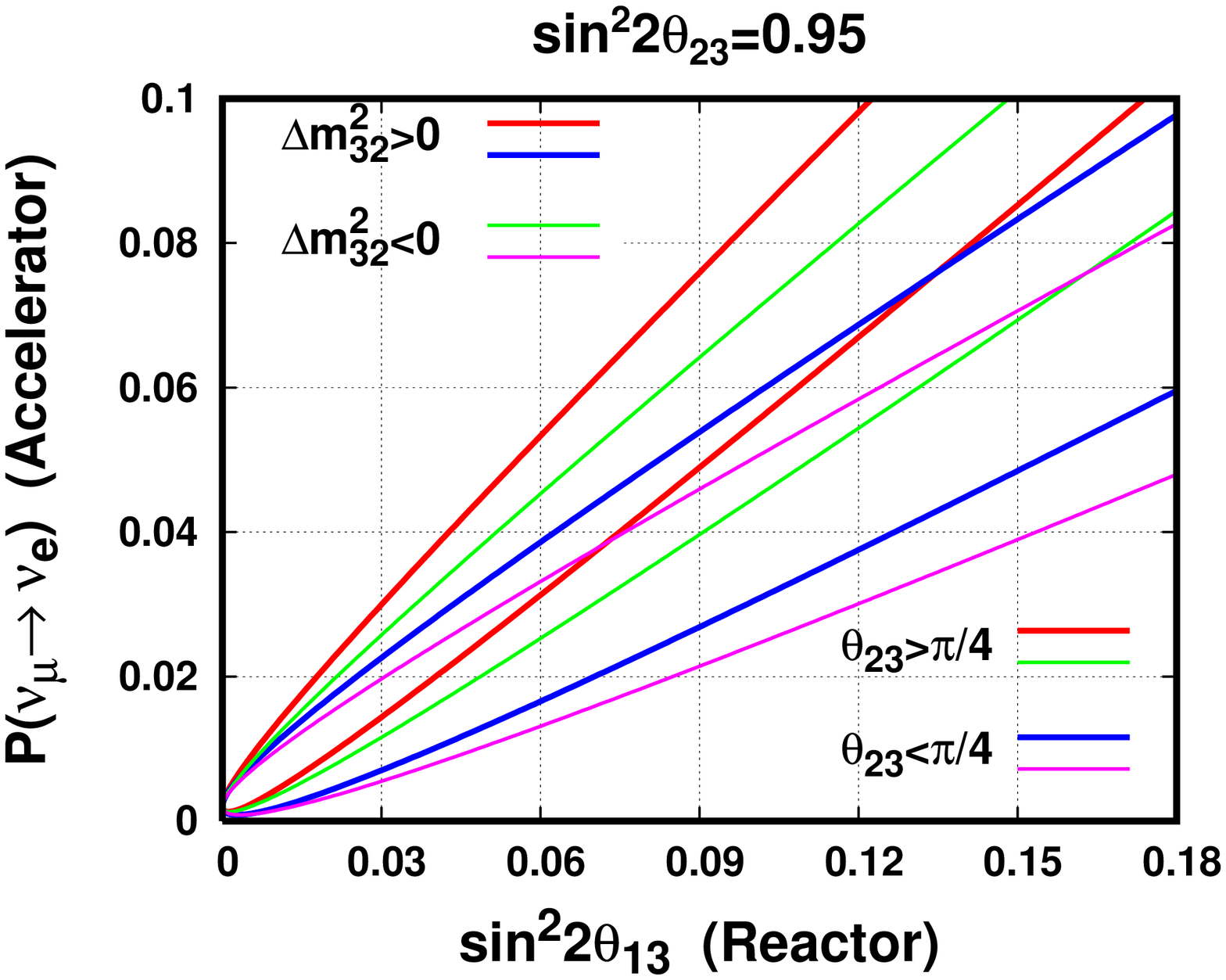}  
\caption{sensitivity of Long Base Line neutrino experiments ($\sin^22\theta_{23}$=0.95)}
\label{fig:LBL_sensitivity}
\end{center}
\end{figure}

\begin{figure}[htbp]
\begin{center}
 \includegraphics[width=0.7\textwidth] {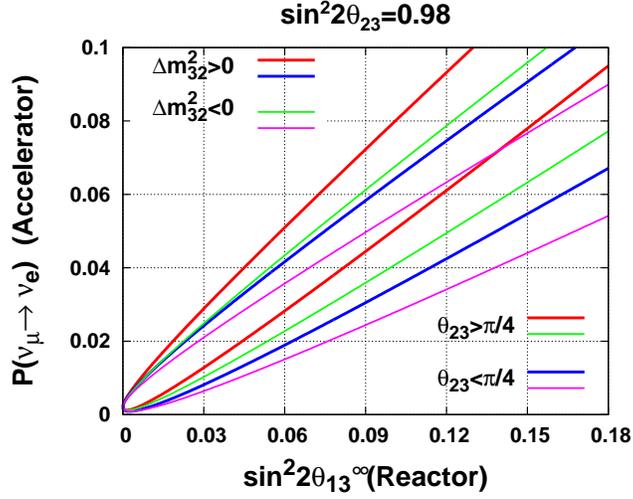}  
\caption{sensitivity of Long Base Line neutrino experiments ($\sin^22\theta_{23}$=0.98)}
\label{fig:LBL_98}
\end{center}
\end{figure}

This degeneracy of  $\theta_{23}$  disturbs the
$\theta_{13}$ determination  by LBL experiments further. 
These situations for the case of $\sin^22\theta_{23}=0.95 $ and 0.98 are depicted in 
Fig.~\ref{fig:LBL_sensitivity} and \ref{fig:LBL_98}.
	 
\begin{figure}[htbp]
\begin{center}
\includegraphics[width=0.7\textwidth] {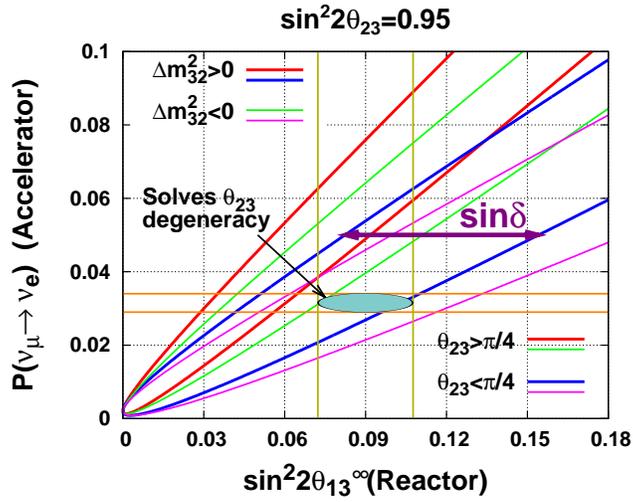}  
\caption{$\theta_{23} $ degeneracy can be solved by combining reactor and accelerator experiment.}
\label{fig:solve_th23_degeneracy}
\end{center}
\end{figure}

The area bounded by each parabola implies the $\sin\delta_l$-term
ambiguity and the $\theta_{23}$ degeneracy corresponds to two
parabolas. Thus even if LBL experiment measures
$P(\nu_{\mu}\rightarrow \nu_e)$ precisely, there is ambiguity in
determining the $\sin^22\theta_{13}$.  On the other hand, the reactor
based $\bar{\nu}_e$ oscillation measurement is a pure
$\sin^22\theta_{13}$ measurement as shown in
Eq.(\ref{eq:reactor_oscillation}). The expected error on
$\sin^22\theta_{13}$ is $\pm 0.01$ and for some parameter regions, the
error of reactor measurement is better than the ambiguities in
$\sin^22\theta_{13}$ determination by LBL experiments.\\ Basically LBL
and reactor experiment measure different parameters and if both
results are combined, there are possibilities to solve the
$\theta_{23}$ degeneracy as shown in the
Fig.~\ref{fig:solve_th23_degeneracy}.
The possibility to remove the $\theta_{23}$ ambiguity
by a reactor experiment was pointed out first in \cite{Fogli:1996pv},
and was elaborated in detail by KASKA group members
\cite{Minakata:2003} using a realistic scenario.
Thus it is important to know the magnitude of $\sin \theta_{13}$ to
evaluate the effect of leptonic CP violation, which may explain the
matter-antimatter asymmetry of the current universe.

\begin{figure}[htbp]
\begin{center}
\includegraphics[width=0.7\textwidth] {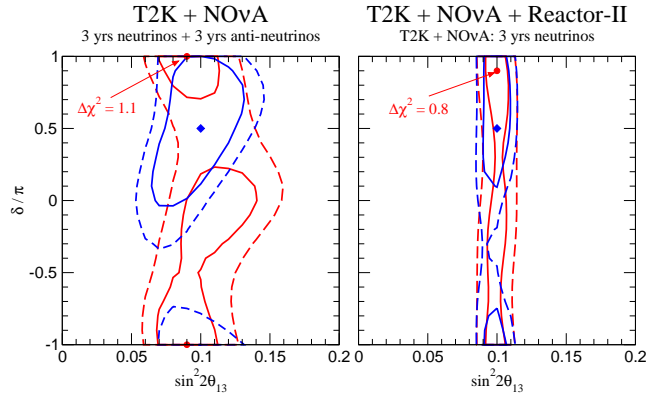}  
\caption{Antineutrino running vs reactor-II ($\delta \sin^22\theta_{13}\sim0.01$). 
The 90\% CL(solid curve) and 3$\sigma$ (dashed curves) allowed regions in the $\sin^22\theta_{13}-\delta_l-$plane for the true values $\sin^22\theta_{13}=0.1$ and $\delta_l=\pi/2$, are shown. 
The blue/dark curves refer tot the allowed regions for the normal mass hierarchy, whereas the red/light curves refer to the sgn($\Delta_{13}^2$)-degenerate solution (inverted hierarchy), where the projections of the minima onto the $\sin^22\theta_{13}-\delta_l-$plane are shown as diamonds (normal hierarchy) and dots (inverted hierarchy). 
For the latter, the $\Delta\chi^2$-value with respect to the best-fit point is also given.
Adopted from~\cite{Huber:2005}. }
\label{fig:anti-vs-react}
\end{center}
\end{figure}
KASKA near detectors will collect 600,000 reactor $\bar{\nu}_e$
events in three years operation.  
This will become the highest statistics so far obtained and the observed
reactor neutrino spectrum will become the world standard.  
The measured spectrum can be used to study fission product.  
Especially
there is no calculation of reactor neutrino spectrum at energy $>$8~MeV
which are produced generally by short lifetime $\beta$-decay isotopes.

In the future, the reactors $\bar{\nu}_e$'s may be used to precisely
measure $\theta_{12}$ by observing the deficit at baseline 50~km, and
$\Delta m_{13}^2$ by observing spectral shape distortion at
baseline $\sim$5~km. Once such parameters are measured precisely, the
electron type neutrinos can be used as a useful tool to probe internal
structure of the sun, supernova explosion mechanisms and radiogenic
structure of the earth.
Because Kashiwazaki-Kariwa nuclear power station is the most intense
neutrino source in the world, it is appropriate to use it to measure
such parameters. In such an occasion, the KASKA-$\theta_{13}$
detectors can be used as precise near detectors as described in the next
section.

%======================================
\section{General Description of the Experiment}
\label{sec:general}

%----------------------------------------------
\subsection{Kashiwazaki-Kariwa Nuclear Power Station }
\begin{figure}[htbp]
 \begin{center}
  \includegraphics[width=1.0\textwidth] {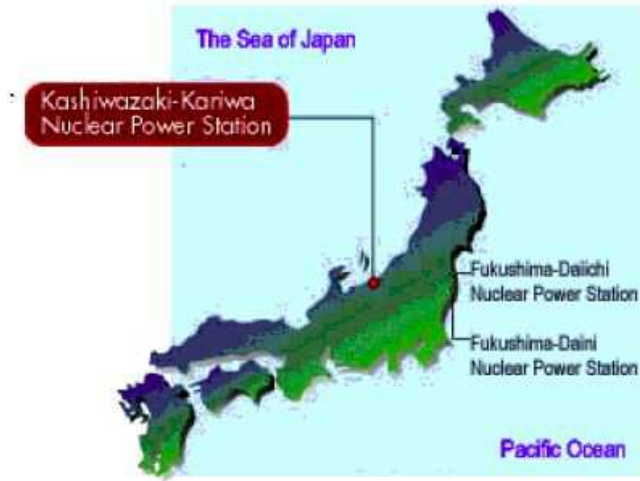}  
 \end{center}
\caption{Location of the Kashiwazaki Kariwa nuclear power station.~\cite{TEPCO}}
\label{fig:kk_location_map}
\end{figure}

KASKA is an abbreviation of $\it{Kas}$hiwazaki-$\it{Ka}$riwa Nuclear
Power Station, following the tradition to name the experiment after the
power station to use.
The Kashiwazaki-Kariwa nuclear power station~\cite{TEPCO}, owned and operated by Tokyo Electric Power Company (TEPCO), 
is located 220~km northwest of Tokyo  and 60~km south-west of Niigata-city 
(Fig.~\ref{fig:kk_location_map}).

\begin{figure}[htbp]
 \begin{center}
  \includegraphics[width=0.9\textwidth] {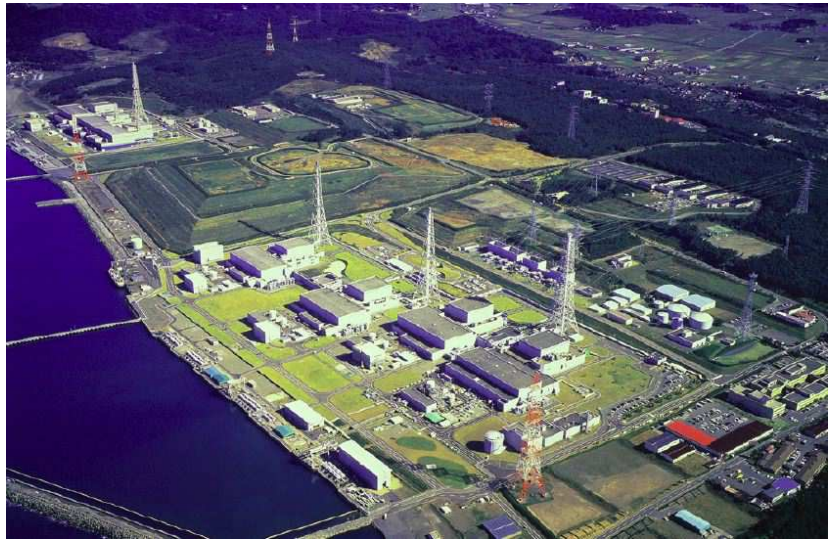}  
 \end{center}
\caption{Aerial view of the Kashiwazaki Kariwa Nuclear Power Station.~\cite{TEPCO_site_photo}}
\label{fig:site_photo}
\end{figure}

\begin{figure}[htbp]
 \begin{center}
  \includegraphics[width=0.8\textwidth] {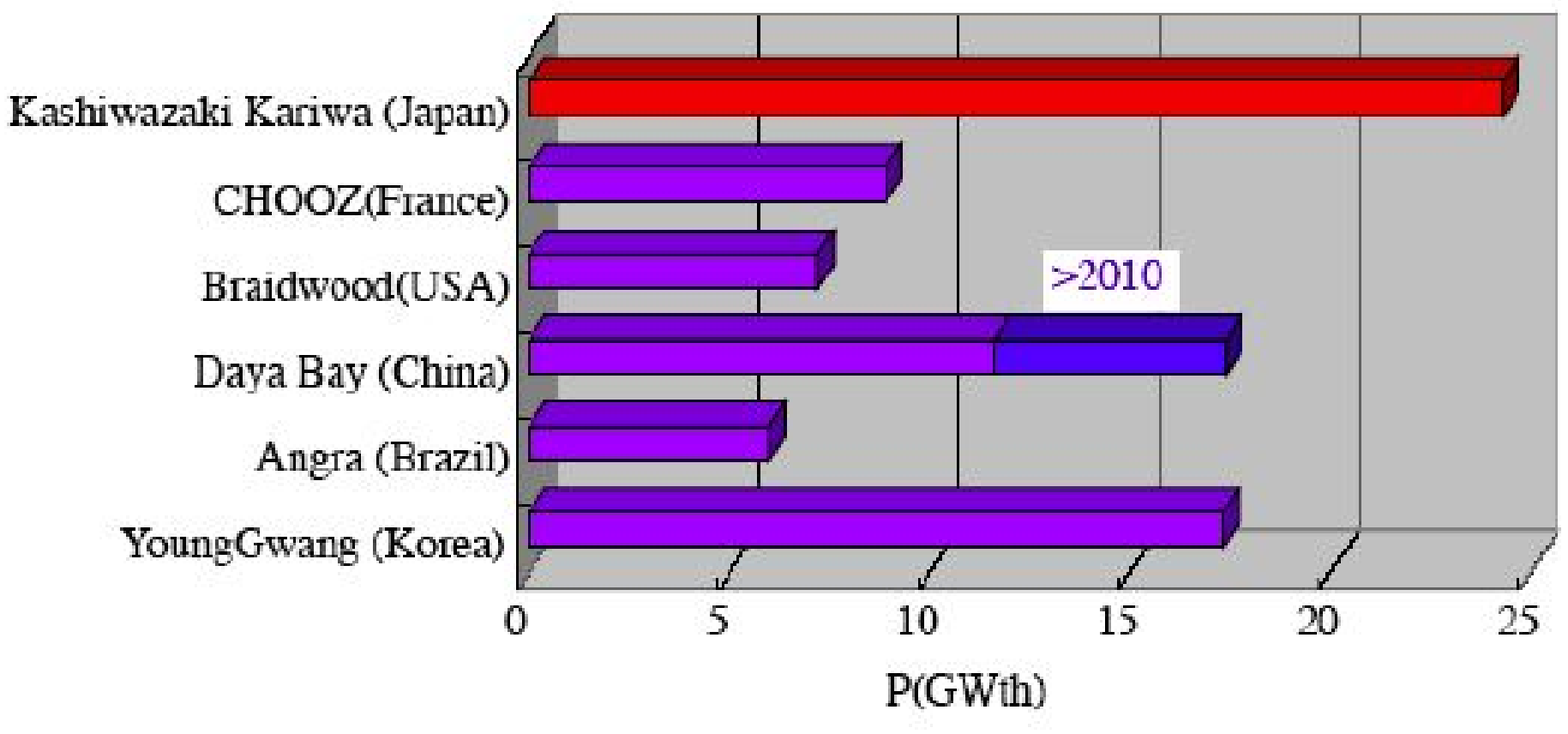}  
 \end{center}
\caption{Site powers for reactor $\theta_{13}$ experiments. }
\label{fig:site_power}
\end{figure}

The name of the plant comes from the names of the local area  on which the plant site is locate, Kashiwazaki-city and Kariwa-village. 
Fig.~\ref{fig:site_photo} shows a picture of the power station. 
The plant site has an area of 4.2 square kilometers, approximately 1.5~km wide and  3~km long. 
The Kashiwazaki-Kariwa power station has 7 reactors which have a total capacity of world's largest thermal energy of 24.3GW. 
The reactors are numbered, from south-west to north-east, as \#1,\#2, \#3, \#4 and \#7, \#6, \#5. 
These reactors are arranged approximately in a line forming two clusters, each consisting of 4
and 3 reactors.    
The distance between the two clusters is $\sim$1.5~km. 
Fig.~\ref{fig:site_power} compares the thermal powers of the proposed reactor $\theta_{13}$ projects in the world.
Kashiwazaki-Kariwa nuclear power station is the most powerful neutrino sources, 
producing an Avogadro number of $\bar{\nu}_e$'s in every 2~minutes  when all the reactors are operating.  
The reactor \#1 through \#5 is the conventional boiling light water reactors (BWR) and \#6, \#7 are  advanced BWRs (ABWR).  
Table~\ref{tab:reactor_parameter_I} and \ref{tab:reactor_parameter_II} summarize the reactor parameters. 
\begin{table}[htbp]
\caption{Kashiwazaki-Kariwa Reactor Parameters\cite{INSC}}
\begin{center}
\begin{tabular}{|c|c|c|c|}
\hline
\#& Type & Thermal Power    & Start Operation \\
{} &            &          (GWth)          &  \\
 \hline
 1 &  BWR5 & 3.293   &1985.09.18  \\
 2 &  BWR5 & 3.293   &1990.09.28 \\
 3 &  BWR5 & 3.293   &1993.08.11 \\
 4 &  BWR5 & 3.293   &1994.08.11 \\
 5 &  BWR5 & 3.293   &1990.04.10 \\
 6 &  ABWR & 3.926   &1996.11.07  \\
 7 &  ABWR & 3.926   &1997.07.02 \\
  \hline
 \end{tabular}
\end{center}
 \label{tab:reactor_parameter_I}
\end{table}
\begin{table}[htbp]
\caption{BWR, ABWR parameters\cite{INSC}}
\begin{center}
\begin{tabular}{|c|c|c|}
\hline
\#& BWR5  &   ABWR  \\
{} &              &               \\
 \hline
 Thermal Power  &  3.293(GWth) &  3.926(GWth)  \\
 Electric Power  &  1.067(GWe) & 1.315(GWe)   \\
 Fuel Inventry &   132tHM &  150tHM  \\
 Active Core Height &   3.71m &  3.71m  \\
 Active Core  Diameter &   4.75m &  5.16m  \\
 Fuel &   UO$_2$ pellet &  UO$_2$ pellet \\
  Cycle length  & 13months &  13months   \\
 Fraction of Core Reloaded Each Cycle &   25\% &  25\%   \\
 Enrichment &   3.4\% &  3.5\%   \\
 \hline
\end{tabular}
\end{center}
\label{tab:reactor_parameter_II}
\end{table}
The newest reactor will be more than 10 years old when  KASKA is expected to start data taking.
Therefore all the reactor fuel is well in equilibrium state and  average  neutrino flux per unit power and energy spectrum are stable. 
 
The reactor operator TEPCO is very much supportive for basic research. 
KASKA group performed a boring study  in the site of Kashiwazaki-Kariwa nuclear power station  at the site of near detector in the fall of 2004. 
TEPCO and the power station helped much of our activities in the site. 

%-------------------------------------------------------------------
\subsection{Reactor Neutrinos}

\begin{figure}[htbp]
 \begin{center}
  \includegraphics[width=1.0\textwidth] {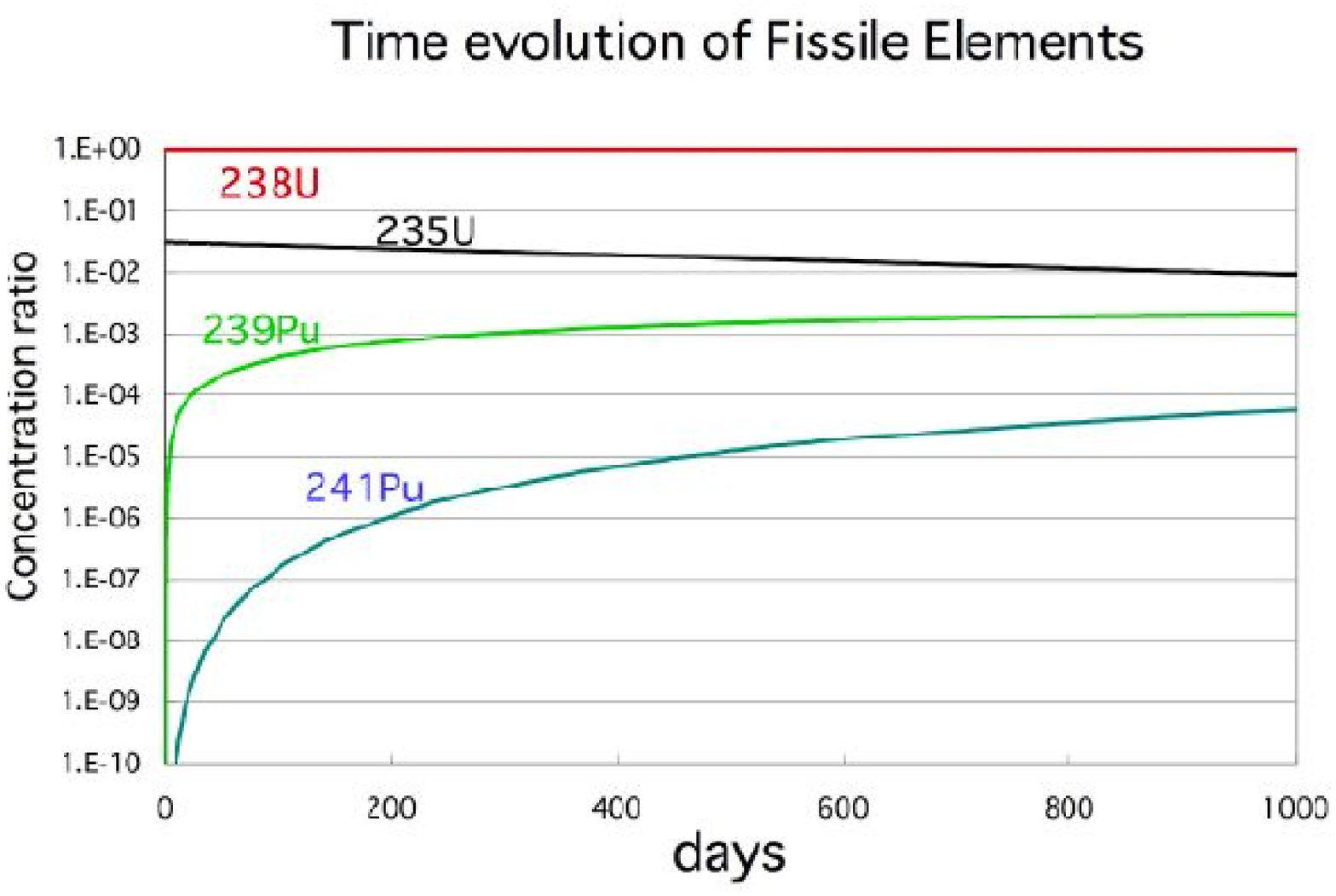}  
 \end{center}
\caption{Reactor fuel burn up.}
\label{fig:time_evolution}
\end{figure}

\begin{figure}[htbp]
\vspace{9pt}
\hspace{10pt}
%\framebox[55mm]{\rule[-21mm]{0mm}{43mm}}
\includegraphics[scale=0.45]{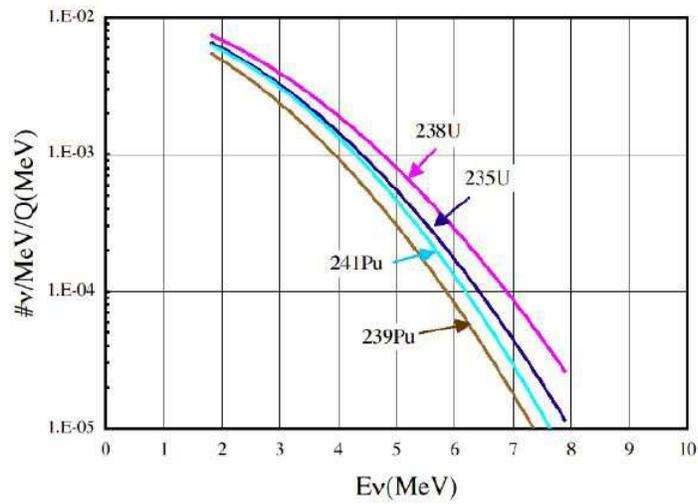}
\caption{Neutrino spectra for main fission elements}
\label{fig:nue_spectra}
\end{figure}

In reactors, the power is generated in fission reactions of
fissile elements such as uranium and plutonium.
An example of fission reaction is shown below.
\begin{equation*}
^{235}\textrm{U}+n \rightarrow ^{140}\textrm{Te}+^{94}\textrm{Rb}+2n. 
\end{equation*}
The fission products are generally neutron rich nuclei and perform several $\beta$ decays
before they become stable nuclei. 6~$\beta$-decays take place in average per fission, which means 6~$\bar{\nu}_e$'s are produced per fission along with 200~MeV of energy release, resulting in production of $5 \times 10^{21} \bar{\nu_e} $ per second for Kashiwazaki-Kariwa nuclear power station.  \\
The component of reactor fuel changes along with 'burn-up' through neutron absorption and beta decay processes as well as  the fission reactions. 
Below shows an example of plutonium production scheme. 
\begin{eqnarray*}
 \begin{split}
  &^{238}\textrm{U}+n   \rightarrow ^{239}\textrm{U} \\
  &^{239}\textrm{U}       \rightarrow ^{239}\textrm{Np}+e^-+\bar{\nu_e} \\
  &^{239}\textrm{Np}     \rightarrow ^{239}\textrm{Pu}+e^-+\bar{\nu_e}\\  
  &^{239}\textrm{Pu}+n  \rightarrow ^{240}\textrm{Pu}\\
  &^{240}\textrm{Pu}+n  \rightarrow ^{241}\textrm{Pu}\\
 \end{split}
\end{eqnarray*}
$^{239}$Pu and $^{241}$Pu produced are also  fissile elements and contribute to the power and neutrino generation. 
$^{238}$U  performs fission reaction after absorbing energetic neutrons.
Fig.~\ref{fig:time_evolution} shows an example of change of the fuel components caused by the burn-up effect for the case of constant neutron flux. 
$^{235}$U, $^{238}$U, $^{239}$Pu and $^{241}$Pu contribute more than 99.9\% of the fission reactions. 
Fig.~\ref{fig:nue_spectra} shows the neutrino spectrum for the four main components\cite{nu_spectrum}.
Because the neutrino spectra depends on fission element,  the energy spectrum of the neutrinos from reactor changes along with the burn-up.  
The fuel exchange takes place approximately once per year, in which 1/4 of old fuel rods are replaced to new ones.
The fuels of all the  Kashiwazaki-Kariwa reactors are in equilibrium stage,  that is 1/4 of the fuel is new and 1/4 of the fuel is 1 year old, 1/4 of the fuel is 2 years old and the rest is 3 years old . 
Thus the component ratio of the fuel and neutrino spectrum is stabilized in time.

In liquid scintillator, $\bar{\nu_e}$'s are detected through inverse
$\beta$ decay reaction with protons,

\begin{eqnarray*}
\bar{\nu_e} + p \rightarrow  e^+ + n & 
\end{eqnarray*}

The energy threshold of this reaction is 1.8~MeV and the energies of the $\bar{\nu_e}$'s
produced in the $\beta$-decay of U, Pu, Np  and neutron produced in the core are too low to
perform the reaction.  
Because a typical kinetic energy of recoiled neutron is only a few tens of KeV, the kinetic energy of the positron is $\bar{\nu_e}$ energy minus 1.8~MeV. 
The positron annihilates with electron and produces two 0.511~MeV $\gamma$'s. 
Although the lifetime of ortho positronium is 140~ns,  the pick-off annihilation takes place within a few nano second~\cite{PickOff}. 
Finally, the visible energy of the event  becomes,
\begin{equation}
 E_{vis}^{e^+}=E_{\nu}-0.78\mbox{MeV}
\end{equation}
with a good accuracy. 
Therefore the neutrino energy and visible energy can be directly connected and the modulation of neutrino energy spectrum due to
neutrino oscillation can directly be measured, unlike $\nu e$ elastic scattering case. \\
Because the reaction is the inverse process of the neutron beta decay, the cross section is closely related to the neutron lifetime $\tau_n$,
\begin{equation}  
 \sigma_{\nu p}=\frac{2\pi^2E_ep_e}{1.7152m_e^5\tau_n} 
 \approx 9.6\times 10^{-44}E_e(\mbox{MeV})p_e(\mbox{MeV/c}) [\mbox{cm}^2]
\end{equation}
where  1.7152 comes from the phase space factor, including Coulomb, weak magnetism, recoil and outer radiative corrections.  
Since the $q^2$ of the reaction is very small compared with nucleon mass scale, the correction is also small and it is known to the precision of 0.2\% \cite{Xsection}.

\begin{figure}[htbp]
 \begin{center}
  \includegraphics[width=0.8\textwidth] {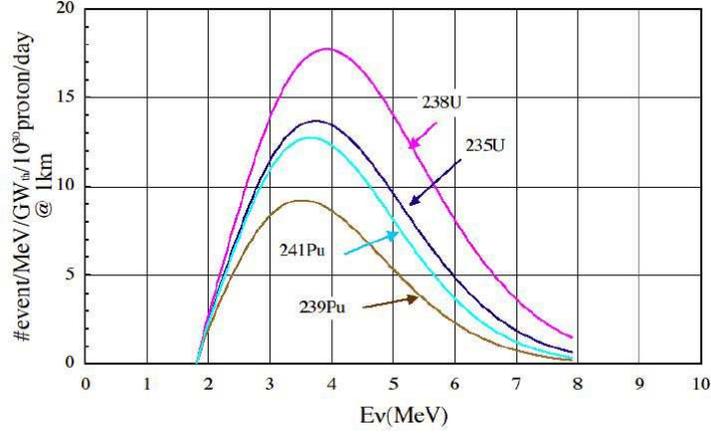}  
 \end{center}
\caption{Energy spectrum for detected neutrino ($f_{\nu} \sigma_{\nu p}$). 
 A simple parametrization of reactor neutrinos is used~\cite{nu_flux_parametrize}. }
\label{fig:event_spectra} 
\end{figure}

The visible energy spectrum in case no neutrino oscillation is calculated as,
\begin{equation}
\mu(E,t)=N_p \frac{P(t)}{4\pi L^2}\sigma_{\nu p}(E)\sum_i f_i(E) n^i(t)
\end{equation}
where $P(t)$ is the power generation of the reactor at time $t$, 
$L$ is the distance between the detector and the reactor,
$ \sigma_{\nu p}(E)$ is the inverse beta decay cross section, 
$f_i(E)$ is the neutrino spectrum per fission of element $i$, 
$ n^i$ is the fission rate per unit power generation, and 
$N_p$ is the number of target protons.
Due to the burn-up effect $n^i$ is a function of time. 
Fig.~\ref{fig:event_spectra} shows the $\bar{\nu}_e$ energy spectrum of the $\bar{\nu}_e p$ signals from the four main elements calculated using data from ref~\cite{nu_flux_parametrize}.  
A typical energy of the detected reactor neutrino is around 4~MeV and typical visible energy is 3~MeV. 
The ratio of event rate is $^{235}$U: $^{238}$U: $^{239}$Pu:$^{241}$Pu$\sim$1:1.39:0.62:0.88.
Therefore, when calculating expected number of events, it is necessary to know the
fission rate  of each fissile element in the reactor cores. 
Typically the uncertainty of the expected neutrino event rate due to the uncertainty of the fuel components is less than 1\%. 
The neutrino spectrum from each fissile elements are known in such a way that the neutrino detection rate agrees with data at 2.5\% of accuracy~\cite{nu_spectrum}. 
The thermal power generation will be provided from the rector operating company.
The error of the thermal power measurement is mainly caused by the flow meter of the cooling water, which is 2\% or less. 
In total, the absolute error of the expected neutrino event is less than 3.4\%. 
However, most of the errors are common for near and far detectors and will cancel out when both data are compared
and relative error on oscillation measurement caused from reactor neutrino flux uncertainty is expected to be 0.2\% as described in Section~\ref{sec:systematics}.

%-----------------------------------------------
\subsection{General Description of Neutrino Detector}
\begin{figure}[htbp]
 \begin{center}
  \includegraphics[width=0.6\textwidth] {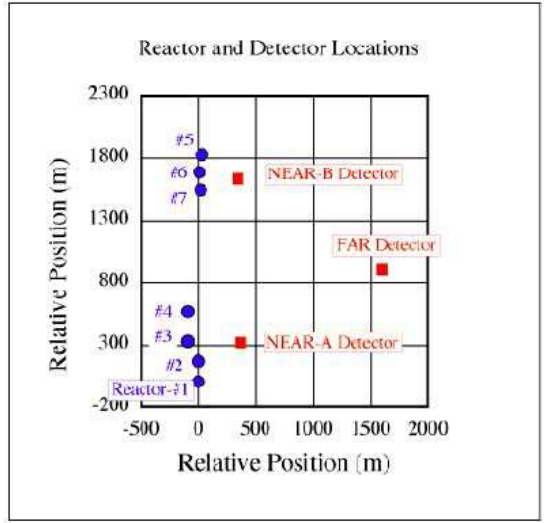}  
 \end{center}
\caption{Reactor and detector relative positions.}
\label{fig:reactor_position}
\end{figure}

In KASKA experiment, four neutrino detectors with an identical structure will be used. 
Two near detectors are placed in shaft halls at approximate distances of 350$\sim$400~m from respective
reactor cluster and two far detectors are placed in a shaft hole at a distance of $\sim$1.6~km from all the reactors.  
Fig.~\ref{fig:reactor_position} shows relative reactor and detector positions. 
By comparing the data of near and far detectors, most of the systematic errors
such as detection efficiency and neutrino flux will cancel out as will be described in Section~\ref{sec:systematics}.  
Table~\ref{tab:detector_baseline} shows event and cosmic-ray rate parameters for  KASKA and CHOOZ detectors.
\begin{table}[htbp]
\caption{Baseline, event rate and cosmic-ray background ratio.
The reactor operation, event selection and live-time efficiencies reduce neutrino signal rate roughly half. }
\begin{center}
\begin{tabular}{|r|c|c|c|c|c|c|c|}
\hline
\                     & target            &  Average  & Thermal           & Event           &               & $\nu/\mu$\\
 Detector       & mass             & Baseline   & Power               & rate              &  Depth   &  ratio     \\
{}                 &  (ton)            &    (km)       & (GW$_{th})$  &  (/day)         & (mwe)    & \\
 \hline
 KASKA-A &  6                  &   0.4           & 13.2                 &  520             & 90           & 1.4    \\
                -B &  6                  &   0.35         & 11.1                & 570              & 90           & 1.6 \\
                -C &  6$\times$2   &   1.6          & 24.3                 & 60$\times$2 & 260        & 1\\
 \hline
 \hline
 CHOOZ     &  5                   &   1.07         & 8.5                  & 40                 & 300         &0.9\\
 \hline
\end{tabular}
\end{center}
\label{tab:detector_baseline}
\end{table}
Although the KASKA-far baseline is significantly longer than CHOOZ one and the depth is shallower, the ratio of  neutrino flux and cosmic-ray rate becomes better  thanks to the high power of the power station.  
Moreover, KASKA detectors have thicker buffer layers and the background is expected to be significantly improves from CHOOZ. 

\begin{figure}[htbp]
 \begin{center}
  \includegraphics[width=0.9\textwidth] {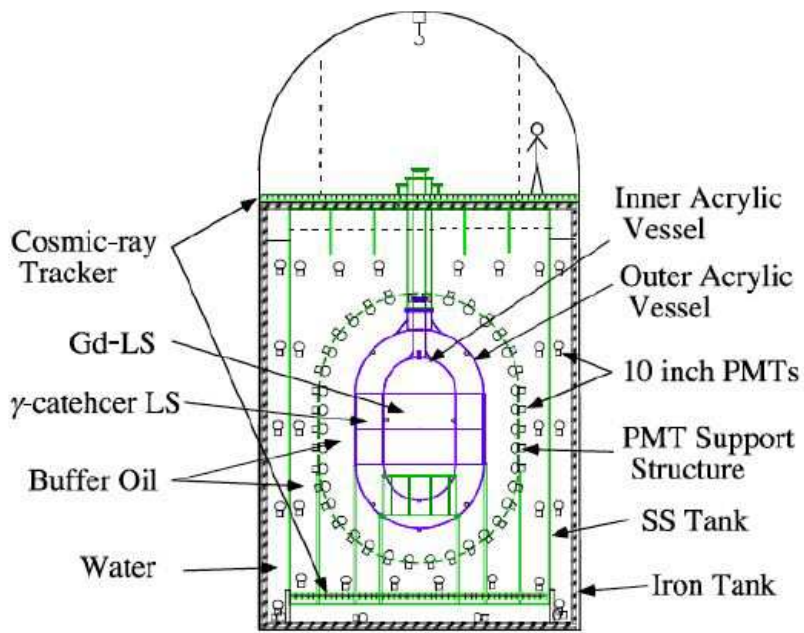}  
 \end{center}
\caption{Schematic of the KASKA detector.}
\label{fig:detector}
\end{figure}

\begin{figure}[htbp]
 \begin{center}
  \includegraphics[width=0.6\textwidth] {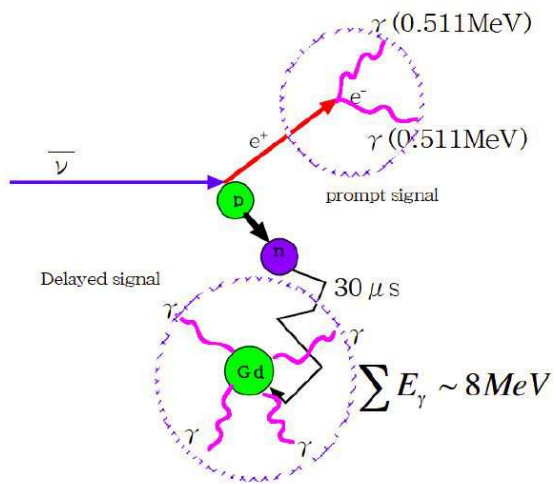}  
 \end{center}
\caption{Schematic of neutrino signal}
\label{fig:delayed_coin}
\end{figure}

Fig.~\ref{fig:detector} shows a schematic view of the detector. 
The central part (Region-I) is 6 tons of Gadolinium loaded liquid scintillator (Gd-LS). 
A candidate  LS is a PaloVerde type~\cite{paloverde:GDLS}, which was successfully used in acrylic containers. 
The liquid scintillator is pseudocumene and mineral oil base with 0.1~\% Gd concentration, with specific gravity 
of $0.8~\mbox{g/cm}^2$ and H/C ratio of 1.8. 
Fig.~\ref{fig:delayed_coin} shows a schematic diagram of the neutrino reaction.  
The reactor $\bar{\nu}_e$ is detected in the Gd-LS with the inverse $\beta$-decay process.
\begin{equation}
\bar{\nu}_e + p \rightarrow e^+ + n
\end{equation}
The positron has kinetic energy of $E_\nu-1.8$~MeV and eventually annihilates with an electron, producing two 0.5~MeV $\gamma$'s as described in the previous subsection. 
The neutron produced has energies of $O(10~\mbox{KeV})$ and is thermalized quickly by colliding with protons and absorbed by Gd. 
The Gd produces 8~MeV cascade $\gamma$-rays after the neutron absorption. 
\begin{equation}
  n+\textrm{Gd} \rightarrow \textrm{Gd'} +\gamma s (\Sigma E=8MeV)
\end{equation}
Because energies of $\beta$ and $\gamma$ rays of natural radio isotopes are lower than 5~MeV, 
the neutron signal is free from such backgrounds. 
The neutron absorption signal occurs typically 30~$\mu s$ after the positron signal.
The $\bar{\nu}_e$ event is identified by requiring that both signals occur within 200~$\mu s$.
With this delayed coincidence technique, the backgrounds are severely suppressed.

\begin{figure}[htbp]
 \begin{center}
  \includegraphics[width=1.0\textwidth] {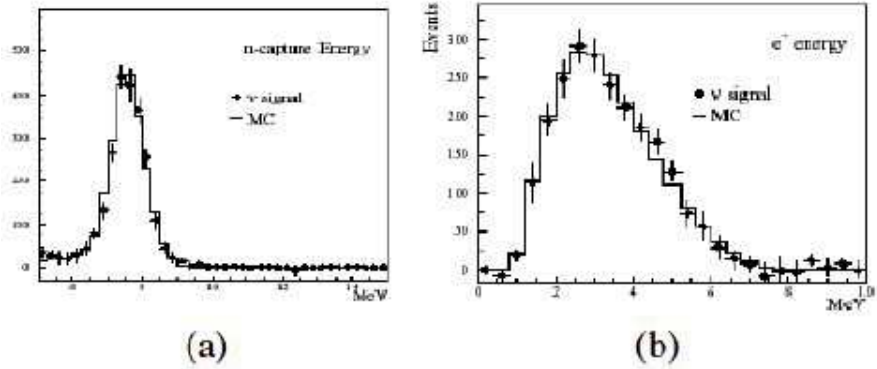}  
 \end{center}
\caption{Energy Spectrum of (a) Gd signal and (b) positron signal~\cite{CHOOZ}.}
\label{fig:energy_spectrum}
\end{figure}

The Gd-LS is contained in an UV-transparent acrylic vessel with a shape of two hemispheres sandwiching a cylinder. 
The radius of the hemisphere and cylinder is 0.9~m and the height of the cylinder is 1.8~m. \\
There is a 70~cm thick unloaded liquid scintillator layer (Region-II) outside of the
Region-I.  
This region is used to catch the $\gamma$-rays which escape from
Region-I, and to reconstruct the original energies of Gd and
positron signals.  
Fig.~\ref{fig:energy_spectrum} shows the energy spectra of the Gd and positron signals measured by CHOOZ experiment which also has 70~cm $\gamma$-catcher region. 
A clear peak is observed in the Gd signal. 
Because the event rate at the energy threshold (=5~MeV) is small the dependence of the efficiencies on the cut parameter is small. 
This LS will be formulated by mixing PC, paraffin oils and flours, and light output and specific gravity are adjusted to same as those of Gd-LS. 
The region-II scintillator is contained in a UV transparent acrylic vessel with the hemisphere-cylinder structure  whose radius and cylinder height are 1.6~m and 1.8~m, respectively.\\
There is a buffer oil layer outside region-II which shields liquid scintillators from $\gamma$-rays  and neutrons  coming from outside.  
A PMT support structure divides the buffer oil at 90~cm distance from the region-II acrylic vessel. 
The inner part (Region-III)  mainly works as $\gamma$-ray shield coming  from the PMT.
The outer buffer part (Region-IV) is used as a $\gamma$-ray and neutron shield from detector walls and outer soils, and works as a cosmic-ray veto.
In order to improve the cosmic-ray veto efficiency, the buffer oil may be made of a weak scintillator having 10\% of light output of Gd-LS.
The base oil will be a paraffin oil having a similar specific gravity as region-II scintillator. 
These oils are contained in a stainless steel tank with a diameter 6.5m and shell thickness 15~mm.\\ 
There are 300 10-inch low background PMTs looking at the scintillator region, whose photo cathode covers 10\% of the surface area. 
Some LEDs and CCD cameras are also arranged in the tank which are used to monitor the position of calibration devices. \\
Outside of the stainless steel tank is  pure water. 
Water proof PMTs are submerged in the water and work as Cherenkov cosmic-ray veto counter. 
The water is contained in a hermetic cylindrical iron tank. 
This tank also works as shield for geo-magnetic field. 
Outside of these tanks, there are thick iron layers to shield gamma rays
from the soils outside.  
This structure has a leak-safe effect.  That is, if a crack should develop in the stainless steel tank, the water outside would come in the oil region and the oil would not come out. 

There are cosmic-ray tracking devices at the top and bottom of the detector. 
The measured cosmic-ray track is used to estimate spallation background
such as $^9$Li and $^8$He based on the relation of the position of the
events and the track.  

In order to reduce the cosmic-ray background, the detectors are placed in deep shaft holes
with a diameter of 5~m.
The depths are at least 150~m  for far detector 50~m for two near detectors.

%================================
\subsection{Sensitivities of  KASKA}

%---------------------------------------
\subsubsection{$\sin^22\theta_{13}$ measurement}
The expected size of the relative systematic error $\sigma_{\text{rel}}$
on the near-far comparison of KASKA experiment
is better than 1\%, possibly be 0.5\% level.
Since the far detectors will collect 70K reactor $\bar{\nu}_e$ with 3 years of data taking and near detectors will 
collect 10 times more data, the statistic error becomes 0.4\%. 

\begin{figure}[htbp]
 \begin{center}
  \includegraphics[width=0.6\textwidth, angle=-90] {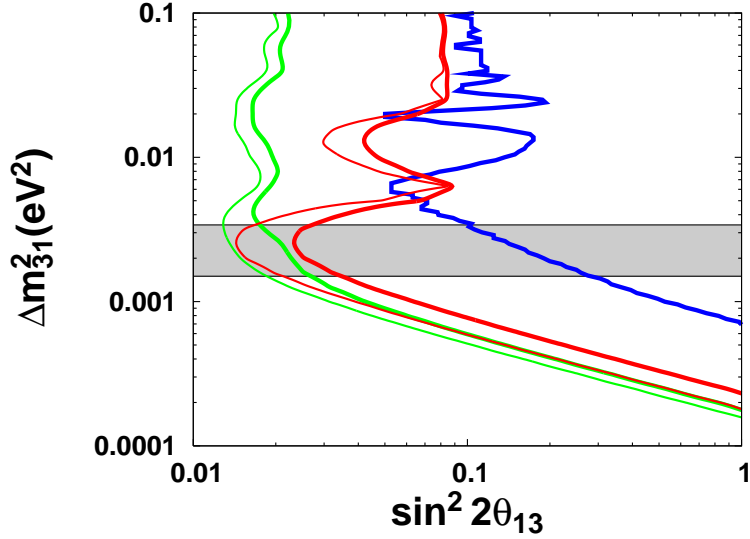}  
 \end{center}
\caption{90\% CL sensitivity to $\sin^2{2\theta_{13}}$ in KASKA is
shown with the analysis for 1 degree of freedom.
 Red curves are of the rate analysis,
and thin and bold ones are obtained with  0.5\% and 1\%
relative errors in the near-far comparison, respectively.
 2.5\% error on the flux is included in the analysis also.
 Green curves are of the rate+shape analysis with 14 bins
within $1.8\mbox{MeV}\leq E_\nu \leq 8.8\mbox{MeV}$.
 $0.5\%$ ($1\%$) uncorrelated relative error between detectors
is assumed to be divided into $0.5\%$ ($1\%$)
uncorrelated error between bins and $0.5\%$ ($1\%$)
correlated one. 
The similar assumption is used for
2.5\% error on the flux also.
 The blue curve shows the bound by the CHOOZ experiment
with the analysis for 2 degrees of freedom.
 The shaded region corresponds to 90\%CL allowed region
by the atmospheric neutrino measurement in SK-I:
$1.5\times 10^{-3}\mbox{eV}^2\leq \Delta m^2
\leq 3.4\times 10^{-3}\mbox{eV}^2$}
\label{fig:sensitivity}
\end{figure}

The expected sensitivities for rate only and rate + shape analysis are shown in the Fig.~\ref{fig:sensitivity}.
If no deficit is observed, it is possible to set upper limit on $\sin^22\theta_{13}$ at 0.015 to 0.020. 
This sensitivity is 10 times better than the CHOOZ limit. 
The sensitivity at $\Delta m^2 > 0.01 \mbox{eV}^2$ comes from test of the deficit of $\bar{\nu}_e$ flux measured by the near detectors.
Because the absolute systematic error  and statistics error are smaller than previous experiments a sterile neutrino search can be performed in this region.  

\begin{figure}[htbp]
 \begin{center}
  \includegraphics[width=0.5\textwidth, angle=-90] {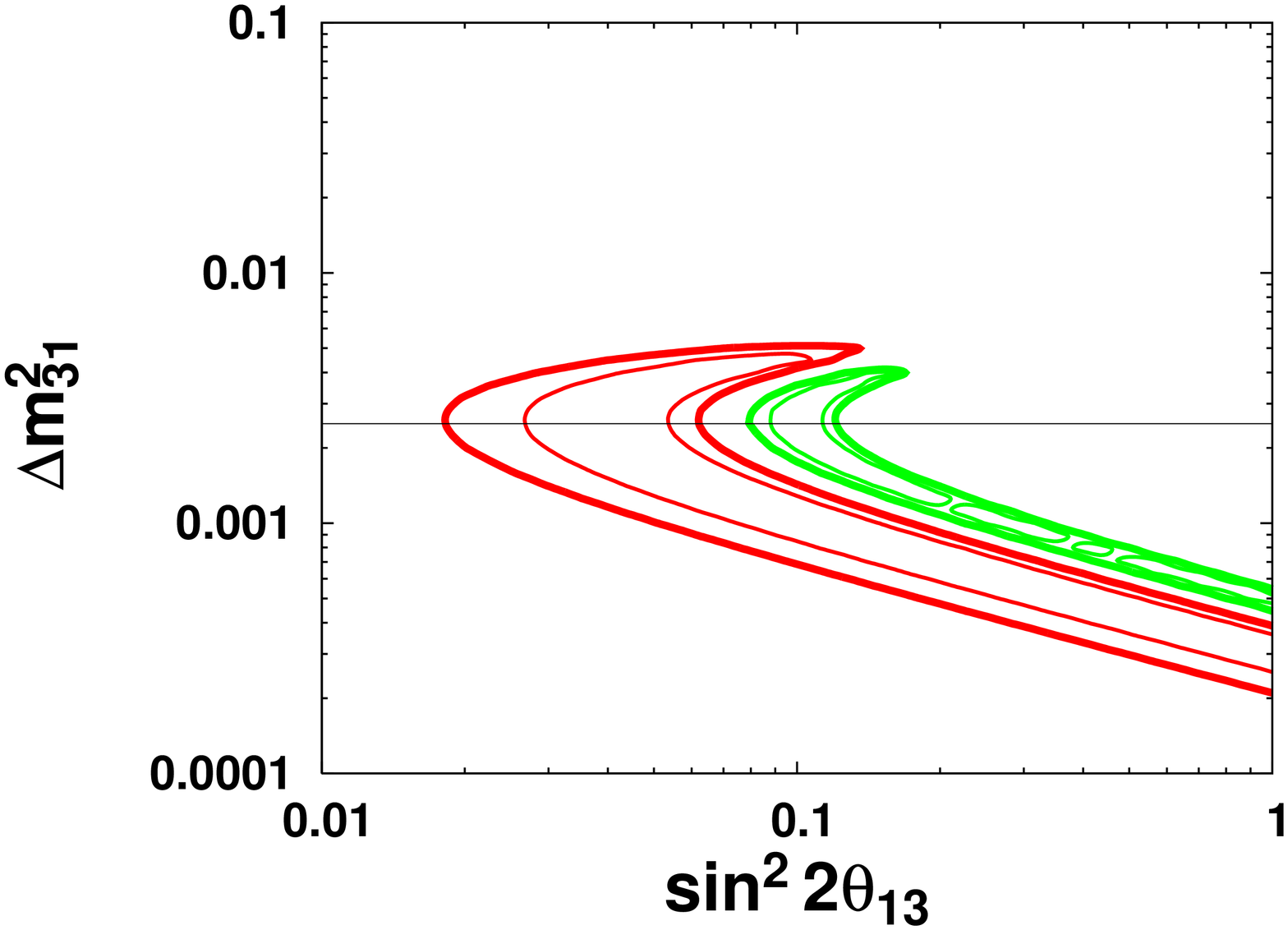}
  \end{center}
    \caption{Showns are the constraints on $\sin^22\theta_{13}$
to be obtained by the rate analysis in the KASKA experiment.
 The true value of $\sin^22\theta_{13}$ is assumed to be 0.04~(red) or 0.1~(green).
 $\Delta m^2_{31}$ is assumed to be known precisely by
the future long baseline experiments
as $\Delta m^2_{31}=2.5\times 10^{-3}\text{eV}^2$.
 The thin and bold curves are obtained by
$\sigma_{\text{rel}}=0.05\%$ and 0.1\%, respectively.}
\label{fig:positive_rate}
\end{figure}

\begin{figure}[htbp]
 \begin{center}
\includegraphics[width=0.5\textwidth, angle=-90] {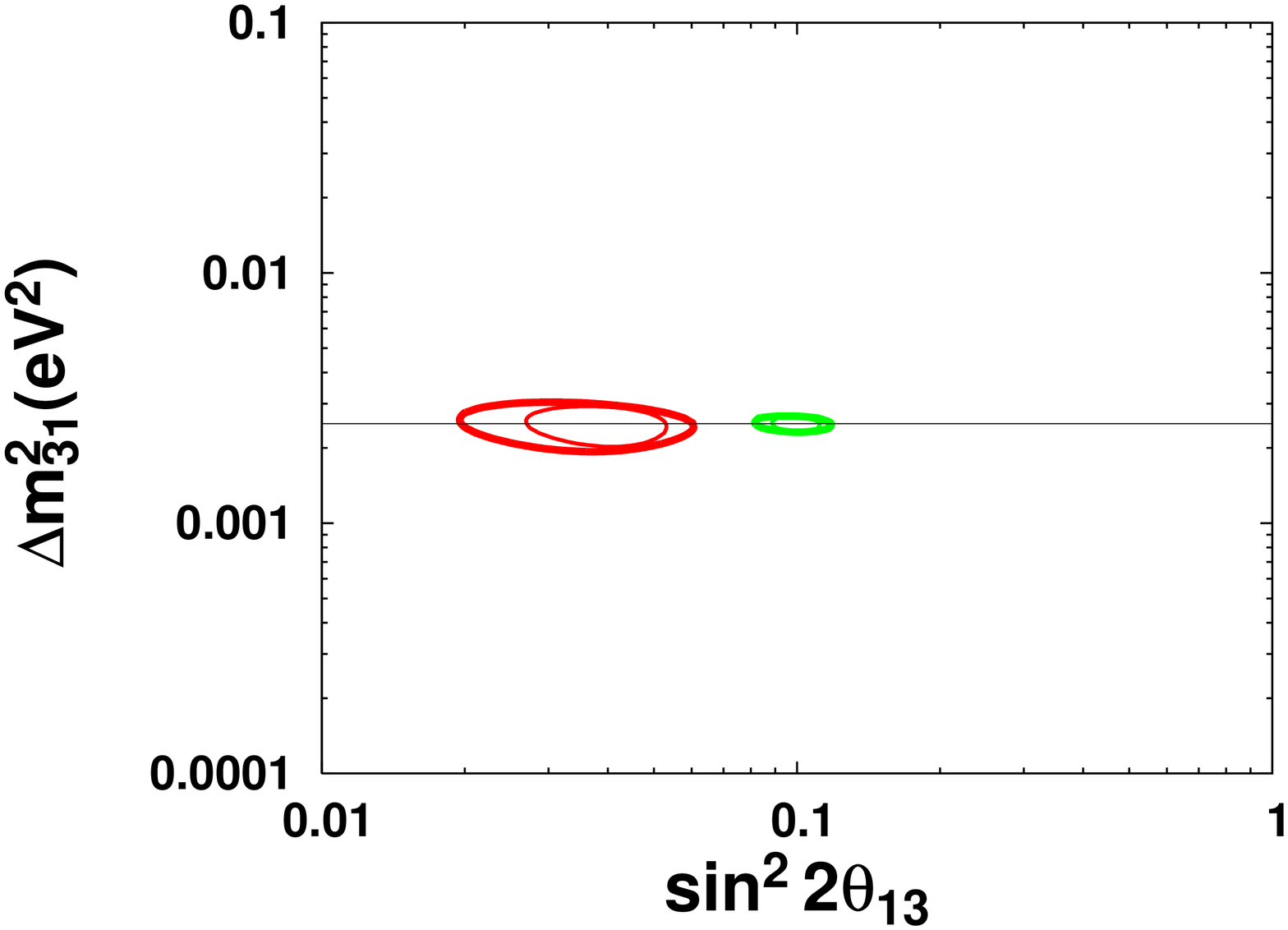}
\caption{Showns are the constraints on $\sin^22\theta_{13}$
to be obtained by the rate+shape analysis in the KASKA experiment.
 The true value of $\sin^22\theta_{13}$ is assumed to be 0.04~(red) or 0.1~(blue).
 $\Delta m^2_{31}$ is assumed to be known precisely by
the future long baseline experiments
as $\Delta m^2_{31}=2.5\times 10^{-3}\text{eV}^2$.
 The thin and bold curves are obtained by the different sizes
of systematic errors which are the same as the values
used in Fig.~\ref{fig:positive_rate}.
}
 \end{center}
\label{fig:positive_shape_rate}
\end{figure}

If the true value of $\sin^22\theta_{13}$ is 0.04 or 0.1,  for example, 
and future long baseline experiments show
$\Delta m^2_{31} = 2.5\times 10^{-3}\text{eV}^2$,
rate-only analysis constrains $\sin^22\theta_{13}$
with  1 degree of freedom (d.o.f.) analysis
as shown in Fig.~\ref{fig:positive_rate}. 
Results with rate+shape analysis are shown
in Fig.~\ref{fig:positive_shape_rate}.
 Note that the width of the region has the meaning on the line of
$\Delta m^2_{31} = 2.5\times 10^{-3}\text{eV}^2$ only
because of 1 d.o.f. analysis. 

If a finite $\sin^22\theta_{13}$ is observed by KASKA experiment it means
that the parameter is reasonably large and there is a good chance to observe leptonic CP violation parameter 
$\delta_l$ from asymmetry of the probability of $\nu_{\mu} \rightarrow \nu_e$ and $\bar{\nu}_{\mu} \rightarrow \bar{\nu}_e$ appearances in future accelerator based long baseline experiment.  
If $\sin^22\theta_{13}$ is too small to be detected by KASKA,
it means it is difficult to observe $\delta_l$ and more sensitive experiments to pin down $\sin^22\theta_{13}$ 
will be necessary before proceeding to accelerator based $\delta_l$ experiments. 

%---------------------------------------
\subsubsection{High statistics reactor $\bar{\nu}_e$ measurement}

The KASKA near detectors will collect 600,000 reactor $\bar{\nu}_e$ events. 
This will be the highest statistics of reactor $\bar{\nu}_e$ detection. 
Table~\ref{tab:high_statistics} compares the statistics and detector related absolute uncertainties of various
experiments. 
The absolute size of the detector related systematic error in KASKA is less than 1.5~\% thanks to the no fiducial volume cut and low backgrounds. 
Therefore the data taken by KASKA will become standard neutrino $\bar{\nu}_e$ spectrum and contribute to other
reactor $\bar{\nu}_e$ experiments. \\

\begin{table}[htbp]
\begin{center}
\scalebox{0.9}[0.9]
{
\begin{tabular}{|l|r|r|r|r|}
\hline
                      &                      &                               & Detector related &  \\
Experiments  & \# events (N) & 1/$\sqrt{N}$ (\%) & Syst. error (\%) & Distance(m)\\
\hline
KASKA Nears  & 600,000 & 0.13 & $<1.5$ & 350, 400\\
CHOOZ \cite{CHOOZ} &     2,700 &   1.9 & $1.8 $ & 1050 \\
PaloVerde \cite{PaloVerde} & 1,700   &  2.4      &   4.9    & 750, 890 \\
Bugey \cite{Bugey}  &   150,000&   0.26  &  4.0 & 15, 40, 95\\
Goesgen \cite{Goesgen}  &   30,000&   0.57  &  6 & 37.9, 45.9, 64.7\\
\hline
\end{tabular}
}
\caption{ Comparison of high statistics measurements of the reactor $\bar{\nu}_e$ flux.}
\label{tab:high_statistics}
\end{center}
\end{table}

A special situation of KASKA is that because 25~\% of KamLAND $\bar{\nu}_e$ come from Kashiwazaki-Kariwa 
nuclear power station, the KASKA near detectors work as near detectors for KamLAND and neutrino flux related
error of KamLAND can partially be cancelled. 
This effect improves $\sin^22\theta_{12}$ error, together with removal of uncertainty of the $\cos^4\theta_{13}$ term
by $\sin^22\theta_{13}$ measurement.

%==================================
\subsection{Possible Future Extensions of the KASKA project}
Because Kashiwazaki-Kariwa nuclear power station generates the highest $\bar{\nu}_e$ luminosity  
in the world, there are a number of opportunities to perform neutrino studies with the best precision.
%-----------------------------------
\subsubsection{KASKA-II: Detecting non-0 $\delta_l$}

\begin{figure}[htbp]
 \begin{center}
  \includegraphics[width=0.6\textwidth] {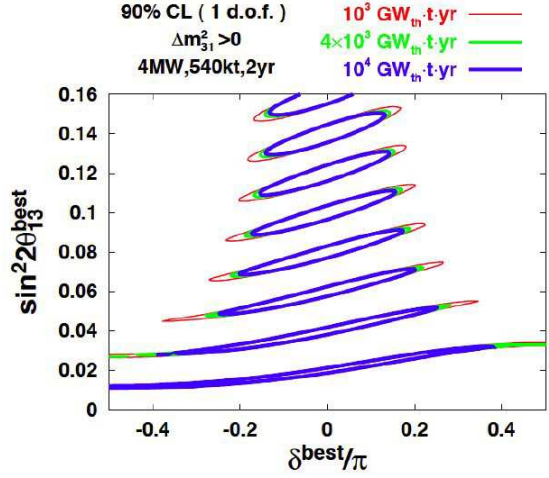}  
 \end{center}
\caption{$\sin\delta_l$ sensitivity by the combination of KASKA-II,
Super-JHF (4MW), and Hyper Kamiokande (540kton fiducial mass)~\cite{Minakata:delta}.
 The curves show the regions in which data mimics $\delta_l = 0$.
 If true values of $\theta_{13}$ and $\delta_l$ exist outside of
all of these regions, we can find that $\delta_l$ has non-zero value.
The sign of $\Delta m^2_{31}$ is assumed to have been determined by then. 
 }
\label{fig:reactor_LBL_delta}
\end{figure}

In order to proceed to an experiment with errors better than 0.5\% level,
12 tons of the far detector mass is not enough. 
A 50 tons of detectors and deeper shaft holes are under consideration for a next generation KASKA experiment (KASKA-II).
By combining KASKA-II and data from Super-JHF or Hyper Kamiokande, there is a chance to identify non-0 $\sin \delta_l$ 
before proceeding to $\bar{\nu}_{\mu}$ beam if $|\sin \delta_l|$ is reasonably large. 
Fig.~\ref{fig:reactor_LBL_delta} shows the regions in which $\sin \delta_l$ is consistent with 0.
Therefore, it is possible to identify non-0 $\sin \delta_l$
if the nature chooses the values outside of these regions
such as $\sin^22\theta_{13}>0.05$ and $|\delta_l|>0.3\pi$~\cite{Minakata:delta}.
%

%---------------------------------------
\subsubsection{KASKA-$\theta_{12}$: Accurate $\sin^22\theta_{12}$ measurement}

 $\sin^22\theta_{12}$ governs oscillation of the electron type neutrinos at large L/E range.
Natural neutrino sources, such as the sun, the earth, supernovae, etc.\ are sources of electron type neutrinos.
Although $\sin^22\theta_{12}$ has been measured by KamLAND
and solar neutrino experiments, further precision measurement is important
because it can be used as an inevitable tool to investigate
inside of the sun etc.

\begin{figure}[htbp]
\begin{center}
 \includegraphics[width=0.5\textwidth, angle=-90] {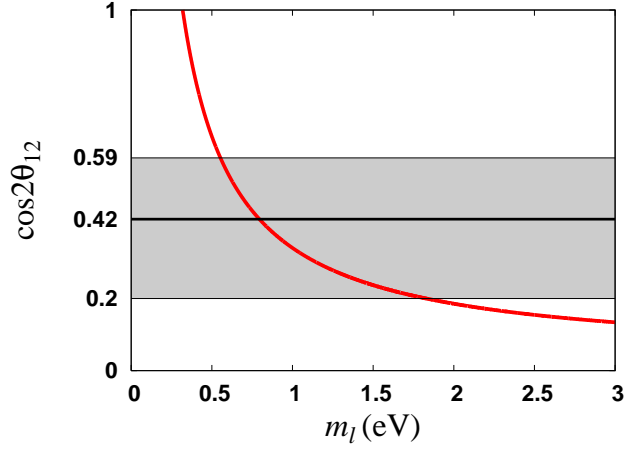}
\end{center}
\caption{The figure shows the relation between $\cos2\theta_{12}$
and the upper bound on the mass of the lightest neutrino
for a given upper bound (0.1eV is used as an example)
on the effective neutrino mass of the neutrino-less double beta decay.
 The shaded region corresponds the 99.73\%CL allowed region
of $\cos2\theta_{12}$ obtained by the combined analysis
with solar neutrino and KamLAND results
under the assumption of CPT invariance.}
\label{fig:double_beta}
\end{figure}

 Another importance of the precise measurement of $\sin^22\theta_{12}$
appears in the relation between the absolute mass of neutrinos
and the neutrino-less double beta decay searches;
 The decay is possible for the Majorana neutrino
but impossible for the Dirac neutrino.
 The absolute mass can not be determined by oscillation experiments
because oscillations depend on $\Delta m^2_{ij}\equiv m_i^2-m_j^2$
but are independent of the absolute mass.
 The information on the absolute mass can be obtained
by the Cosmological Microwave Background measurements~\cite{CMB},
the tritium beta decay measurements~\cite{me},
and the neutrino-less double beta decay searches~\cite{doublebeta}.
 The upper bound on the effective neutrino mass
of the neutrino-less double beta decay
can be translated into the upper bound on the absolute mass
of Majorana neutrinos.
 The translation, however, depends on the value of
$\theta_{12}$ very much as we see in Fig.~\ref{fig:double_beta}.
 Thus, it is important to know the precise value of $\sin^22\theta_{12}$
in order to extract better information on the absolute mass
from the result of experiments of the neutrino-less double beta decay.
If we put KamLAND size detector (1~kton) at a distance $\simeq$50~km
from Kashiwazaki-Kariwa nuclear power station,
which corresponds to the oscillation maximum of $\Delta m_{12}^2$,
it is possible to measure $\sin^2 2\theta_{12}$ with the highest precision
compared with solar-pp neutrino experiments. 
The KASKA-$\theta_{13}$ detectors can be used as the precise near detectors. 
Table~\ref{tab:precise_th12} shows a table taken from the reference
~\cite{Minakata:th12} in which KASKA-$\theta_{12}$ is referred to as
the SADO experiment.
It shows that  with KASKA-$\theta_{12}$ configuration, it is possible to measure $\sin^2\theta_{12}$ with 2.4\% precision. 
In this measurement, background from geo-neutrino is a main issue to obtain the highest sensitivity because energy cut introduces large systematics. 
Note that large detector size does not help to reduce geo-neutrino background
and to obtain a good S/N ratio.
One merit of KASKA-$\theta_{12}$ is a good S/N ratio
by virtue of the high reactor power.
Another merit of KASKA-$\theta_{12}$ is that
the error on the estimation of the geo-neutrino background
can be reduced by the help of the KamLAND experiment 
which will have measured the geo-neutrino flux
of which large part of the sources is common.

\begin{table}
\scalebox{0.9}
{
\begin{tabular}{c|cc}
\hline
\ Experiments\  & \ $\delta s^2_{12}/s^2_{12}$  at 68.27\% CL \ 
                & \ $\delta s^2_{12}/s^2_{12}$  at 99.73\% CL \  \\
\hline
Solar+ KL (present)  & $ 8 $ \%  
                     & $ 26  $ \%   \\
\hline
 Solar+ KL (3 yr)  & $ 7 $ \% 
                   & $ 20 $ \%   \\
\hline
Solar+ KL (3 yr) + pp (1\%) &  $ 4 $ \%  
                 & $ 11$ \%  \\
\hline
\multicolumn{3}{c}{54 km}\\
\hline
 SADO  for 10 $\mbox{GWth} \cdot$kt$\cdot$yr  &  4.6 \% \   (5.0 \%)  
                 &  12.2 \%  \ (12.9 \%) \\
\hline
 SADO for 20 $\mbox{GWth} \cdot$kt$\cdot$yr
                 &  3.4 \% \  (3.8 \%)
                 &  8.8 \% \  (9.5 \%) \\
\hline
 SADO for 60 $\mbox{GWth} \cdot$kt$\cdot$yr
                 & 2.1 \%  \   (2.4 \%)
                 & 5.5  \% \   (6.2 \%) \\
\hline
\end{tabular}
}
\caption{
%Comparison with other $\sin^2 \theta_{12}$ measurements.~\cite{Minakata:th12}
}
\label{tab:precise_th12}
\end{table}

%\begin{figure}[htbp]
%\label{fig:precise_th12}
% \begin{center}
%  \includegraphics[width=0.9\textwidth] {precise_th12.eps}  
% \end{center}
%\caption{Comparison with other $\sin^2 \theta_{12}$ measurements.~\cite{Minakata:th12}}
%\end{figure}

%---------------------------------------
\subsubsection{KASKA-$\Delta m_{13}^2$: Independent $\Delta m_{13}^2$ measurement}

For the case of three generations neutrino,
we have $\Delta m_{13}^2 \approx \Delta m_{23}^2$. 
However it is an important item which has to be tested.
If a middle size detector is placed at around 5~km from Kashiwazaki-Kariwa nuclear power station, several cycles of oscillation can be observed in its energy spectrum. 
\begin{figure}[htbp]
\begin{center}
 \includegraphics[width=0.5\textwidth, angle=-90]
% {eps_bit/5km_positive_shape_rate.eps}
 {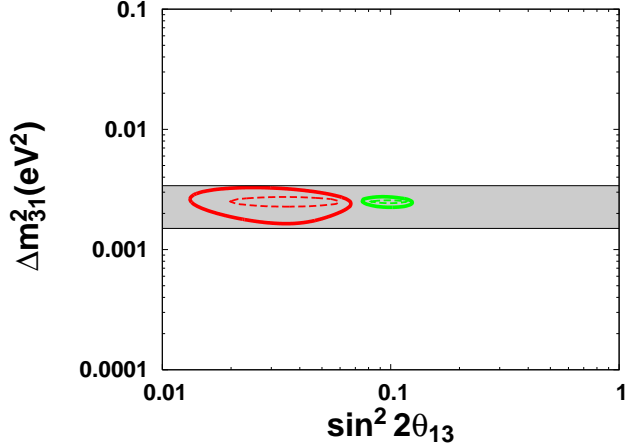}
\end{center}
\caption{The solid and dashed lines show the 90\%CL allowed regions
which can be obtained by the KASKA experiment
without and with an additional 120~ton detector
at 5~km from Kashiwazaki-Kariwa nuclear power station, respectively.
 The allowed regions are obtained in the analysis for 2 degrees of freedom
in order to extract the information on $\Delta m^2_{31}$.
 Red and green curves are for the case
that the true value of $\sin^2{2\theta_{13}}$ is 0.04 and 0.1,
respectively.
 The true value of $\Delta m^2_{31}$ is assumed to be
$2.5\times 10^{-3}\mbox{eV}^2$.
 The shaded region corresponds to 90\%CL allowed region
by the atmospheric neutrino measurement in SK-I:
$1.5\times 10^{-3}\mbox{eV}^2\leq \Delta m^2_{32}
\leq 3.4\times 10^{-3}\mbox{eV}^2$.
 }
\label{fig:5km_positive_shape_rate}
\end{figure}
It is possible to measure $\Delta m_{13}^2$ precisely
from the spectrum analysis of the modulation. 
Fig.~\ref{fig:5km_positive_shape_rate} shows the allowed regions
for the positive result in the KASKA experiment
with an additional 120~ton detector at 5~km
from Kashiwazaki-Kariwa nuclear power station.  
The precision of $\Delta m_{13}^2$ is comparable or better than that of the current $\Delta m_{23}^2$ measurement if 
$\sin^22\theta_{13}$ is relatively large. 
Again, the KASKA-$\theta_{13}$ data can be used as precise reference spectrum. 
\\

\begin{figure}[htbp]
\begin{center}
 \includegraphics[width=0.9\textwidth] {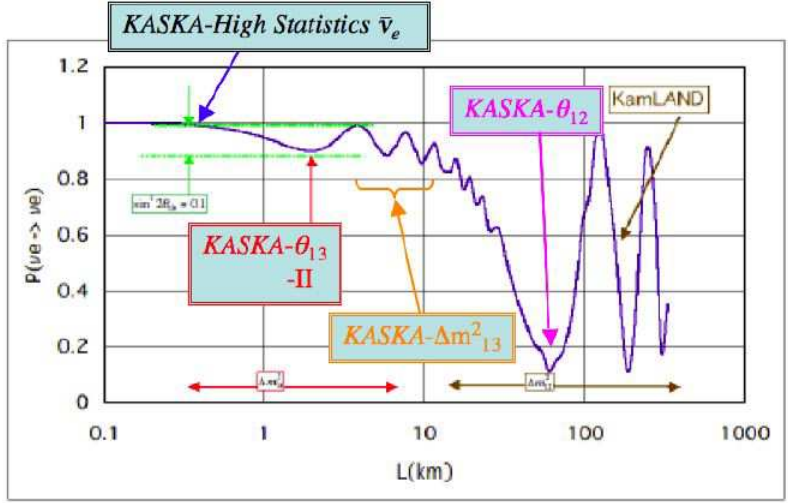}
\end{center}
\caption{Scope of KASKA experiments.  
High statistics measurements of reactor $\bar{\nu}_e$ spectrum with near detectors.  
High accuracy $\theta_{13}$ measurements at $L\sim1.6$~km.
Precise measurement of $\Delta m_{13}^2$ at $L\sim 5$~km and precise measurement of 
$\theta_{12}$ at $L\sim 50$~km.}
\label{fig:future}
\end{figure}

Thus, extensions of KASKA experiment make it possible to measure the oscillation parameters of electron type neutrinos  by exploiting the large power of Kashiwazaki-Kariwa nuclear power station as depicted in Fig.~\ref{fig:future}.
%
%----------------------------------------

%==================
\section{Detector}
\label{sec:design}

%---------------------------
\subsection{Mechanical Structure}
\begin{figure}[htbp]
 \begin{center}
  \includegraphics[width=0.6\textwidth] {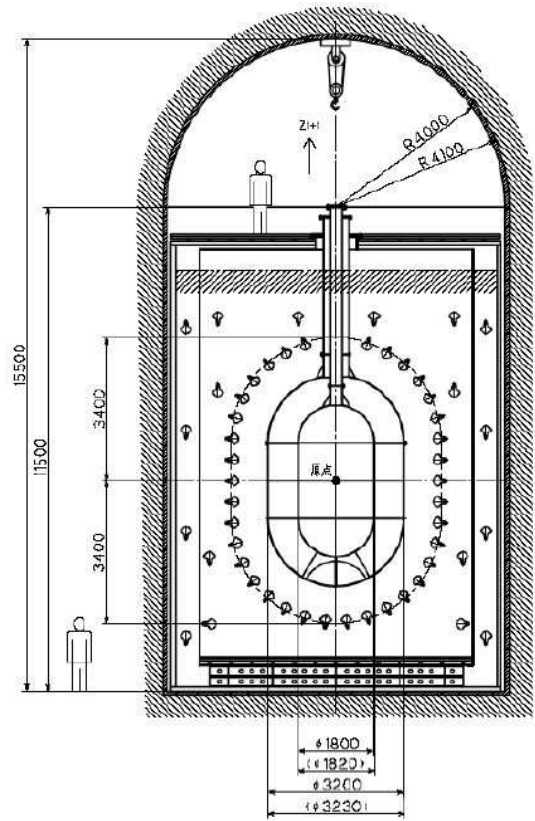}  
 \end{center}
\caption{Detector Structure of one unit.}
\label{fig:detector_structure}
\end{figure}

\begin{figure}[htbp]
 \begin{center}
  \includegraphics[width=0.8\textwidth, angle=-90] {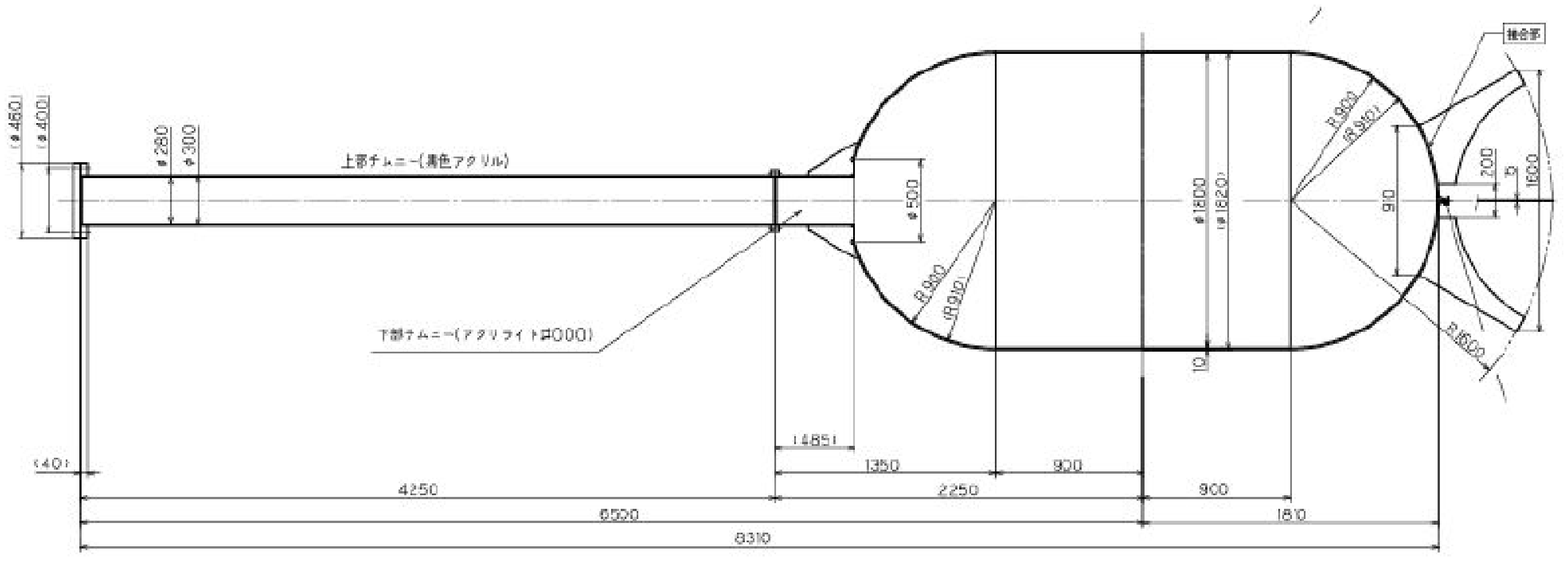}  
 \includegraphics[width=0.9\textwidth, angle=-90] {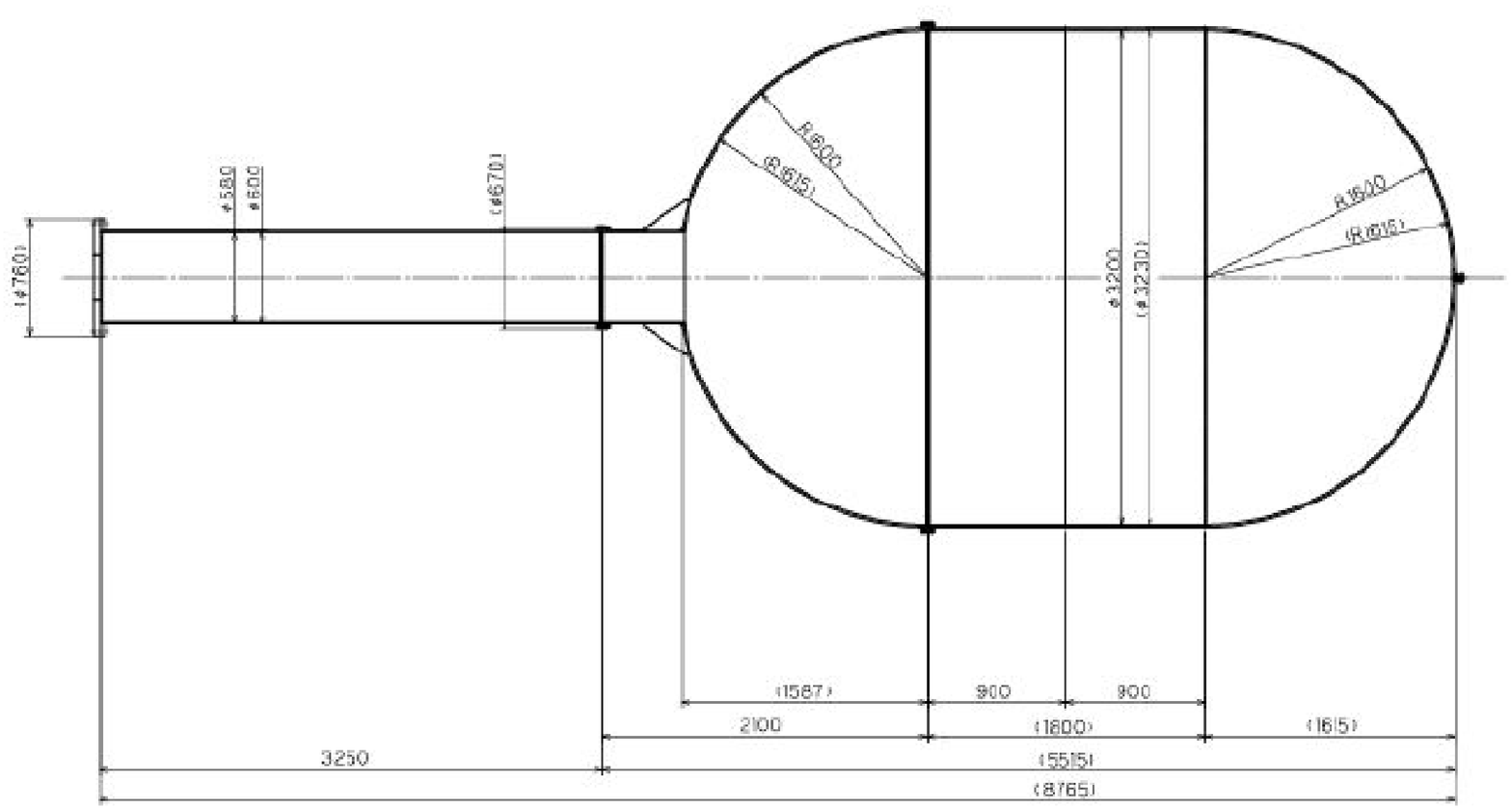}  
 \end{center}
\caption{Acrylic Vessels}
\label{fig:acrylic_vessels}
\end{figure}

Fig.~\ref{fig:detector} shows a schematic view of the KASKA detector and 
Fig.~\ref{fig:detector_structure} shows mechanical structure and size. 
The detector consists of inner acrylic vessel to hold the gadolinium loaded liquid scintillator, outer acrylic vessel to hold the $\gamma$-catcher
scintillator,  PMT support frame made of stainless steel, cylindrical stainless steel tank to hold the buffer oil and cylindrical iron tank to hold the water. 
The shape of acrylic vessels and PMT support frame consists of two hemispheres sandwiching a cylinder part. 
The radius of the hemisphere is 90~cm, 160~cm and 250~cm for inner acrylic, outer acrylic and PMT support, respectively. 
The height of the cylinder part is common, 180~cm.  
Fig.~\ref{fig:acrylic_vessels} shows the structure of the acrylic vessels.
The thickness of the acrylic is 1~cm for inner vessel and 1.5~cm for outer vessel. 
The acrylic is a UV transparent type. 
The weight of the acrylic is 0.24~ton for inner and 0.90~ton for outer. 
The inner volume is 7.6~m$^3$ for inner vessel and 32~m$^3$ for outer vessel. 
Compatibility of acrylic with liquid scintillator is being tested intensively. 
The inner acrylic vessel is made in a factory and transported to the experimental site and brought down to the detector hall.
The outer acrylic vessel is divided into two hemispheres and two 90~cm-high cylinders. 
Those units are carried  to the experimental site by a truck and brought down the shaft hole and assembled in the detector pit. 
There are  chimneys with 30~cm diameter (inner vessel) and 60~cm (outer vessel), which are to be used to put calibration devices in and to control the liquid level. 
The chimneys are made of black acrylic. 

The PMT support frame has a jungle-gym type structure. 
The number of PMT is $\sim$300 and 1.5~tons of buoyancy force of the PMTs will pull the structure up when the oil is full. 
The PMT support structure has counter weights to cancel this buoyancy force. 

The diameter of the stainless steel cylindrical tank is 6.5~m and the height is 990~cm and the the volume is 330~m$^3$.
The thickness of stainless steel cylindrical tank is 15~mm and its total weight is 32~tons. 
This tank holds 260~tons of oils. 
The tank has to strong enough to self-stand when the water around is empty while endure the buoyancy force of $\sim20$~ton when the water is full.  

The diameter of the cylindrical iron tank is 7.5~m and the height is 1,060~cm and the inner volume is 470~m$^3$. 
The thickness is 11~cm to shield $\gamma$-rays coming from outside.  
It is considered to make this iron tank as  geo-magnetic shield by de-magnetizing the iron.
Table~\ref{tab:structure_parameter} summarizes detector parameter structures. 
\begin{table}[htbp]
\caption{Detector Structure Parameters. }
 \begin{center}
  {\footnotesize
   \begin{tabular}{|c|r|r|r|r|r|}
    \hline
    {} &{}&{}&{}&{}&{}\\[-1.5ex]
    Item &  R & H & t & V & W\\[-1.5ex]
 {} &{}&{}&{}&{}&{}\\[-1.5ex]
   &  (cm)& (cm) & (cm) & (m$^3$) & (ton)\\[-1.5ex]
    {} &{} &{}&{}&{}&{} \\[-1.5ex]
    \hline
  \hline
    {} &{}&{}&{}&{}&{} \\[-1.5ex]
    Inner acrylic     & 90      & 180    & 1.0  & 0.20 & 0.24   \\[1ex]
    Outer acrylic     & 160    & 180   & 1.5  & 0.75 & 0.90   \\[1ex]
     PMT support  & 250    & 180    & -     &  -      & 2       \\[1ex]
    S.S. tank          & 325    & 990    & 1.5  & 4.0   & 32    \\[1ex]
      Iron tank          & 375   & 1060  & 11  & 37     & 290   \\[1ex]
       \hline
   \end{tabular}
     \label{tab:Detector_Parameter} }
 \end{center}
 R: Radius of cyliner/hemisphere, H: Height of cylinder/cylinder part, t: Thickness of the vessel, V: Volume of the structure material and W: Weight of the vessel.
 \vspace*{-13pt}
\label{tab:structure_parameter}
\end{table}

%==================
\subsection{Liquid Scintillators}

The KASKA detector uses  four kinds of liquid. 
From inner to outer, $\nu$ target scintillator (Region-I), $\gamma$ catcher scintillator (Region-II), 
buffer oil (Region-III, IV) and water for cherenkov cosmic-ray anti-counter (Region-V). 
Table~\ref{tab:LS_component} shows components of each candidate liquid.

\begin{table}[htbp]
\caption{Liquid Scintillator Components}
\begin{center}
\begin{tabular}{|c|l|l|l|}
\hline
Region& Mixture & Volume (m$^3$)   & Weight (t)  \\

 \hline
  I &  PC(20v\%)+TD(60v\%)+BC521(20v\%) & 7.6   & 6.1        \\
  II &  PC(15v\%)+TD(85v\%) &  24   & 19\\
     &    \hspace{10pt} +PPO(0.6w\%)+bis-MSB(0.01w\%) & &\\
  III+IV &  Paraffin  oil (+PPO) &    300 & 240 \\
  V &  pure water  &   160 & 160   \\

\hline
\end{tabular}
\end{center}
\label{tab:LS_component}
\end{table}

In the table, BC521 is a commercial liquid scintillator of a Saint-Gobain product~\cite{Saint-Gobain} with Gd concentration of 0.5\%. 
PC stands for  pseudocumene (1,2,4 tri-methyl-benzene (C$_6$H$_3$(CH$_3$)$_3$) ), TD stands for Tetradecane (C$_{14}$H$_{30}$).
 v\% means \% in volume and w\% means \% in weight.
Table~\ref{tab:Liquid_Property} shows properties of  these liquids along with the mixed ones.

 \begin{table}[htbp]
\caption{Liquid Properties}
\begin{center}
\begin{tabular}{|c|c|c|c|c|c|}
\hline
liquid    & mass   & specific             & flash                  & H/C  & light    \\
             &            & gravity              & point                  &         & output            \\
{}         & (ton)    & (g/cm$^3$)      &  ($^{\circ}$C)   &         & (\% anthracene) \\
 \hline
BC521~\cite{BC521}  & -          & 0.89                  &   44                   & 1.314  & 68 \\
PC        & -          & 0.880                &   54                   & 1.33    &  -   \\
TD        & -          & 0.763                &   99                   &  2.14   &  -  \\
\hline
  I           & 6.1        & 0.79               &   49.5               & 1.8       & 53    \\
  II          &  19       & 0.78                &  76                 & 2.0       & 53   \\
  III+IV  &  240     & 0.76               &   99           &  2.14        &  5    \\
  V         &  160     & 1.00                & -                       &  -          &  -    \\
 \hline
\end{tabular}
\end{center}
\label{tab:Liquid_Property}
\end{table}

%---------------------------------
\subsubsection{Neutrino  Target  (Region-I)} 
The active neutrino target is  Gd Loaded liquid scintillator (Gd-LS) with Gd concentration of 0.1~\%.  
A candidate is a PaloVerde type liquid scintillator~\cite{PVLS}.  
This LS is formulated by mixing 20~\% of BC521, 20\% of PC and 60\% of TD. 
The molecular formula of the Gd in BC521 is 'gadolinium 2-ethylhexanoate'; 
Gd(CH$_3$(CH$_2$)$_3$CH(C$_2$H$_5$)CO$_2$)$_3$) xH$_2$O].~\cite{paloverde:GDLS}.
The light output is 53~\% anthracene. 
Along with 10~\% of PMT cathode coverage with its 20~\% quantum efficiency, the light yield will be 150~p.e./MeV, which corresponds to statistical energy resolution of 
\begin{equation}
 \frac{\delta E}{E} \approx \frac{8.2\%}{\sqrt{E(\mbox{MeV})}}
\end{equation}
H/C ratio is 1.8 and proton density is $7.7\times10^{28}/\mbox{m}^3$. 
The PaloVerde experiment used this kind of LS contained in 7.4~m long acrylic cells for a few years without problems.
TD is chosen as the  base solvent  because it is mono-molecular paraffin oil  and proton density is exactly known. 
Paraffin oil does not have double bond in its molecular structure and does not absorb scintillation light, and it is  immune to oxidization. 
Because TD and PC are mono-molecular material, the temperature window of the distillation in refinery process is narrow and contamination is intrinsically small. 
This strategy was very successful in the KamLAND liquid scintillator. 
TD has higher flash point  and slightly larger H/C ratio and specific gravity than dodecane which is successfully used in KamLAND detector. 
30~m$^3$ of the Gd-LS for four detectors will be formulated at the same time and  stored in a storage tank  in order to make the components of the four detectors same. 
When filling in the detector, the LS is carried by 200 liter stainless steel drum cans.
The weight of each drum can is measured with relative error of $<$0.3~\%  using carefully inter-calibrated scales
before and after emptying into the detector and the total amount of Gd-LS introduced is measured precisely. 
The acrylic vessel containing the Gd-LS has a chimney with 30~cm diameter. 
1~cm of head level difference corresponds to 0.01\% of LS volume difference. 
By monitoring the head level, precision monitoring of LS volume in the fiducial region can be performed.

%------------------------------
\subsubsection{$\gamma$-ray catcher (Region-II)}
The purpose of this liquid scintillator is to detect the escape gamma rays from region-I and reconstruct original energies of positron annihilation and neutron capture on Gd signal. 
The light output is adjusted to be the same as Region-I LS. 
A candidate of this LS is,
\begin{center}
PC(15\%)+TD(85\%)+PPO(0.6w\%)+bis-MSB(0.01w\%)   
\end{center}
The specific gravity is slightly smaller than  Region-I LS to keep safety of the acrylic vessel.
The concentration of PC is reduced from that of Region-I LS to raise the flash point for safety reason. 

%-----------------------------------
\subsubsection{Buffer oil (Region-III\&IV)}
The purposes of the buffer oil is to shield the liquid scintillators from $\gamma$-rays from the PMT glass and outside soils, and to tag the cosmic-rays. 
300~m$^{3}$ of paraffin oil will be used as the buffer oil. 
In order to detect stopping muons whose energy is less than the Cherenkov threshold and to improve cosmic-ray veto efficiency a slight light output ($\sim 10~\%$ of Gd-LS light output) may be  given to the oil by adding PPO. 
The specific gravity is slightly  smaller  than  Region-II scintillator. 

%---------------------------------
\subsubsection{Cosmic-ray veto (Region-V)}
Outside of the stainless steel spherical tank is filled with 160~m$^3$ of pure water.
The purpose of this region is to tag the cosmic-rays to veto the cosmic-ray related background and is to shield the $\gamma$-rays and neutrons from the soil outside of the cave and concrete to form the cave. 
The water is kept purified by circulating. 

%======================
% Sumiyoshi version

%\subsection{PhotoMultiplier}
\subsection{Photosensor}

\begin{figure}[htbp]
 \begin{center}
  \includegraphics[width=0.45\textwidth] {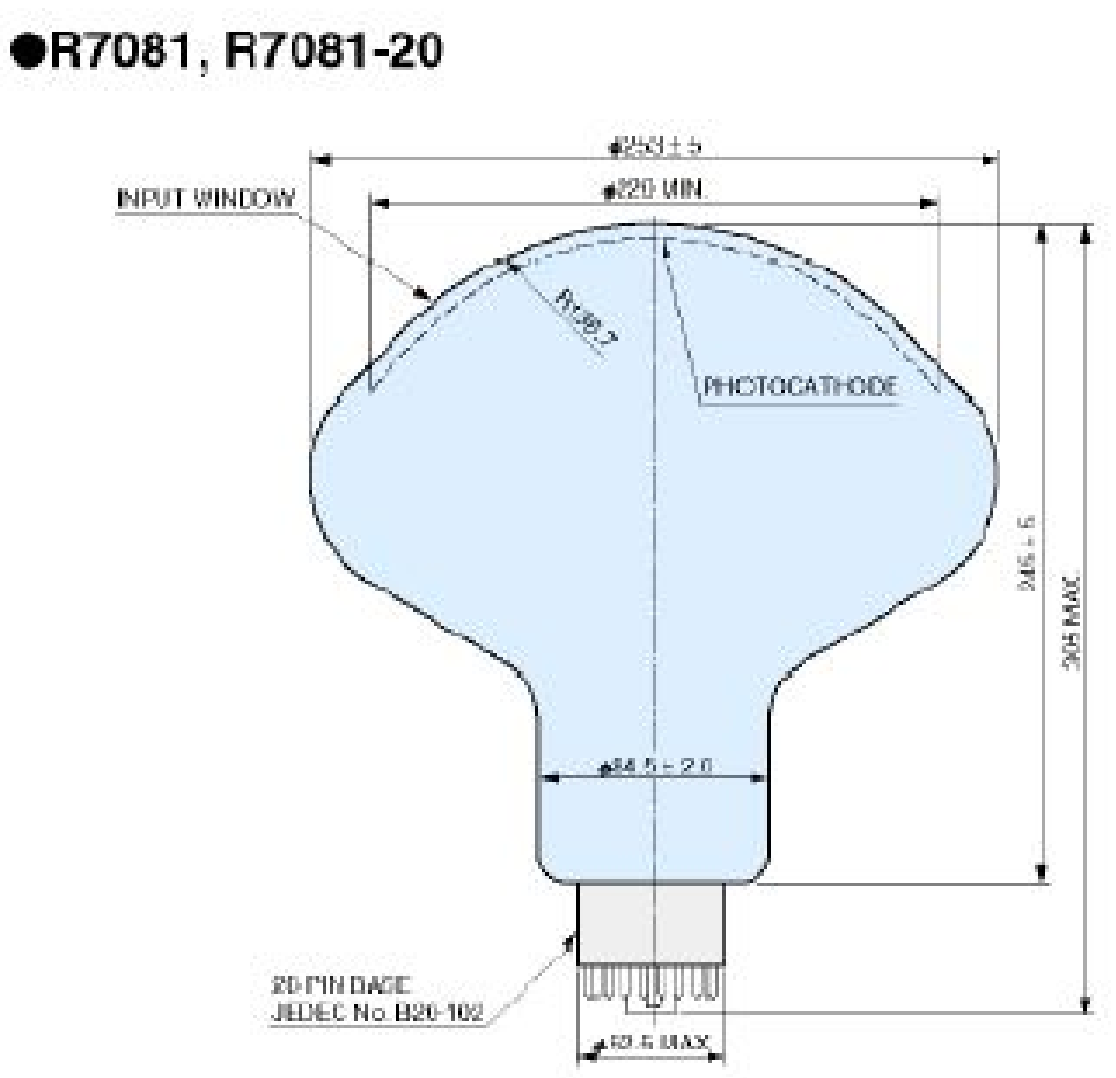}  
 \includegraphics[width=0.38\textwidth] {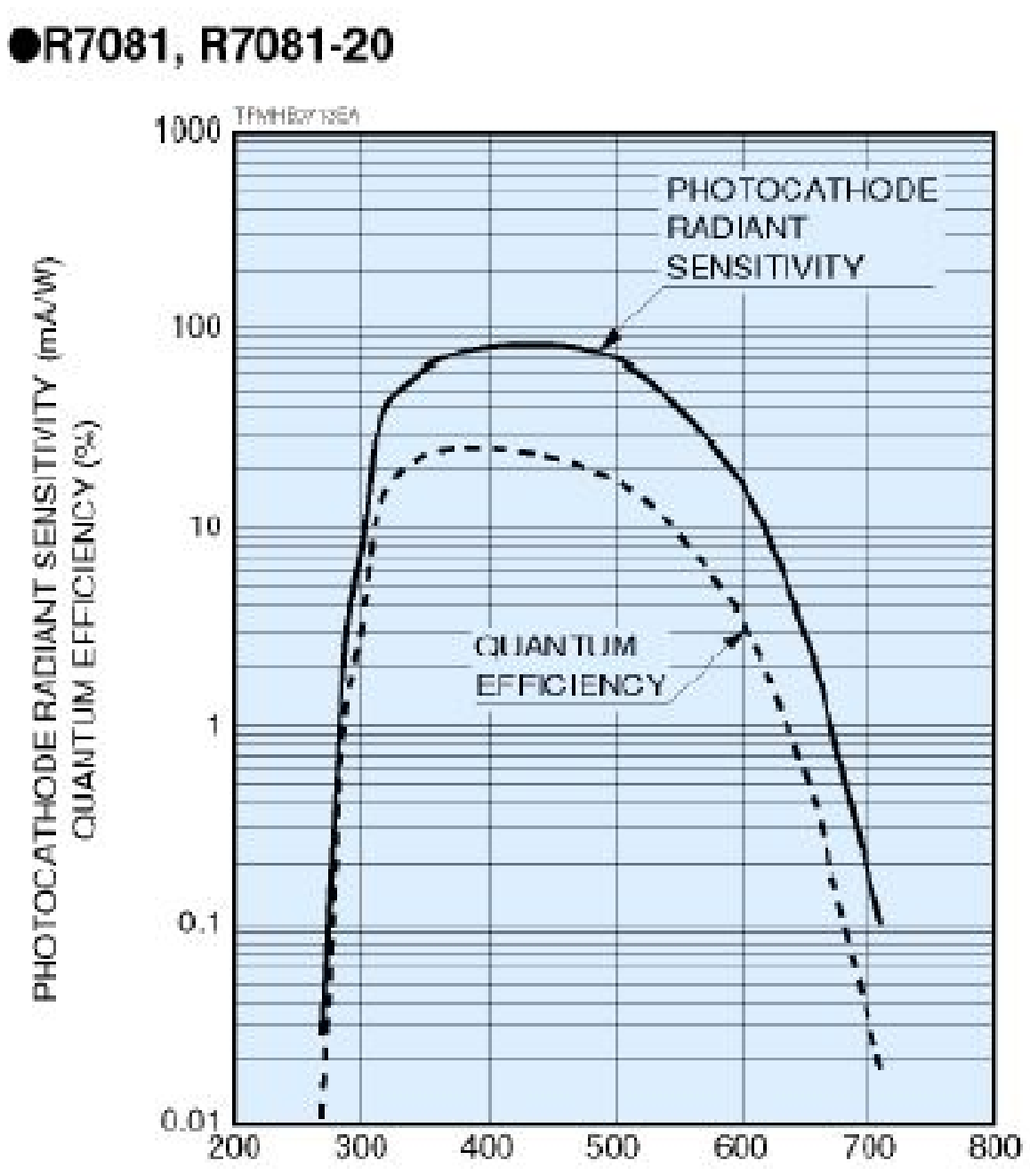}  
 \end{center}
\caption{Shape of the PMT envelop and quantum efficiency.~\cite{HPK_catalog}}
\label{fig:PMT_R7081}
\end{figure}

\begin{figure}[htbp]
\begin{center}
\includegraphics[width=10cm]{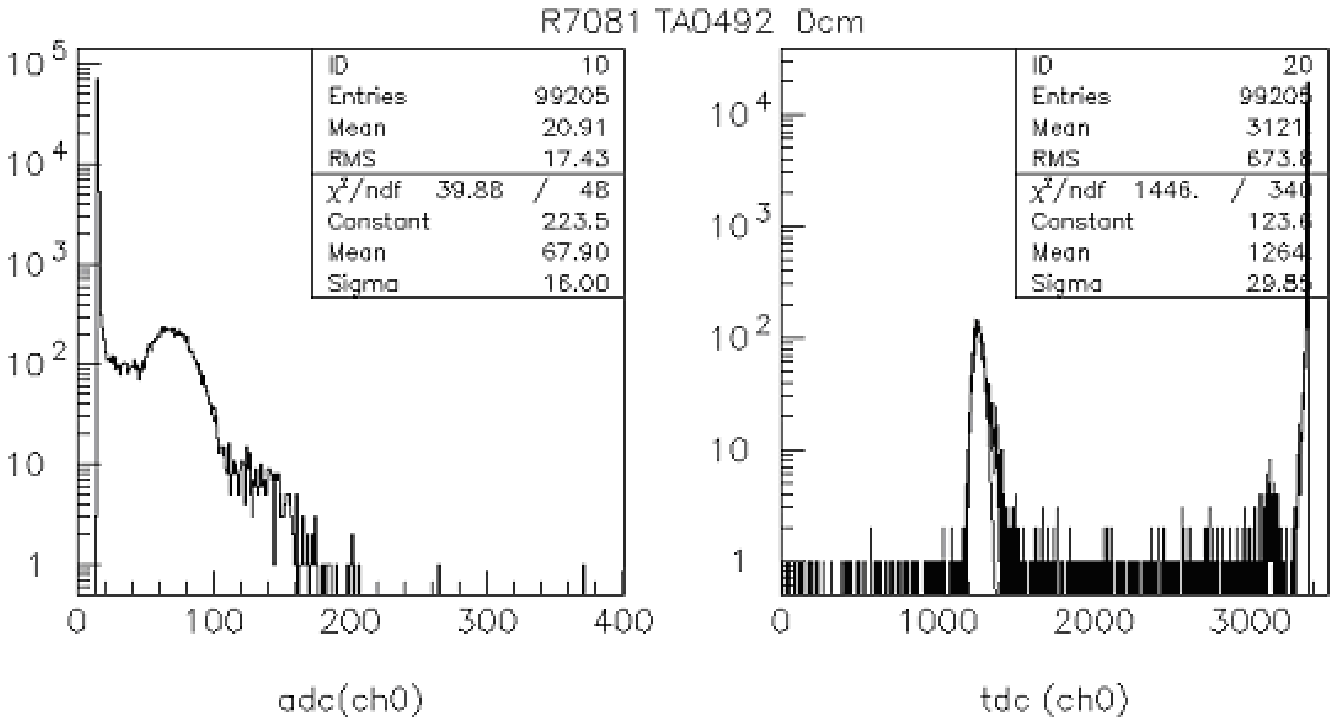}
\includegraphics[width=3.5cm]{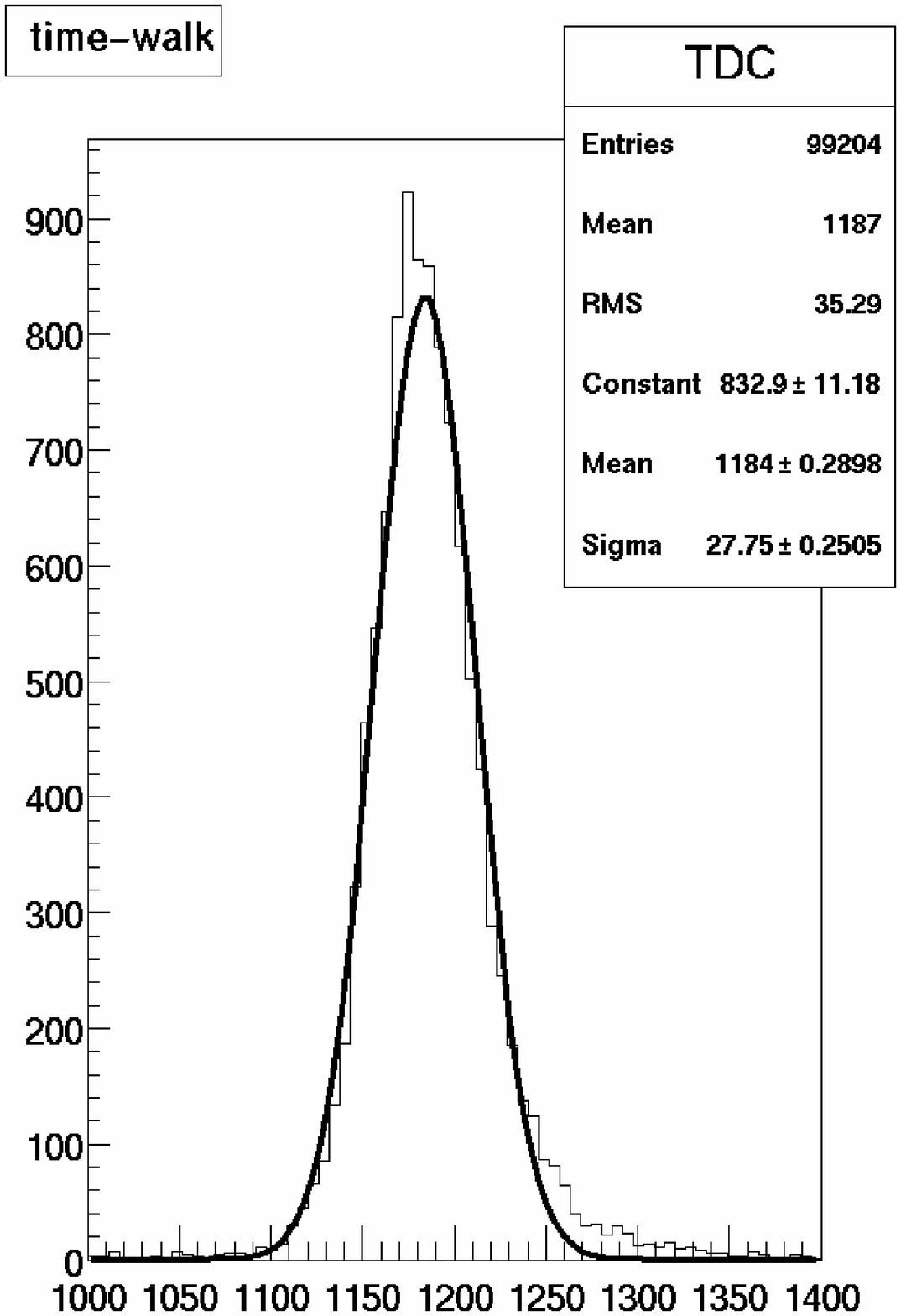}
\end{center}
\caption{a).  Pulse height spectrum and b) leading edge timing distribution for the single photoelectron signals obtained with R7081. 
c) Timing distribution for R7081 after the time walk correction. 
The TTS of 2.5~ns is obtained from this figure.}
\label{fig:PV_TTS_7081}
\end{figure}

In order to detect the scintillation photons from liquid scintillators, large aperture photomultiplier tubes (PMT's) are used. 
One of the candidates is Hamamatsu Photonics K.K. (HPK) 10 inches PMT (R7081). 
In order to obtain necessary photoelectron yield, more than 10\% of the photo cathode coverage is necessary and $>$300 10 inch PMT's are necessary. 
Fig.-\ref{fig:PMT_R7081} shows the shape of the PMT and table~\ref{tab:PMT_property} shows its specifications.
\begin{table}[htbp]
 \caption{PMT specifications~\cite{HPK_catalog}}
 \begin{center}
  \begin{tabular}{|c|c|c|c|c|c|}
   \hline
     Type No. & minimum              & Approx.      &  Q.E.            & Rise   & TTS  \\
                       &  effective area   & Weight        & at 390nm  & Time &           \\
                       &       (mm)               & (g)               &  (\%)      &  (ns) & (ns)  \\
    \hline
    R7081      & 220                        &   1150        &  25             & 4.3       & 2.9\\
     \hline
  \end{tabular}
 \end{center}
 \label{tab:PMT_property}
\end{table}
Fig.-\ref{fig:PMT_R7081} also shows the quantum efficiency for the PMT.
The photo cathode is bi-Alkali type with 25\% quantum efficiency around 400~nm wavelength.\\
Single photoelectrons response for the PMT was measured as follows.
As a light source for this measurement, we used a PiLas picosecond laser light system (¦Ë= 438.7 nm) having a high precision pulse timing ($<$30~ps). 
In order to obtain a single photoelectron response we used neutral density filters to reduce the laser light. 
For the digitalization of the pulse amplitude we used a LeCroy 2249A ADC. 
Fig.~\ref{fig:PV_TTS_7081} a) shows the single photoelectron spectrum for R7081. 
A clean single photoelectron peak can be seen at around 70 ADC channels. 
Fig.~\ref{fig:PV_TTS_7081} b) shows the leading edge timing for the single photoelectron signals. 
We used a TDC (Houshin Co. Ltd) for the timing measurement. 
After a time walk correction using the pulse height information (assuming a square root dependence on the pulse height), a transit-time -spread (TTS) has been obtained to be around 2.5~ns, which is better than the catalogue value of 3.7~ns in which the time walk correction may not be applied. 
Fig.~\ref{fig:PV_TTS_7081} c) shows the timing distribution after the time walk correction. 

\begin{figure}[htbp]
\begin{center}
\includegraphics[width=9cm]{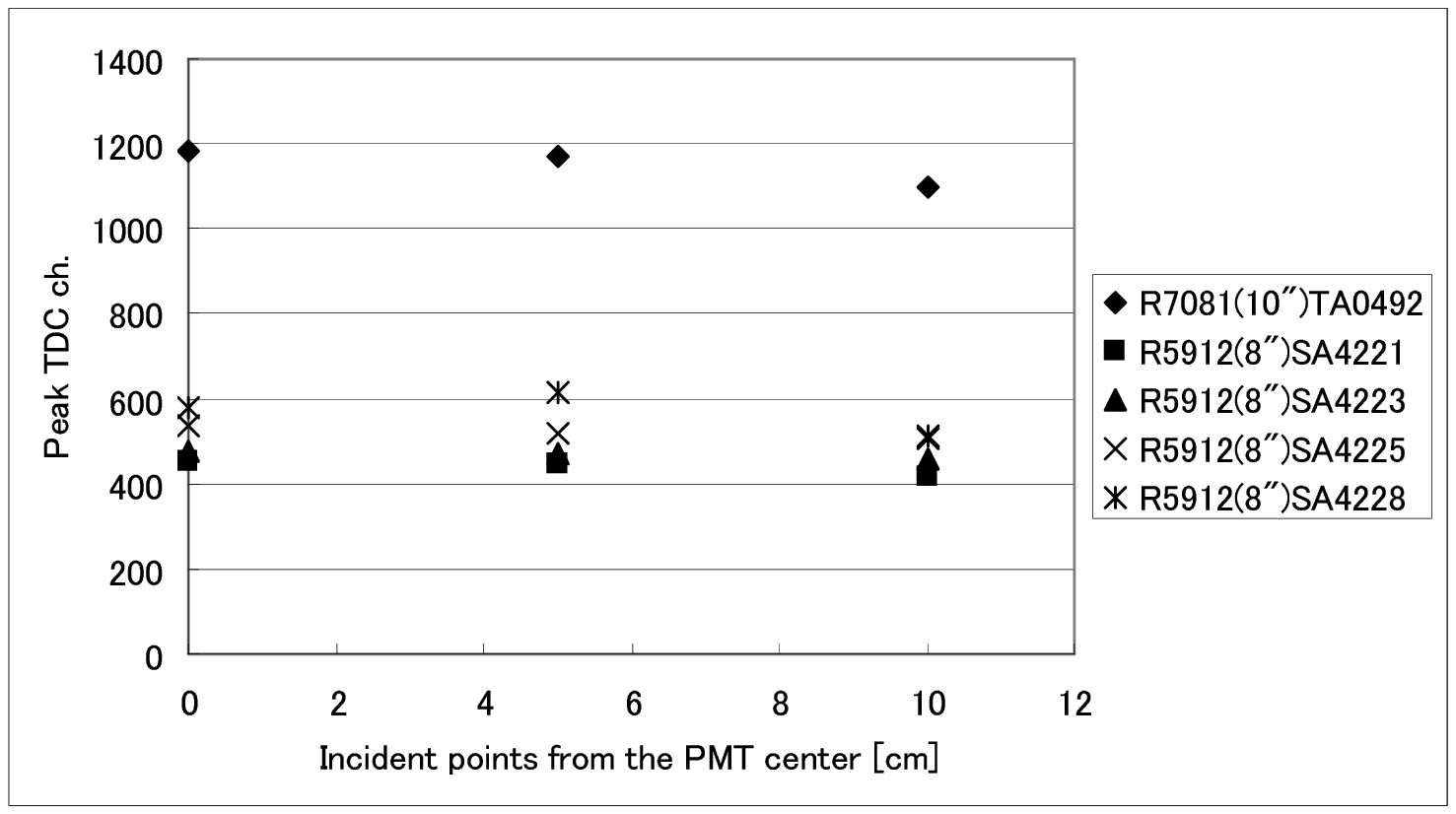}
\includegraphics[width=9cm]{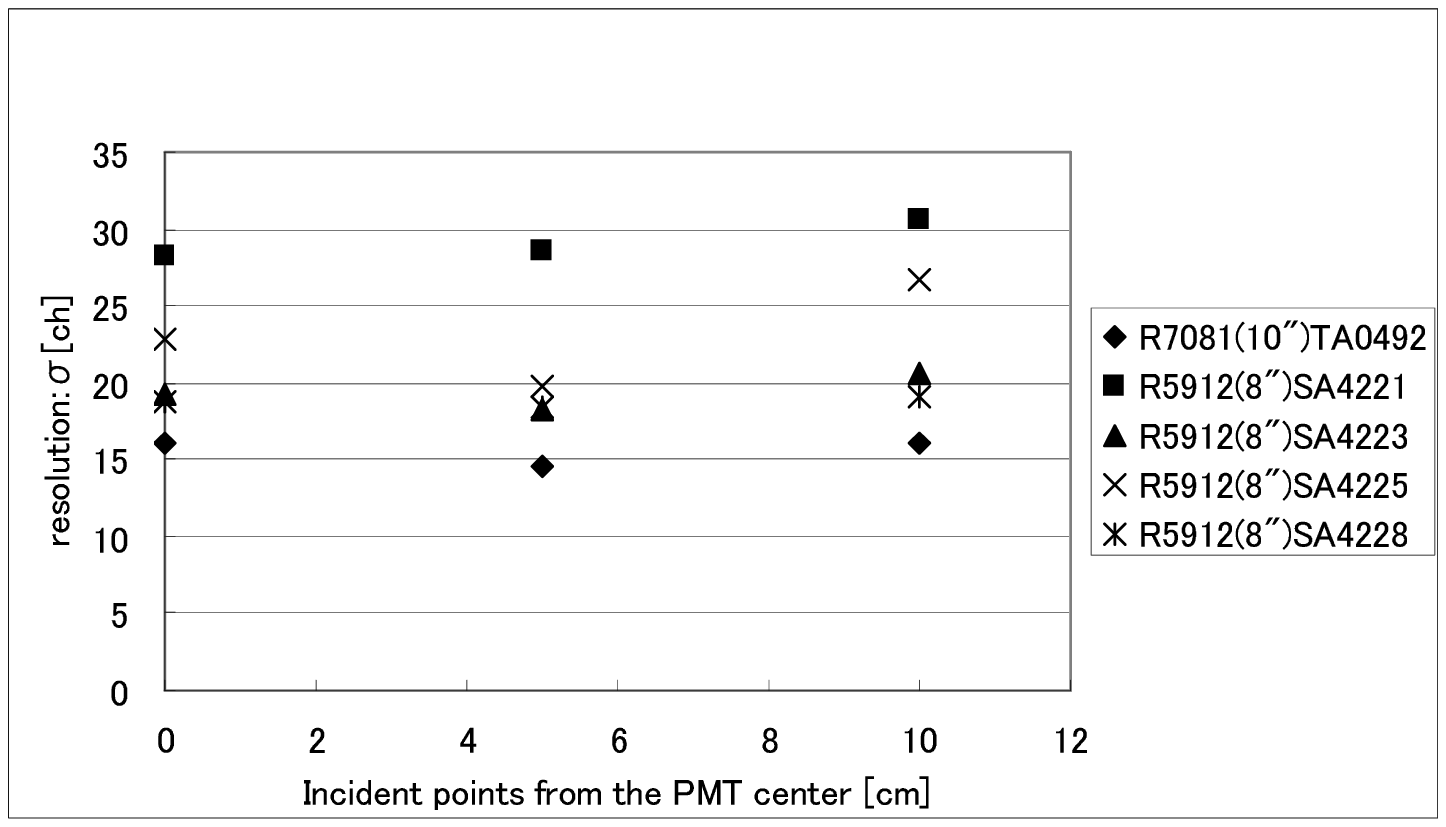}
\end{center}
\caption{a) Uniformity of the response for three R5912 and one R7081. The incident positions of the laser light were at 0 (center), 5~cm and 10~cm from the PMT surface center. 
Peak values were stable within 10\%.
b) Uniformity of the resolution for three R5912 and one R7081. }
\label{fig:PMT_uniformity}
\end{figure}

One of our worry for the large size PMT's is response uniformity over the large surface area. 
By changing the laser light incident position, we have measured the pulse height spectra for three different positions. Fig.~\ref{fig:PMT_uniformity} a) shows the peak ADC channels for single photoelectron spectra measured at three incident positions: 0 (center), 5~cm and 10~cm from the PMT surface center. 
Data for R5912 are also shown for comparison. 
Both R5912 and R7081 have a good uniformity within 10 \% variance. 
Fig.~\ref{fig:PMT_uniformity} b) shows the uniformity of the resolutions (standard deviations). 
They are also very uniform over the measured surface area. 
We can conclude from the above measurements that both R5912 and R7081 have enough performance for the detection of the scintillation light. 

Generally PMT glass contains significant amounts of radioactive elements and causes single background hits.
In order to reduce this background,  a low background glass is developed. 
Table~\ref{tab:radio_purity} compares the U/Th/K contents of the PMT glass used in various experiments.
With this glass and 90~cm of buffer thickness between glass and region-II scintillator, 
the single rate of the detector will be a few Hz at above 0.7~MeV threshold, which is satisfactory value. 

\begin{table}[htbp]
 \caption{radio purity of PMT glasses}
 \begin{center}
  \begin{tabular}{|c|c|c|c|c|}
   \hline
     Type                            & $^{238}$U    & $^{232}$Th      &  $^{40}$K    & Reference     \\
                                         &    (ppm)           &     (ppm)         &  (ppm)      &    \\
    \hline
    KASKA candidate       & 0.046              &  0.12              &  0.01         &  \cite{Ooura}\\
    CHOOZ                       & 0.07                &   0.1               &  0.018       & \cite{CHOOZ}     \\
    SuperK/KamLAND     & 0.48                &   0.47            &  0.08          & \cite{Piepke}      \\
    Kamiokande                 & 0.25                &  0.32             & 0.016          &  \cite{Piepke}   \\
     \hline
  \end{tabular}
 \end{center}
 \label{tab:radio_purity}
\end{table}
  
    There is an optional idea to install some 2" PMTs besides the 10" PMTs to measure the 
  total energy of cosmic ray signals without saturation problem.  
  
\begin{figure}[htbp]
 \begin{center}
  \includegraphics[width=0.45\textwidth] {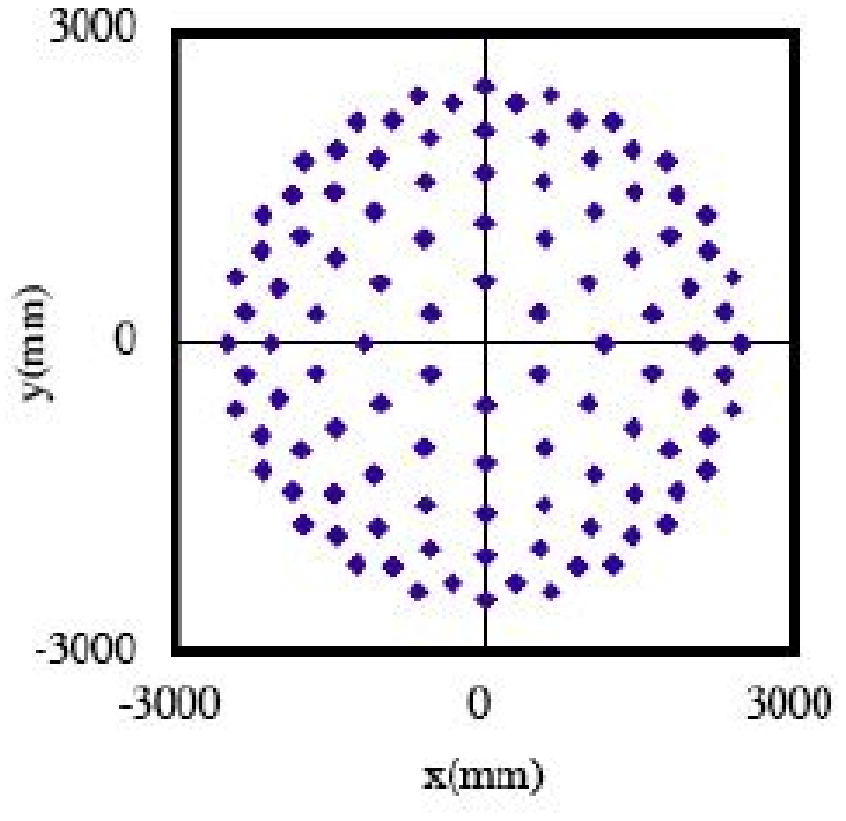}  
 \includegraphics[width=0.445\textwidth] {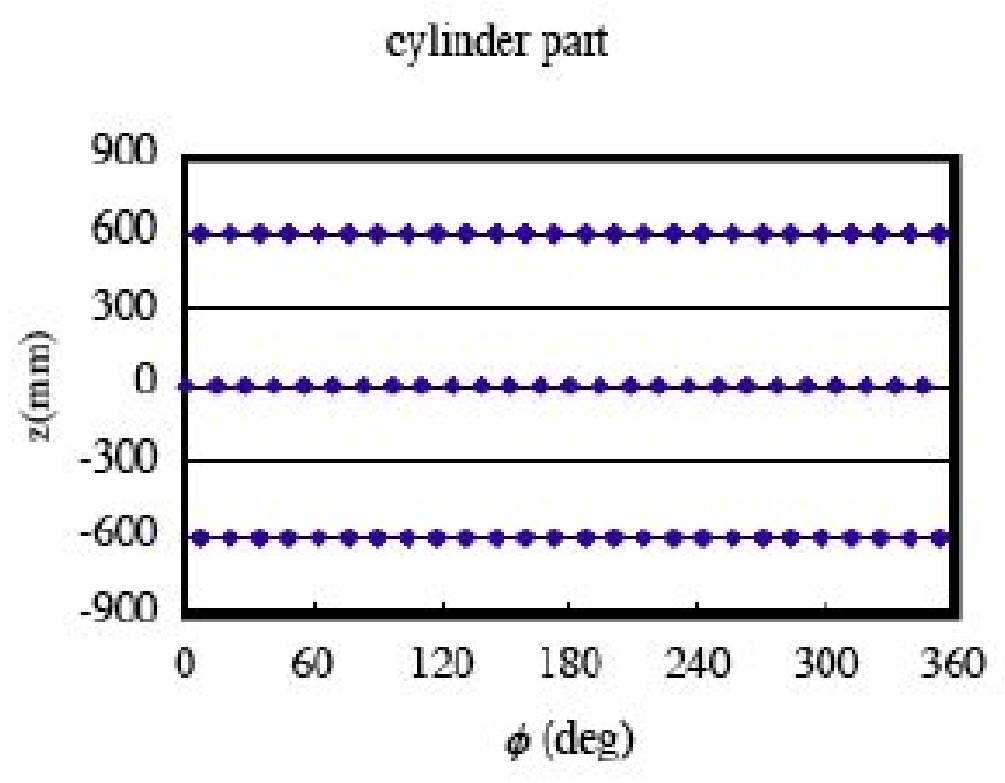}  
 \end{center}
\caption{PMT positions. 
Left: top/bottom view of the hemisphere part. 
Right: unfolding view of cylinder part. }
\label{fig:PMT_position}
\end{figure}

  Fig.~\ref{fig:PMT_position} shows the PMT positions. 
  300 PMTs are uniformly distributed. 

  %===============================
  \subsection{Cosmic-ray Tracking Device}
A cosmic-ray tracker system will be installed in the detector. 
The role of the cosmic-ray tracker is to track the cosmic muon passage  which goes through the detector.
Spallation background events are estimated by its association to the cosmic-ray track. 
The position resolution required is a few centimeters,
comparable to or  better than the resolution of the vertex reconstruction
for the neutrino events in the target region.

To achieve these purposes, several options can be considered for the hardware
implementation: use of liquid-scintillator(LS)-based, plastic-scintillator(PS)-based or gaseous detectors such as RPC.
In the case of a gaseous detector, building a gas-supplying system down in the shaft hole would be complicated, and some safety issues might arise, too. Not much attention has been paid on this option at this moment.

\begin{figure}[htbp]
\begin{center}
\includegraphics[width=8cm]{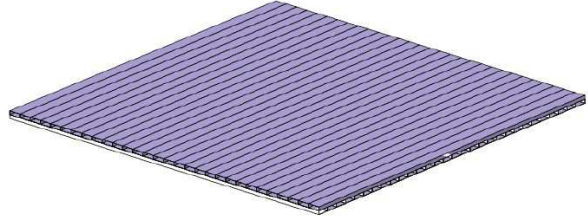}
\end{center}
\caption{Schematic drawing of a layer of cosmic-ray tracker.}
\label{fig:scinti-array}
\end{figure}

In cases of LS-based or PS-based detectors, the detector will consist of
scintillator strips with a segmentation to match the position resolution.
In a design investigated, the scintillator is a rectangular with 10~cm in width, 3~cm in height, and about 5~m in length.
In order to cover the whole area of the main detector, about 50 scintillator strips are arranged.
A layer consists of two planes, X and Y, which give a 2-dimensional position.
Fig.-\ref{fig:scinti-array} shows the scintillator layer.
These two layers are set up on the top and bottom of the main detector.
In reality, there have to be some dead spaces to accommodate the mechanical
support structure.

\begin{figure}[htbp]%
\begin{center}%
\includegraphics[scale=1.5, width=8.5cm]{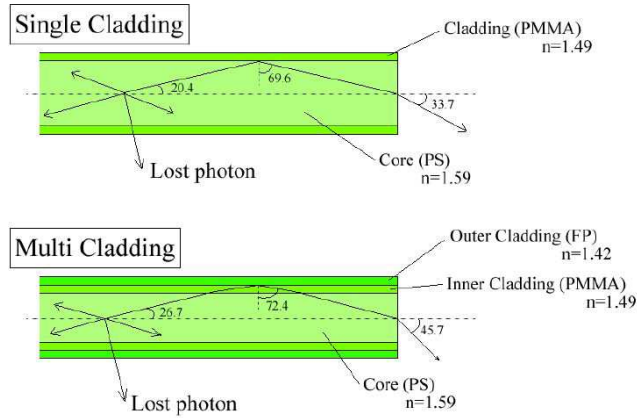}%
\end{center}%
\caption{Sketch of Y-11 WLS fiber.}%
\label{fig:fiber-cladding}%
\end{figure}

A wavelength-shifting (WLS) fiber is used to transport the scintillation light to the photodetectors such as PMTs. 
By using the WLS fiber, the photodetector can be set up in a convenient place.
The scintillation light is read from both sides. 
The time difference between both sides will also give a redundant
position measurement.
The WLS fiber considered is about 1.5 mm in diameter, and a multi-clad type.
Because its angle of the total reflection compared with a single-clad type is larger, the light yield is larger as shown in Fig.-\ref{fig:fiber-cladding}.
It is necessary to consider the attenuation length of the WLS fiber.
For example, the attenuation length of a multi-clad type fiber, Y11(200)MS made by Kuraray, is about 3.6 meters.
The light through the WLS fiber is read by Multi-Anode PMT (MAPMT) such as Hamamatsu H7546 series. 
The advantage to use MAPMTs is to save the cost, and a smaller occupied space compared with PMTs installed for all cells.

%------------------------------------
\subsubsection{A case of LS-based detector}
\begin{figure}[htbp]
\begin{center}
\includegraphics[width=5cm]{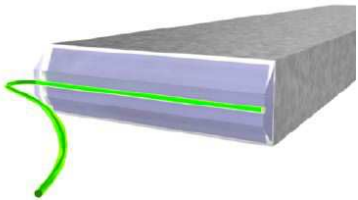}
\end{center}
\caption{Drawing of an acrylic cap and inserted WLS fiber at the end
 of a LS-based detector.}
\label{fig:acryl-cap}
\end{figure}

The advantage of this option is  low cost scintillator. 
An aluminum or SUS case and an acrylic cap can be used for the container. 
UV-transparent acrylic should be used for the cap. 
There is a ditch on an acrylic cap, and the WLS fiber is inserted as shown in Fig.-\ref{fig:acryl-cap}.
In this case, it is necessary to consider the attenuation length and
light yield because the solid angle of the fiber to absorb the scintillation light is small.
Multiple readout fibers may be necessary to collect enough photons.

%-----------------------------------
\subsubsection{A case of PS-based detector}
\begin{figure}[htbp]%
\begin{center}%
\includegraphics[width=6.75cm]{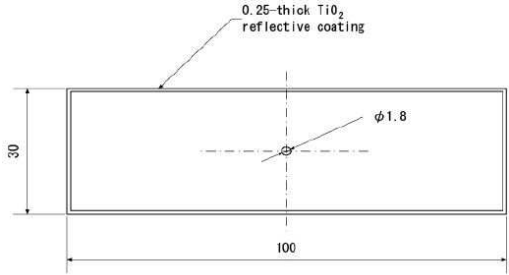}%
\end{center}%
\caption{Drawing of an extruded PS strip. The unit is mm.}%
\label{fig:plastic-scinti}%
\end{figure}

In this case, extruded plastic scintillator~\cite{Anna:NIM2001} can be used.
This scintillator has inferior optical property, but costs much less than usual PS.
The scintillator will be painted on the surface to reflect the light, and a hole will be made to pass the WLS fiber.
Extruded PS is used by the SciBar detector at K2K\cite{Nitta:NIM} and
by the MINOS experiment~\cite{Minos}.
Basic characteristics of this scintillator is shown in table~\ref{tab:cos-track:extruded-scinti}.
The structure of the plastic scintillator is shown in Fig.~\ref{fig:plastic-scinti}.
WLS fiber is inserted in the center of the scintillator. 
The advantage is that it is not necessary to consider the attenuation length of scintillator.
However, it is necessary to consider the attenuation length of the WLS
fiber compared with the case of a LS-based detector.

\begin{table}[htbp]
\caption{Basic characteristics of the scintillator used in SciBar detector}%
\begin{center}
\begin{tabular}{ll}%
\hline
Scintillator material & polystyrene with PPO (1\%) and POPOP (0.03\%)\\
Emission wavelength & 420~nm (blue)\\
Dimensions & 2.5 $\times$ 1.3 $\times$ 300~cm$^{3}$\\
Hole diameter & 1.8 mm\\
Reflector material & TiO$_{2}$ (15\%) infused in polystyrene\\
Reflector thickness & 0.25~mm\\
\hline
\end{tabular}
\end{center}
\label{tab:cos-track:extruded-scinti}%
\end{table}

 %======================================
 \section{Shaft Hole}
\label{sec:hole}

\begin{figure}[htbp]
 \begin{center}
 \includegraphics[width=1.5\textwidth, angle=-90] {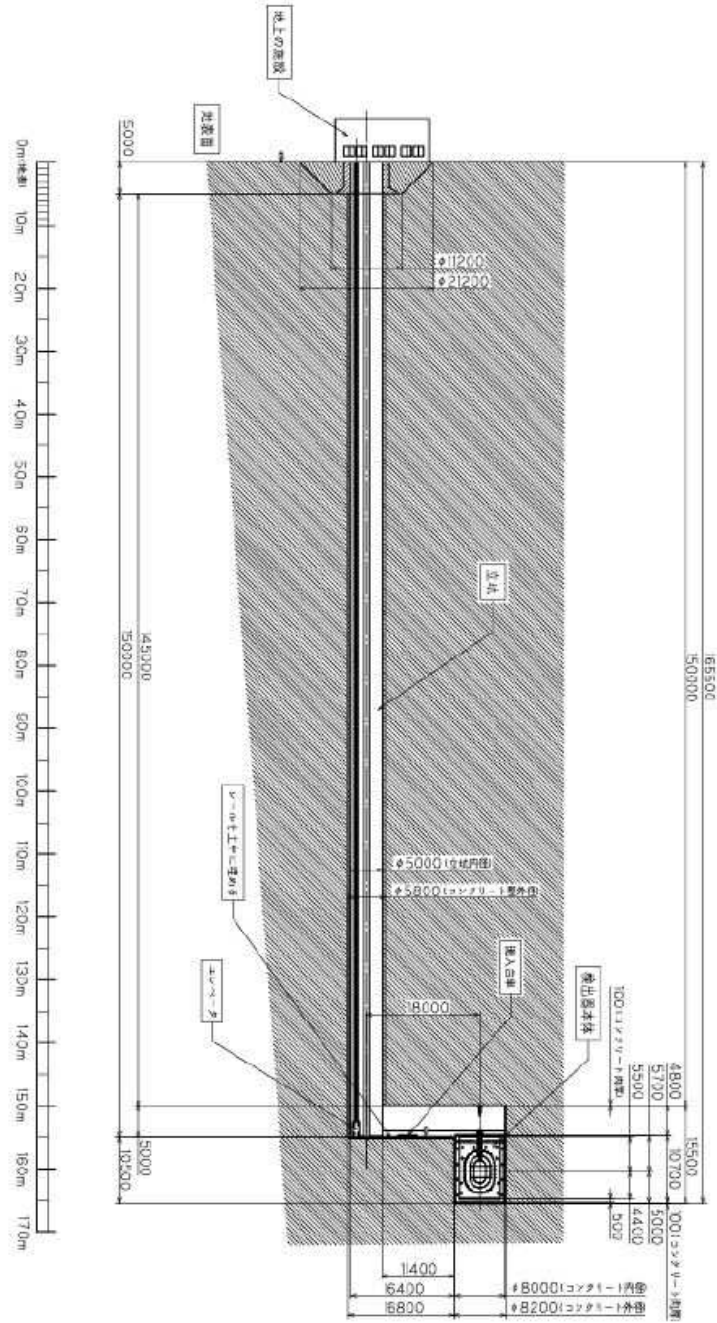} 
 \end{center}
\caption{Dimension  of a shaft hole for the far detector (one detector case).}
\label{fig:hole}
\end{figure}

\begin{figure}[htbp]
\begin{center}
  \includegraphics[ width=1.5\textwidth, angle=90] {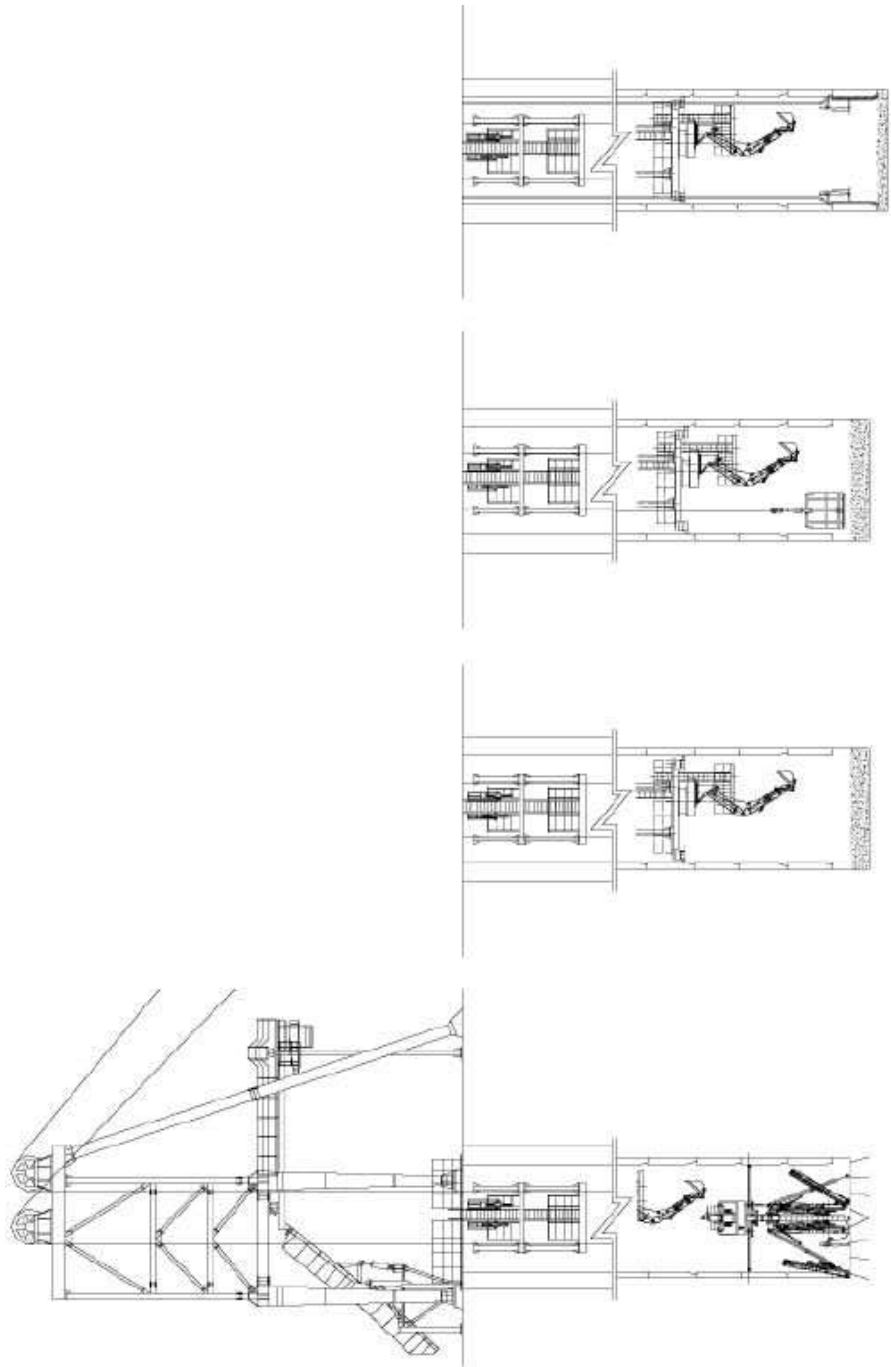}  
  \caption{Example of short step excavation procedure. From left to right, drilling/setting an explosive, blasting, taking out debris, moving down/concreting. }
 \end{center}
\label{fig:construction}
\end{figure}

\begin{figure}[htbp]
 \begin{center}
 \includegraphics[width=0.3\textwidth] {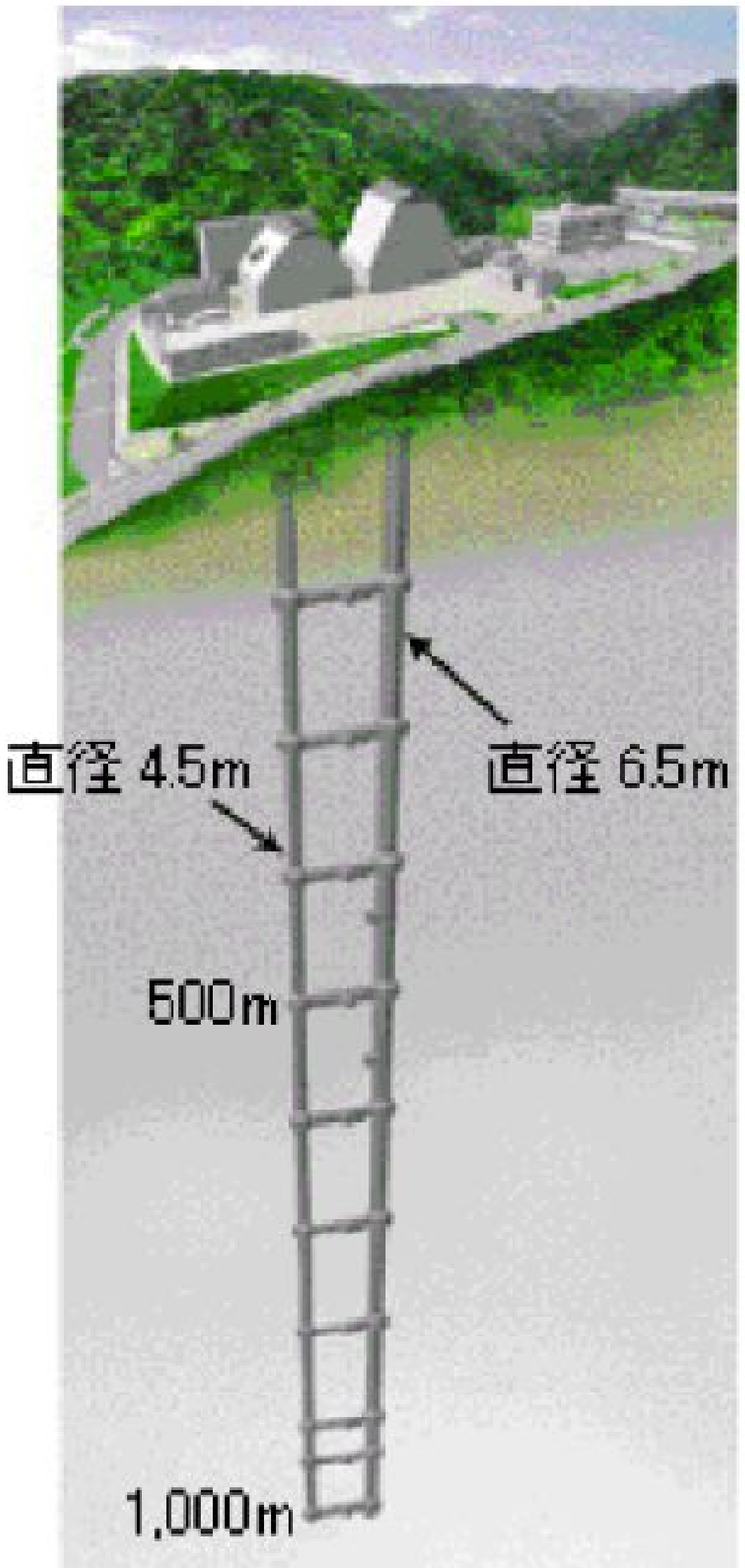}
  \includegraphics[width=0.5\textwidth] {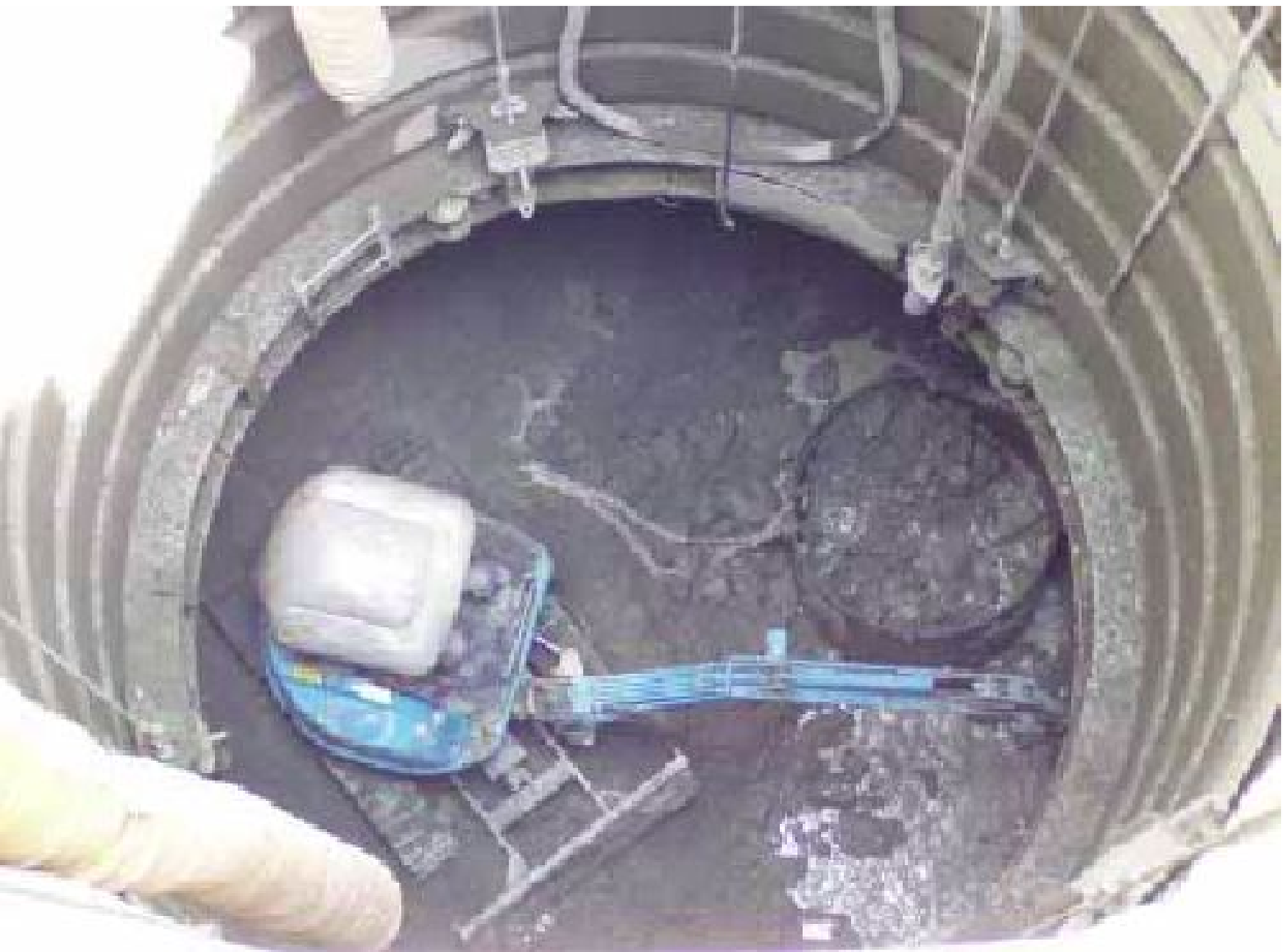} 
 \caption{1000~m depth twin holes and its construction sciene~\cite{JAEA_mizunami_home}}
 \end{center}
 \label{fig:shovel}
\end{figure}

\begin{figure}[htbp]
\begin{center}
\includegraphics[width=9cm]{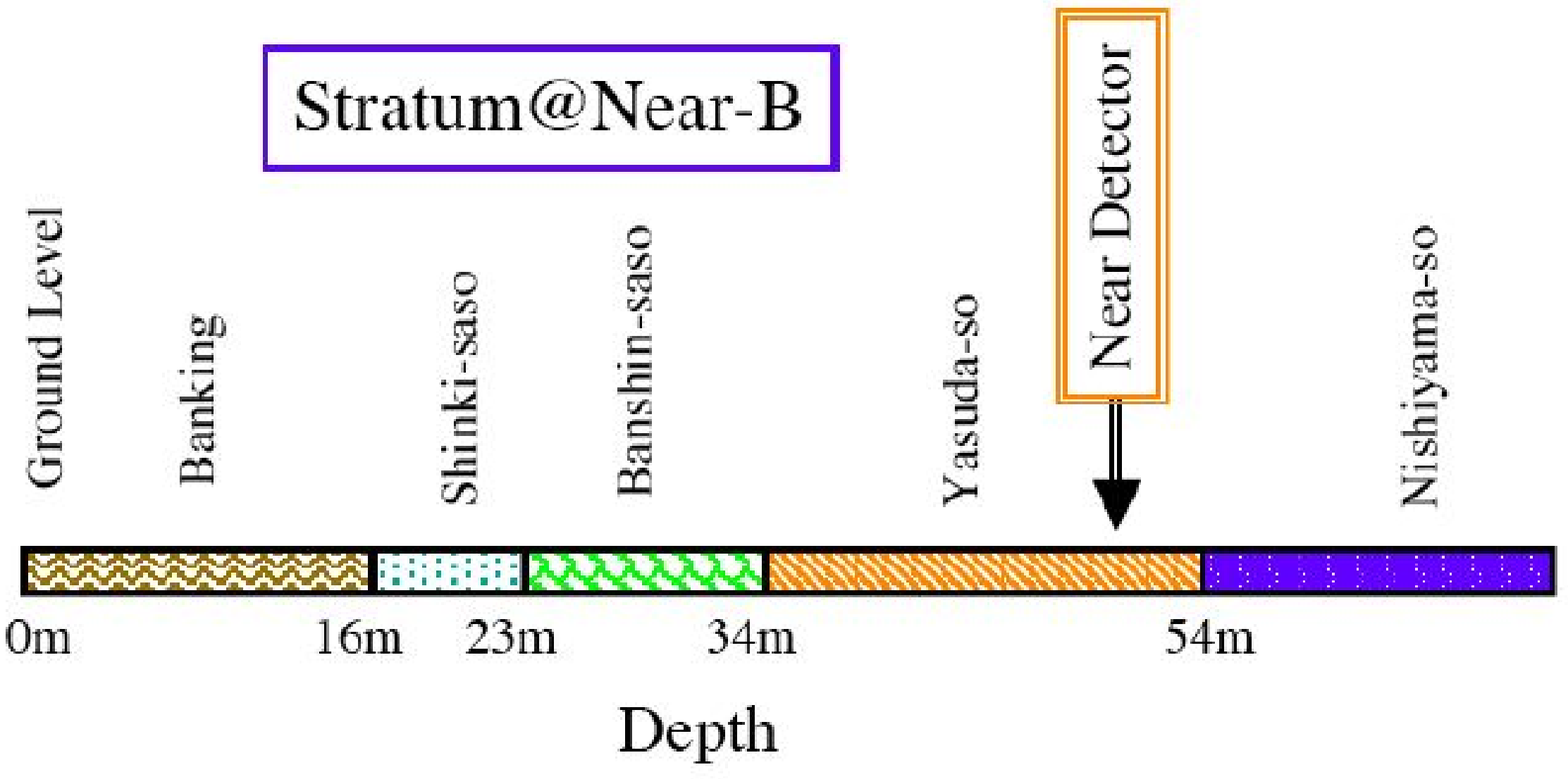}
\end{center}
\caption{Stratum at near-B detector position, measured by boring study.}
\label{fig:stratum_at_B}
\end{figure}

The required accuracy on $\sin^2 2\theta_{13}$ measurement is a fraction of  percent.
It is necessary to suppress the cosmic-ray  related backgrounds significantly to realize such accuracy by placing the detectors deep underground.
For KASKA case the far detector should be placed at least 250 meter water
equivalent (m.w.e.) deep to sufficiently suppress cosmic ray muons and near detectors should 
be placed at around 80m.w.e depth.
Because there is no mountain around Kashiwazaki-Kariwa nuclear power station for such distances less than 2~km, KASKA uses vertical shaft holes.
As the density of the soil in this area is 1.75~g/cm$^3$, the physical depth of the shaft holes are
150m for far detector and 50m for near detectors. 
Fig.~\ref{fig:hole} shows the schematic of the shaft hole and detector pit under consideration. 
The diameter of the shaft hole is 5m.
A 18m long horizontal corridor connects the shaft hole and the detector pit. 
The detector pit has a cylindrical  hole of 8m diameter and 10m height. 
The neutrino detector will be placed in the pit. 
The construction method of the holes will be a short-step method. 
A few meters are dug down at one time by blasting,  then debris are taken out by a crane and then the wall is reinforced by concrete. 
These procedures are repeated until the hole reaches the target depth.
Fig.~\ref{fig:construction} shows an example of how such construction proceeds. 
Constructions of such a vertical shaft hole have been performed in several locations.
For example, Japan Atomic Energy Association has been digging 1,000m-depth shaft holes with diameter of 4.5m and 6m in parallel for the purpose of studying deep underground storage of nuclear waste. 
Fig.~\ref{fig:shovel} shows a figure of the twin shaft holes and a picture taken while a shovel car is cleaning the debris. 
R\&D for digging the KASKA holes  has been under way. 
Extensive studies of the geology were performed by the electric power company when the reactors were constructed and overall structure of the geology around this area is well known. 
Fig.~\ref{fig:stratum_at_B} shows stratum at near-B detector position which was measured by our boring study.
Table~\ref{tab:stratum} shows the properties of strata. 

\begin{table}[htbp]
% \tbl{strata properties}
\caption{Properties of strata}
 \begin{center}
  {\footnotesize
   \begin{tabular}{|c|c|c|c|}
    \hline
    {} &{} &{} &{} \\[-1.5ex]
     Name & main constituents & density & water component  \\
    {}         &      {}                  &  (g/cm$^3$)  &(\%)                      \\
    \hline
    {} &{} &{} &{}  \\[-1.5ex]
    Shinki-Saso   &fine sand                  & 1.64&5.6   \\[1ex]
    Bansya-Saso    &rough sand             &1.83 &15.1  \\[1ex]
    Yasuda-so  &silt, clay                       &1.76  &47.0  \\[1ex]
    Nishiyama-so    &silt rock/clay rock &1.74   &47.4   \\[1ex]
    \hline
   \end{tabular}
    }
 \end{center}
 \vspace*{-13pt}
\label{tab:stratum}
\end{table}

\begin{figure}[htbp]
 \begin{center}
  \includegraphics[width=1.0\textwidth] {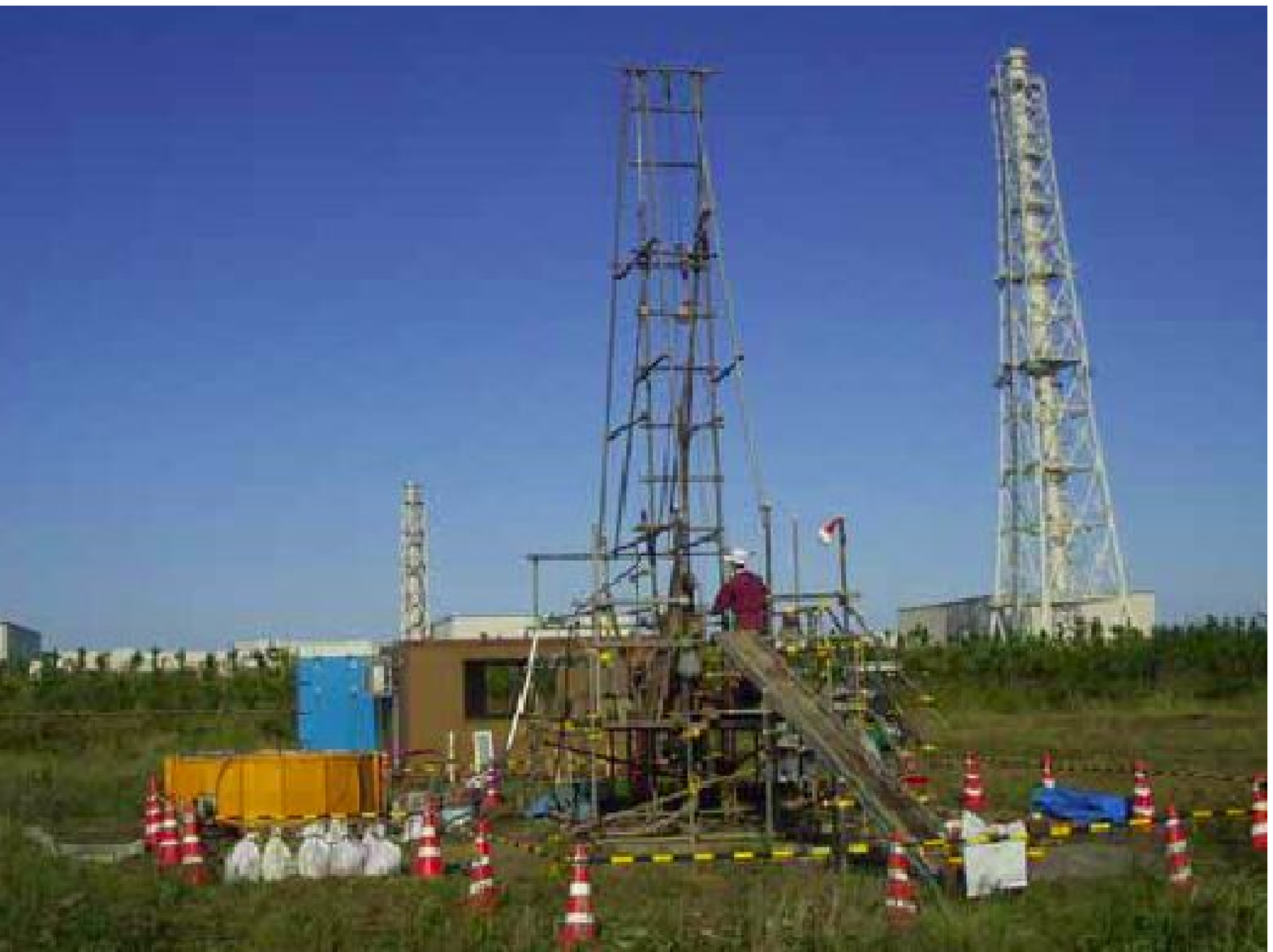}  
 \caption{Boring study at the near-B site. The towers behind the boring scaffold are ventillation towers of  reactor buildings.}
 \end{center}
 \label{fig:boring_scene}
\end{figure}

The general property of the geology is low density mud with abundant water component. 
A boring study for the KASKA-B detector was performed in fall 2004. 
Fig.~\ref{fig:boring_scene} show a picture taken during the boring study.  
Sample cores were taken from several depths and properties of the soils, as well as U/Th/K contaminations, have been studied. 
The cosmic-ray and $\gamma$-ray background were also measured at various depths 
 using NaI and plastic scintillators. 
 These information will be used for detailed design of the KASKA neutrino detector.

	%==================================
\section{Readout Electronics}
\label{sec:electronics}
\subsection{Property of Signals}
\label{subsec:signals}

\begin{figure}[htbp]
  \begin{center}
   \centerline{\includegraphics[width=0.8\linewidth]{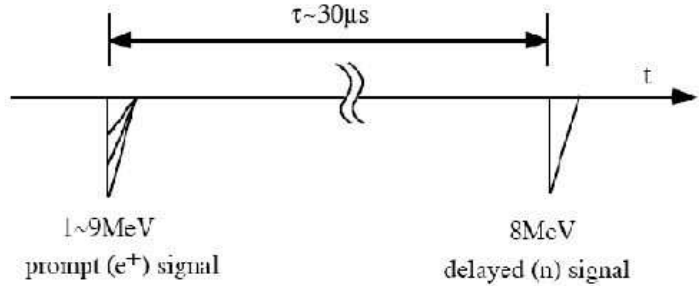}}
   \caption{Characteristics of $\bar{\nu}_e$ signal.}    
   \label{fig:signal} 
  \end{center}
  \end{figure}

 The KASKA experiment is composed of four identical detector units
 each having 300 photomultiplier tubes (PMT), to view the Region-I and II liquid scintillators.
Because there are 4 detectors, the total number of PMTs is 1,200.
Those signals from the PMTs are read out by flash ADCs (FADC). 
There also will be hundreds of signals from cosmic-ray anti-counter and cosmic-ray tracking devices. 
Signals from those devices are readout with conventional ADC/TDC systems.
The following description  concentrates on the readout electronics for the central PMT data, 
which are being developed for this experiment. 
Signal sources are anti-neutrino interaction, cosmic rays and other backgrounds and calibration signals. 
Detection and identification of anti-neutrino is performed  by a prompt and a delayed signal 
generated in an inverse $\beta$-decay as described in section (\ref{sec:general}).
 The former signal is produced by a positron and the latter is 
 created by $\gamma$ rays from gadolinium nucleus after absorbing neutron. 
Fig.~\ref{fig:signal} shows the conceptual picture of the $\bar{\nu}_e$ signal. 

The energy of the prompt signal is 1 to 9~MeV and that of the delayed signal is 8~MeV. 
The mean interval time of the two signals is about $30~\mu$s.
The rates of neutrino signals per detector are $\sim$1/hour for far detector and $\sim$10/hour for near detector.
The dominant background source is cosmic-ray muons.
The maximum energy deposit of the cosmic-ray in the scintillator is 800~MeV for single muon and much more for showering muons. 
The rate of the cosmic-rays are  $\sim$15~Hz for far detectors and $\sim$170~Hz for near detectors, respectively. 
The cosmic-ray signal has to be separated from the neutrino signal with very high efficiency for the suppression 
of backgrounds. 
An important effect to be considered  is production of radioactive elements with spallation reaction. 
It leaves some radio activity which remains after the cosmic-ray passing. 
These activities have to be eliminated from the neutrino signals also. 
As the energy deposit of the cosmic-ray signals is huge compared with the reactor neutrino signals , 
the electronics system will be saturated. 
It is, therefore, necessary to realize fast recovery from cosmic 
ray signals for the deadtime-less measurement. 
In addition, there are low energy single background signals which come from radioactive elements in the detector 
material and soils outside of the tunnel, such  U and Th, and their decay chains and also from  
$^{40}$K and $^{60}$Co, whose visible energy is less than 5~MeV. 
It is, therefore, expected to be reduced 
to less than 10~Hz with 0.7~MeV energy threshold by putting a thick absorbing buffer. 
Finally, the positron produced by stopping muon (Michel positron), whose maximum energy is 53~MeV, will be used 
for the calibration in the energy range up to a few 10~MeV. 
\begin{equation*}
\mu^+ \rightarrow e^+ + \nu_e + \bar{\nu}_{\mu}
\end{equation*} 

As the lifetime of muon is 2.2$\mu$s, the electronics has to recover quickly from the saturation caused by 
the initial $\mu^+$ signal.  

Under these circumstances, the requirements for electronics are as follows.

 \begin{itemize}
  \item Measurement of output charge  of each PMT hit to reconstruct energies of the signal.
  \item Measurement of the timing of each PMT hit to reconstruct positions.
  \item Sufficiently wide dynamic range for neutrino signals and cosmic ray signals.

  \item Dead-time-less Multi-hit readout capability. 
  \item Measurement of the timing separation between two contiguous events. 

  \item A fast recovery from the saturation of large cosmic-ray signals and monitoring after-muon activities. 
 \end{itemize}

In KASKA experiment, most of those information are measured by FADCs, with multi-bank 
memories for deadtime-less measurement with precise time stamps used to measure the
time interval of the prompt and delayed signals.

%===================================
\subsection{Electronics Scheme}
\begin{figure}[htbp]
  \begin{center}
  \begin{minipage}[htbp]{8cm}
   \centerline{\includegraphics[width=1.2\linewidth]{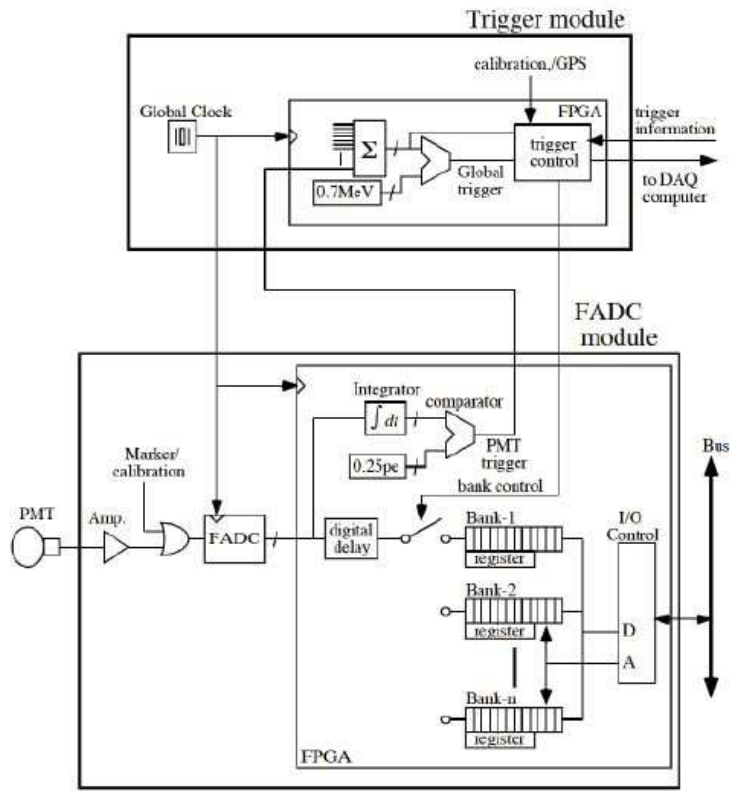}}
   \caption{Schematics of KASKA electronics.}   
   \label{fig:electronics} 
  \end{minipage} 
  \end{center}
  \end{figure}

 Fig.~\ref{fig:electronics} shows a schematic diagram of the readout electronics. 
 A signal from the PMT is amplified and put into a 500MHz FADC together with the marker or calibration signals. 
 The output from the FADC is divided  into a storage circuit and a trigger circuit. 
 The FADC data going to the storage circuit are recorded in a bank memory of 300~ns equivalent depth, after 
 a digital delay which compensates the latency of the trigger decision. 
 There will be two, or more if necessary, bank memories to eliminate the dead time. 
 When a trigger is issued, the direction of the FADC data flow is switched from a bank to the next one. 
The newly
 coming data being digitized are recorded/shifted in the new bank memory, while the stored data in the first 
 bank memory are being readout by a computer.  
 Each bank has a register which holds the data of absolute timing when the trigger is issued. 
 The FADC data for triggering is digitally time-integrated to get the total charge information of the signal.
 If  the total charge exceeds 0.25~p.e. equivalent, the PMT is considered to have a signal and a PMT trigger is 
 issued. This scheme has been provn to work with latency of $\sim$80~ns for a prototype. 
Then a PMT trigger information is sent to a trigger decision module, where the number of hits from 
 all the PMTs is counted up. If the number of hit PMTs exceeds 0.7~MeV threshold, a global trigger is issued, 
 which is sent to a trigger control circuit. 
 The trigger control combines the amplitude and timing information of the data and decides if the currently 
 active bank is switched  and send information about which bank to read out to the data acquisition computer. 
 
Followings are assumed KASKA detector parameters and the properties of signals for the rough estimate 
of necessary gain and dynamic range which are required for the electronics.
\begin{enumerate}
 \item  KASKA detector parameters
  \begin{itemize}
  \item Required energy loss for one photon in the scintillator: $150~\mbox{eV}$\\
              $\Rightarrow$ 46\% Anthoracene equivalent.
  \item PMT photocathode diameter: 22~cm.
  \item Radius of photocathode position in detector: 3~m \\
              $\Rightarrow$ photocathode coverage=10~\%. 
   \item Radius of the region-I scintillator: 1.4~m. 
   \item Quantum efficiency of PMT: 20~\%.
  \item amplification gain of PMT: $5 \times 10^6$.
  \item Rise/Fall time of single photoelectron signal: $10/15$~ns.
  \item Specific impedance of the signal cable: $50~\Omega$.
  \item Load impedance of PMT output including a back termination: $25~\Omega$.
  \item Sampling rate of the FADC: 500~MHz.
  \item Input voltage range of the FADC: $\pm 0.3$~V.
  \item Resolution of the FADC: 8~bits $\Rightarrow$ 2.3~mV/ch.
  \item Amplifier gain: 10.
  \item Attenuator ratio: 5:1.
  \item Trigger threshold for one PMT: 0.25~p.e. equivalent.
  \end{itemize}
         
\item  One photoelectron signal \\
Approximating the signal shape to a triangle with 10~ns rise time and 15~ns fall time, the peak voltage of 
the one photoelectron signal is 1.6~mV.  
After the amplifier of gain 10, it corresponds to 7 FADC counts. 
The signal is distributed to 13 bucket wide. 
The trigger discrimination is applied to the output data of digital integrator. 
The integrated  FADC counts for one photoelectron is 44, and 0.25~p.e. PMT trigger 
threshold corresponds to 11 counts. 
In actual signals, the shape of the tail is exponential and the signal 
will distribute to much wider time span than 25~ns and the integration period will be 50~ns. 

 \item  Low energy signals \\
An average number of photoelectrons created at one PMT for 1~MeV signal is 0.45~p.e. 
Thus half of PMT do not have a hit and 0 suppression is effective to reduce the data size. 
The maximum energy for the reactor neutrino signal is 9~MeV, which produces 4.1~p.e./PMT of scintillation light if the event occurs at the center of the detector.  
Because events occurred near the acrylic wall deposit 3.5 times more light to the nearest PMT than those occurred at the center of the detector, the nearest PMT will generate 14~p.e./PMT for such peripheral events with energy 9~MeV. 
The dynamic range of the FADC corresponds to 37.5~p.e.,  which safely covers the maximum reactor neutrino energy.

 \item  High-energy signals \\
A penetrating cosmic muon causes as much as $750$~MeV of energy deposit, because the maximum track length in the scintillators is 5~m. 
750~MeV of energy deposit produces 350~p.e./PMT of scintillation light.  
This is out of dynamic range of the FADC system and at least 1/5 of attenuation is necessary. 
 If the scintillation light comes at the same time, the pulse height from the PMT becomes  1.5~V. 
 The output of the amplifier will saturate for this kind of signals and fast recovery from the saturation is required to the amplifier.
 A clipping diode can be installed to the amplifier input.

% Fig.~\ref{fig:frontend} shows the schematics of frontend part of the electronics. 

%  \begin{figure}[htbp]
%  \begin{center}
%   \centerline{\includegraphics[width=0.6\linewidth]{eps_bit/front_end.eps}}
%   \caption{Schematic of front end part.}    
%   \label{fig:frontend} 
%  \end{center}
%  \end{figure}
  
Another idea to treat this issue is to install smaller diameter PMTs in addition to 10"PMT, which is now
our baseline scheme. 
If 2"PMT is used, the aperture of the PMT is 1/25 of the 10"PMT and 
the number of photoelectron obtained by PMT is as much smaller and the PMT will not saturate even 
for cosmic-ray signals. This information is used to judge the cosmic-ray is showering or not  and set 
additional dead time for showering muons. 

\end{enumerate}

%--------------------------------------------------------
  \subsection{Trigger}
The basic trigger is issued for events which have energy greater than 0.7~MeV. 
All the events with the energy greater than the threshold are recorded.
For this energy the number of expected photoelectron is 0.3~p.e./PMT. 
Thus the energy of the signal approximately corresponds to the number of hit PMTs and
the global trigger is issued  based on the multiplicity of the hit PMTs. 
The hit multiplicity for 0.7~MeV signal is about  90~PMT hits. 
A PMT hit is defined by  'PMT trigger', which is issued when the 50~ns-wide integrated FADC hits becomes 
0.25~p.e. level or more. 
The PMT hit information is also used for 0-suppression procedure. 
The integration of the FADC data and the summation of PMT hits are performed by field programmable gate 
arrays (FPGA) as shown in Fig.~\ref{fig:electronics}. 
The information of the PMT trigger from all the PMTs are sent to the global trigger module 
and total number of hits with integration time of 100~ns is calculated. 
The trigger circuit can generate more sophisticated trigger, such as delayed coincidence +  low threshold trigger, 
pre-scaled trigger, random trigger, GPS trigger for absolute time stamp,¡¡forced trigger for calibrations, etc.

  \subsection{Readout modules}
The FADC module is composed of a daughter board, which is a main part with a FADC chip, and 
a mother board which works mainly  as an interface to the readout bus.

\begin{figure}[htbp]
  \begin{center}
  \begin{minipage}[htbp]{8cm}
   \centerline{\includegraphics[width=1.2\linewidth]{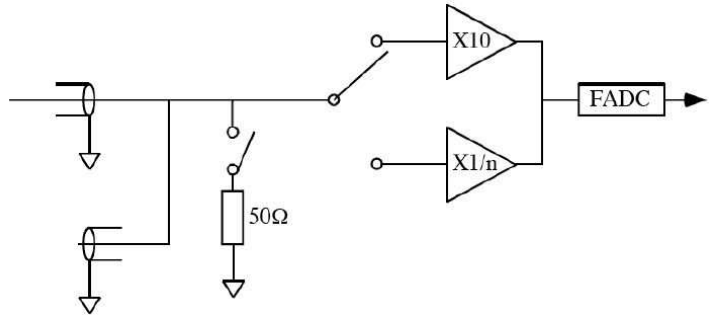}}
   \caption{FADC modules input}   
   \label{fig:FADC_module} 
  \end{minipage} 
  \end{center}
  \end{figure}

 \begin{enumerate}
 \item Daughter board: It includes a buffer amplifier whose gain is 
       selected either 10 or 1/5 by a jumper.  
      The input impedance is selectable either high or low by a jumper, so that two FADC units 
       can be connected by a daisy chain of one input as a high gain 
       FADC and a low gain one. 
      Fig.~\ref{fig:FADC_module} shows a schematic of the input circuit of FADC module. 
       It is also usable as independent FADC channels, as well. 
       The amplifier is followed by a FADC chip, which is 8-bit 500 MSPS ADC as a baseline. 
       It will be driven by a 500~MHz clock in the present design. 
       The input voltage range is $\pm 0.32~V$.
 \item Mother board: It includes FPGAs which work as controllers of the module and an interface with 
       rear-end electronics which will be either CompactPCI or VME. 
 \end{enumerate}
 
 Timing depth of one peace of data is 300~ns, which is determined by 
both pulse duration for the fast component and the expected latency of the 
trigger. 
 The FADC chip is always active and the digitized data are written 
on a bank of memories in a FPGA. 
 When the module receives a trigger from the global trigger circuit,  the currently active 
bank and associated counter is frozen and the module activate the next bank in turn. 
 The frozen memories are then read out to 
the rear-end electronics, while the next data are independently read into 
another bank. 
 This feature of multi-buffer memories is adopted for deadtime-less DAQ, that is essential 
for the present experiment. 
 The time 
interval between a buffer and a successor is measured by the absolute time stamp.\cite{fadc-yk}

 %============================
\section{Calibration}
%---------------------------
\subsection{Introduction}

Because reactor $\theta_{13}$ measurement needs a high precision and reliability, intense
calibrations are necessary throughout the run time. 
Calibration sources are classified into two categories. 
One category is optical and electronic sources, such as diffusive laser ball, optical fibers, LED and so on.  
The sources are used to measure responses of PMT and electronics and performs relative calibration. 
The other category is energy and position calibration sources. 
These are radioactive sources (gamma source, neutron source, positron source). 
Gamma sources are used for determination of absolute energy scale and energy 
resolution. 
Neutron and positron sources are used for measurements of detection 
efficiency and response. 
In addition, in order to  locate a source at aimed 
positions, a calibration manipulator system will be developed.

\begin{center}
\begin{table}[htbp]
\begin{tabular}{||c||c||c||}
\hline
& Source & Feature\\
\hline\hline
Optical	& Diffusive laser ball (SNO) & Isotropic laser\\
\cline{2-3}
 & UV  laser (CTF-Borexino) & Pulsed laser\\
 & LED & simple \\
\hline\hline
Neutron source & $^{241}$Am-$^{9}$Be & with 4.4~MeV gamma ray\\
\cline{2-3}
 & $^{252}$Cf & spontaneous fission \\
\hline\hline
Gamma source & $^{137}$Cs & 0.662~MeV\\
\cline{2-3}
 & $^{54}$Mn & 0.835~MeV\\
\cline{2-3}
& $^{60}$Co & 1.173, 1.333~MeV\\
\cline{2-3}
 & $^{65}$Zn & 1.35~MeV\\
 \cline{2-3}
 & $^{40}$K & 1.5~MeV\\
\cline{2-3}
 & H neutron capture & 2.22~MeV\\
\cline{2-3}
 & $^{12}$C neutron capture & 4.95~MeV\\
\cline{2-3}
 & Gd neutron capture & 8~MeV\\
\cline{2-3}
 & p-T source (SNO) & 19.7~MeV(Q value)\\
\hline\hline
Positron source & $^{22}$Na & 2.8y, 0.511~MeV, 1.275~MeV\\
\cline{2-3}
& $^{68}$Ge & 280d, alternative energies\\
\cline{2-3}
& Michel positron & $E_{max}$=53~MeV\\
\hline
\end{tabular}
\caption{various sources}
\label{tab:calib_source}
\end{table}
\end{center}

%--------------------------------
\subsection{Optical and electronic calibrations}

The purpose of optical and electronic calibrations is to check the response of each PMTs and to monitor the attenuation length of liquid scintillators. 
An  UV  pulsed laser can be used for monitoring of performances such as one p.e. gain, uniformity, timing and pedestal,  of each PMT. An isotropic light source like a diffusive laser ball (SNO) 
can be used for monitoring the attenuation length of the detector components.

%--------------------------
\subsection{Energy calibration}
In an antineutrino event,  the positron produced by inverse-$\beta$-decay interaction loses its energy and annihilates with electron 
in LS and the total energy deposit is detected as a prompt signal by the PMTs. In addition, 
neutron produced by the inverse-beta interaction is captured by Gd after about 
30$\mu$sec, and gamma rays are emitted that are detected as a delayed signal by the PMTs. 
Therefore $\gamma$ ray source, positron source and neutron source are used for energy 
calibration.\\

%---------------------------
\subsubsection{$\gamma$ ray sources}

The purpose of using $\gamma$ ray sources for calibration is to determine the absolute 
energy scale and energy resolution. 
Gamma ray sources can be classified into low energy region corresponding to the prompt signal response (0.7$\sim$9.0~MeV) and high energy region corresponding to the delayed signal response (5.0$\sim$11~MeV). 
There are many kinds of gamma ray sources in the low energy region. 
For example, $^{137}$Cs(0.662MeV) and $^{60}$Co(1.173MeV+1.333MeV) are useful sources in the low energy region. 
On the other hand, there are little kinds of gamma ray sources in the high energy region, because it is difficult to produce long life and high rate sources. 
Therefore, if we need gamma source in high energy region, one option is to develop an original source like pT source used by SNO experiment. 

%----------------------
\subsubsection{Positron sources}

The purpose of calibration using positron sources is measurement of positron response. 
$^{22}$Na is a very useful source because of having longer life (2.8 years) than other positron 
sources. 
Positron produced by $\beta^+$ decay annihilates with electron in the source package, 
and then two gamma rays (0.511~MeV) emit back to back. In addition, one more gamma ray 
(1.275~MeV) emits. 
These annihilation gamma rays are used as mimic signals. 
$^{68}$Ge is a useful 
source, too, because of emitting gamma rays with alternative higher energies, but the half 
life is relatively short (287 days).
Another way is to use cosmic rays.
If $\mu^+$ stops in the LS, it decays and produce energetic positrons in the LS. 
\begin{equation}
 \mu^+ \rightarrow e^+ +\nu_e + \bar{\nu}_{\mu}. 
\end{equation}
The energy distribution of this positron (Michel positron) is well known and it can be used to
calibrate energy range up to 50~MeV. \\
  
 %-----------------------
\subsubsection{Neutron sources}

The purpose of calibration using neutron sources is the measurement of neutron capture response 
by the nuclei (H, $^{155}$Gd, $^{157}$Gd) in LS. 
Gamma rays produced by neutron capture of H, 
$^{155}$Gd and $^{157}$Gd give a total energy of 2.223~MeV, 8.536~MeV and 7.937~MeV respectively. Especially, the neutron capture by $^{155}$Gd and $^{157}$Gd with large cross sections is the main response of the delayed signal in KASKA experiment, and we measure absorption probability and  timing by using neutron sources. 
$^{241}$Am-$^{9}$Be source is useful for this purpose. 
In Am-Be sources, $\alpha$ particle produced by $^{241}$Am decay is irradiated to $^{9}$Be nucleus, and then 
it forms excited state of $^{13}$C and decays to neutron plus 4.4~MeV $\gamma$. 
$^{252}$Cf  is a spontaneous fission source and can produce high rate neutrons.
	
%--------------------
\subsection{Calibration device manipulator}
\textbf{Region-I}\\
A very basic calibration equipment is a vertical string suspendable in the LS. 
Calibration sources are put in the known position along the $z$-axis.
This type of calibration will be most frequently performed. 
In KASKA experiment, three dimensional calibration is necessary to study the properties of signals 
near the acrylic walls. 
In order to accurately locate sources at target positions, we develop a calibration manipulator. 
We have two candidates of manipulator types, wire type like SNO and robot arm type. 
For the wire type, the area of the device that shadows the scintillation photons is small.
However, for the wire type, the anchors hanging the wires may damage the acrylic vessel if the load to the anchor is too strong. 
In addition the anchors and wires must be kept in the LS. 
On the other hand, by the robot arm type, all components 
can be taken out of the LS for physics running, and the direction of sources can be controlled. In addition, by replacing the source with a CCD camera, we can perform visual inspections, during the oil filling and to measure absolute positions of the 3 dimensional calibration device by comparing with the positions of LEDs which are distributed around acrylic sphere or PMT sphere.\\

%--------------------
\subsubsection{Wire type manipulator}
Fig.-\ref{fig:manipulator}-left shows how to locate the calibration source in the detector by a wire type manipulator. 
The concept is to use a wire manipulator as in SNO detector. 
Four ring-blocks are attached inside the inner acrylic sphere and a wire goes through the ring. 
One end of the wire is tied to the 
calibration device and the other end goes all the way up to the top of the chimney and the 
length are controlled by a stepping motor. 
By controlling the length of the three wires, the 
position in the horizontal plane is determined. 
The 4th wire directly comes down the chimney 
and is attached to the top of the calibration device. 
The weight of the calibration device is heavier than that of LS and by controlling the length of the 4th string, the 
vertical position of the calibration device is defined. 
The basic calibration is frequently performed to put a radioactive source along the vertical line. Then less frequently 3 dimensional calibration will be performed.\\

%\begin{figure}[h]
%\begin{center}
%\includegraphics[scale=0.35]{wire.eps}
%\caption{Wire type manipulator}
%\end{center}
%\end{figure}

%----------------------------
\subsubsection{Robot arm type manipulator}

\begin{figure}[htbp]
\begin{center}
\includegraphics[scale=0.4]{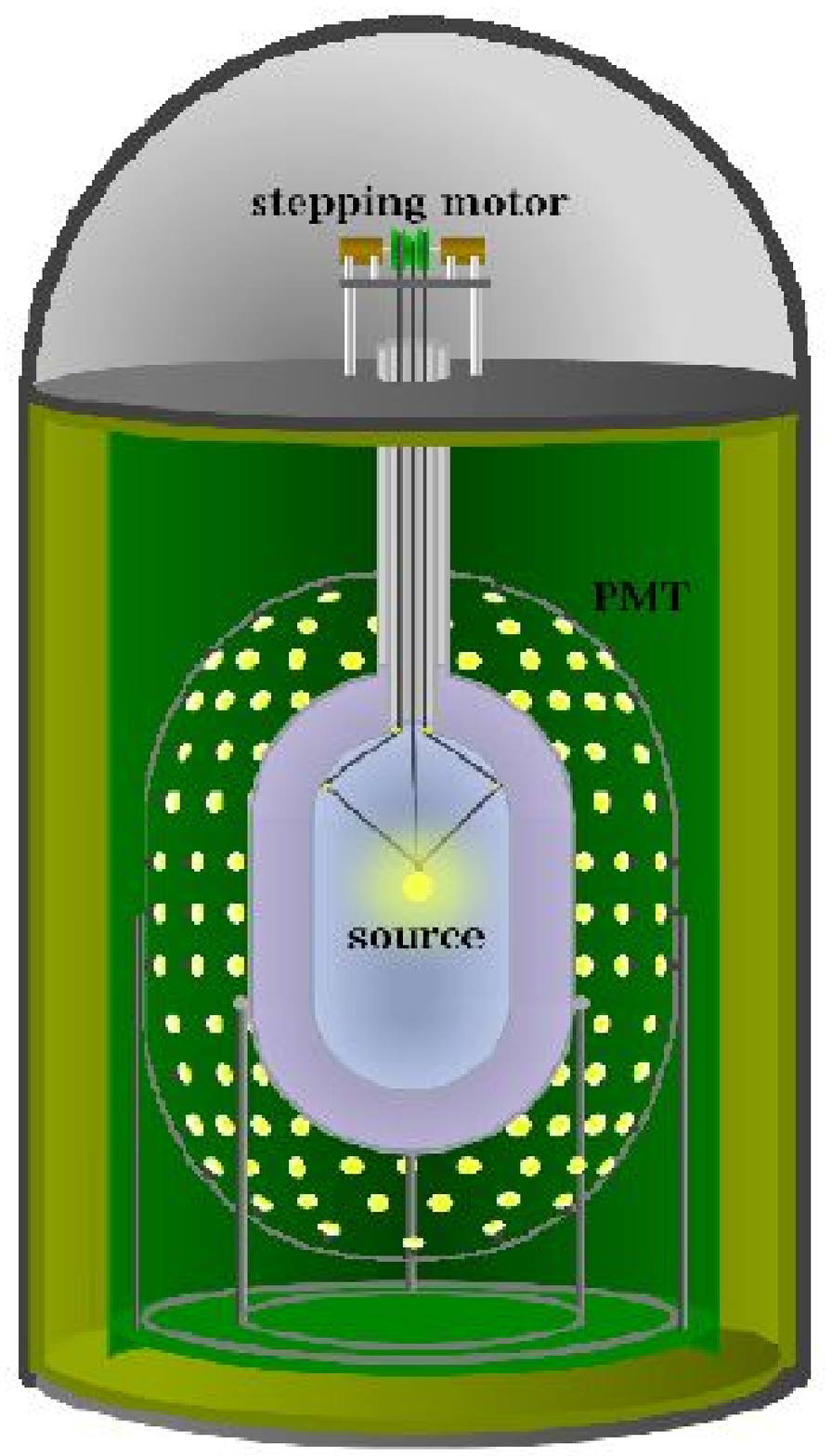} 
\includegraphics[scale=0.37] {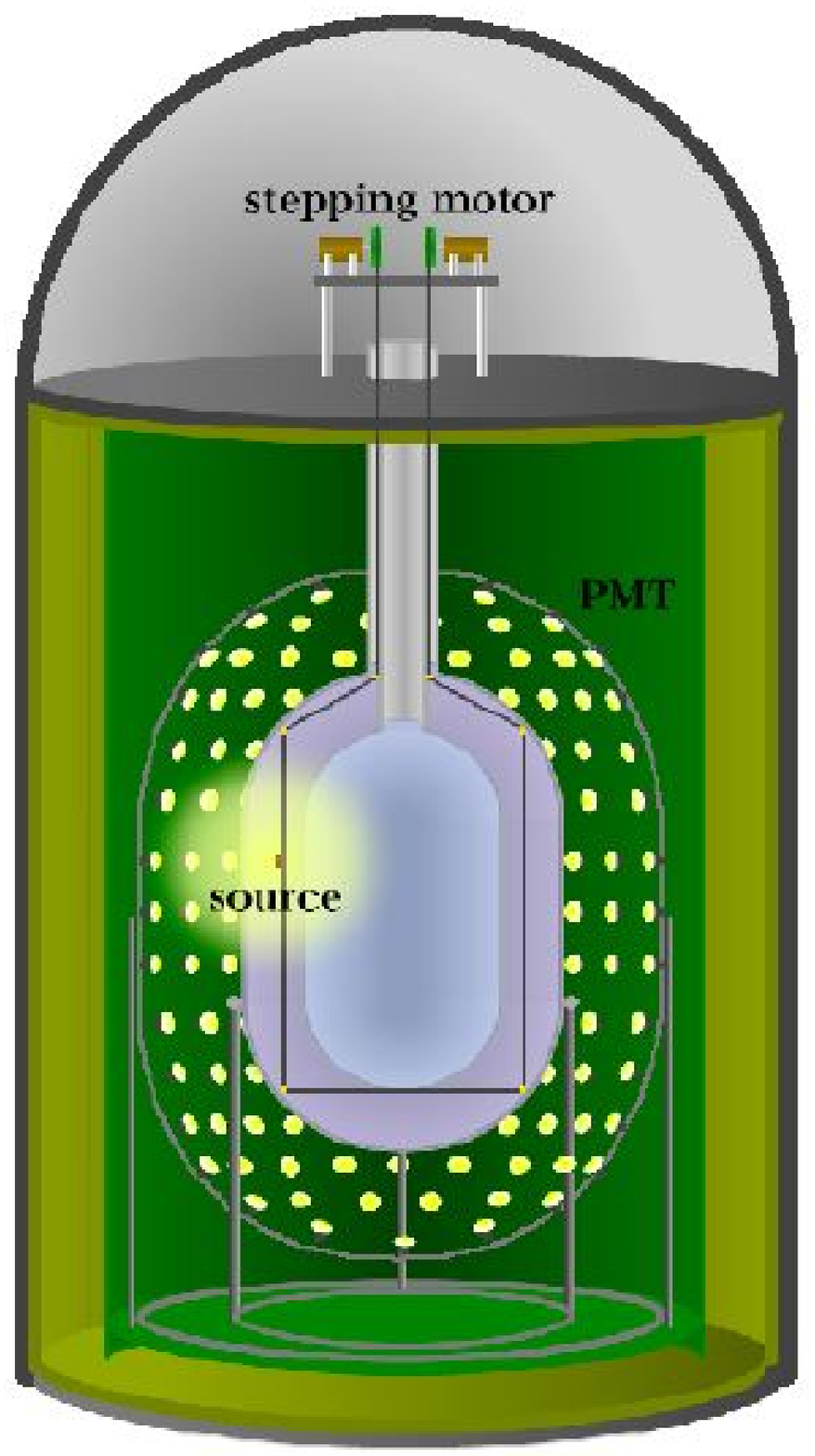}
\caption{Source manipulator}
\label{fig:manipulator}
\end{center}
\end{figure}

Fig.-\ref{fig:manipulator}-right shows how to locate the calibration source in the detector by a robot arm type manipulator. 
By using two joints and a rotating bar going through the chimney and a connecting arm, sources can be located at target points. 
In addition, by using the joint connecting a source to arm, the direction of the source can be controlled. 
By making air chambers inside the arm, it is possible to make it zero-buoyancy to make the force to move the joint small. \\

%\begin{figure}[htbp]
%\begin{center}
%\includegraphics[scale=0.35]{robot.eps}
%\caption{Robot arm type manipulator}
%\end{center}
%\end{figure}

\begin{figure}[htbp]
\begin{center}
\includegraphics[scale=0.4]{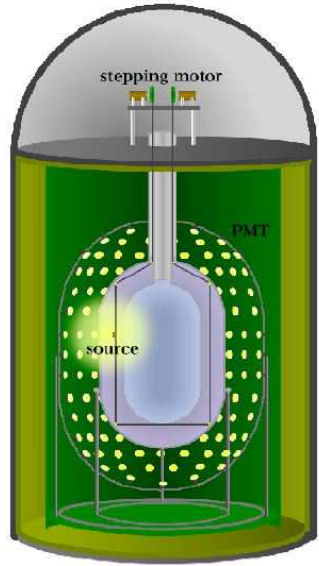}
\caption{Region II calibration}
\label{fig:Region_II_Calib}
\end{center}
\end{figure}

\textbf{Region-II}\\
As for the Region-II, the main purpose of the calibration is to monitor the change of light output and attenuation length. 
String guides will be equipped in the Region-II as shown in Fig.-\ref{fig:Region_II_Calib}. 
A string runs through the string guides.
Radioactive sources are connected to the string and moved along the vertical direction by controlling the length of the strings.\\

\textbf{Region-III}\\
Low luminous LEDs, high luminous infrared LEDs and CCD cameras are mounted on the inner surface of the stainless steel tank. 
The low luminous  LEDs are used to provide one photon level signals to PMTs to monitor the PMT gain easily. 
High luminous LEDs are placed at known positions and become a fiducial marks for the CCD cameras.
The CCD cameras are used to watch the condition of the detector and to measure the position of the 
calibration sources making use of LEDs as reference positions.

%======================================
\section{Backgrounds}
\label{sec:BKG}
 
The background reduction and its precise estimation is a main issue of KASKA experiment because 
the uncertainty associated with background estimation is one of the most severe
sources of the systematic error and the detector system has to be designed carefully to eliminate it. 
 
\begin{figure}[htbp]
 \begin{center}
  \includegraphics[width=0.8\textwidth,clip] {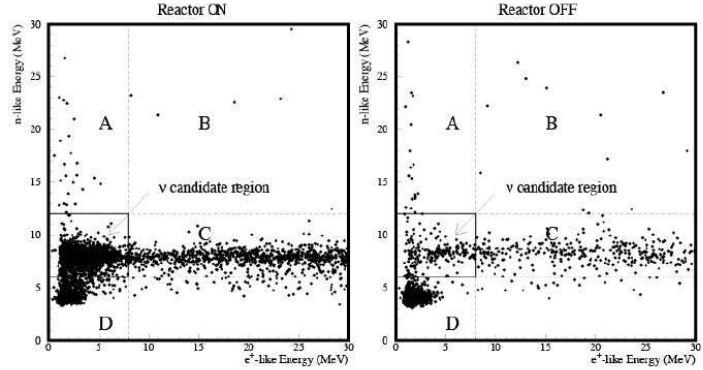} \\
 \end{center}
\caption{Correlation of prompt energy and delayed energy measured by CHOOZ  group~\cite{CHOOZ}. 
The left data were taken during reactor ON and the right data, during reactor OFF, showing backgrounds. 
The events in Region-A and D are caused by accidental coincidences, Region-B are caused by a Michel electrons from stopping $\mu$ and Region-C are supposed to be caused by the fast neutron background. 
}
\label{fig:chooz_reactor_off}
\end{figure}

Fig.\ref{fig:chooz_reactor_off} shows an example of the background measured in CHOOZ experiment.

The events in region-D are caused by accidental coincidence of $\gamma$-rays from detector material and rocks.  
Those in Region-C are interpreted as correlated background caused by fast neutron. 
The events in region-A are caused by accidental coincidence between  $\gamma$-rays and high energy signals such as recoil protons. 
The events in region-B can be associated with decaying muon after stopping in the detector.  
Those backgrounds leak in the neutrino candidate region and CHOOZ experiment observed 4.0\% of correlated background and 1.7\% of accidental backgrounds. 

The Kashiwazaki-Kariwa nuclear power station has 7 reactors and it is
impossible to take  reactor-off data. 
The strategy of the KASKA experiment for the background evaluation is to 
reduce the absolute background rate to be less than a few\% level and then to make the absolute error of the background estimation smaller than a half \%, by making use of the following methods. 

(1) Powerful reactors: 
The Kashiwazaki-Kariwa nuclear power station is the world most powerful reactor complex whose thermal power is 24.3GW, which produces  3 times more neutrinos than CHOOZ power station. 
With this high neutrino flux, the S/N ratio and statistics become large at a given detector configuration. 

(2) Underground laboratory:
  The neutrino detectors are placed in a tunnel excavated at the bottom of deep shaft holes
   to reduce the cosmic-ray rate. 
   The most severe backgrounds, such as fast neutron and correlated signals from spallation products are originated by cosmic rays. 
The water-equivalent depth of the far detector is 260~mwe, which reduces the cosmic ray rate down to 1/240 of that on the surface. 
   For near detectors,  the depth is roughly 90~mwe which reduces the cosmic-ray rate down to $\sim$1/30.  
     
(3)  Gd loaded liquid scintillator: 
The total energy of the $\gamma$-rays from the neutron captured Gd is $\sim$ 8~MeV. 
This is much larger than the highest natural $\gamma$-ray energies of 2.6~MeV
and natural highest $\beta$-decay energy of 5~MeV.
Thus the delayed signal is free from background of natural radio isotopes.
 
\begin{figure}[htbp]
 \begin{center}
  \includegraphics[width=0.8\textwidth] {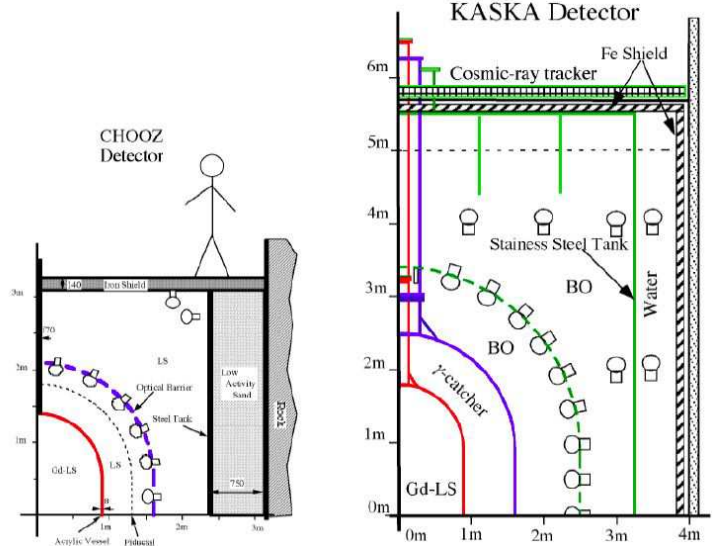}  
 \end{center}
\caption{Comparison of the CHOOZ and KASKA detector.}
\label{fig:KASKA_vs_CHOOZ}
\end{figure}

(4) Thick shields:
   The external backgrounds such as $\gamma$-rays and fast neutrons can be reduced by thick shields. 
Fig.-\ref{fig:KASKA_vs_CHOOZ} compares the KASKA detector and the CHOOZ detector. 
The thickness of the KASKA shield is much thicker than that of CHOOZ detector and the external  backgrounds 
are significantly reduced compared with the CHOOZ case. 
An option is to make the buffer oil as a weak scintillator having a light output of 1/10 of Gd-LS. 
90~cm thick buffer  between PMT array and the liquid scintillator reduces the $\gamma$-rays from PMT glass. 
There is at least 2.2m thick buffer and water barriers between the liquid scintillator and iron tank.¡¡
The $\gamma$-ray and neutron background from outside of the detector is reduced to negligible level with these buffers. 

(5) Cosmic-ray tracker: 
When cosmic-ray hit $^{12}$C, radioactive elements are produced by spallation reactions. 
Among them, long lived $\beta$-neutron emitters, such as $^9$Li or $^8$He are most significant background sources. 
Because spallation products are produced along the cosmic-ray track, the rate of such kind of background can be estimated by the distribution of distances between the cosmic-ray track and signal candidate. 
A cosmic-ray tracker system is implemented to estimate this kind of spallation backgrounds. 

%=====================================
\subsection{Background Sources}
%The accidental background is the events in which independent background signals accidentally happens within the delayed coincidence timing window and regarded as neutrino signals. 

%The accidental background rate $f_{AB}$ is expressed as, 
%\begin{equation}
%f_{AB}=f_{e^{+}}f_n\omega
%\end{equation}
%where $f_{e^{+}}$ and $f_n$ are rates of $e^{+}$ like signals and neutron like signals. 
%$\omega$ is the coincidence window. 
%For KASKA experiment, the energy windows for positron and neutron signals are, 0.7~MeV to 9~MeV and 5~MeV to 11~MeV, respectively.
%The timing window is 200~$\mu$s. 

\begin{figure}[htbp]
 \begin{center}
  \includegraphics[width=0.7\textwidth] {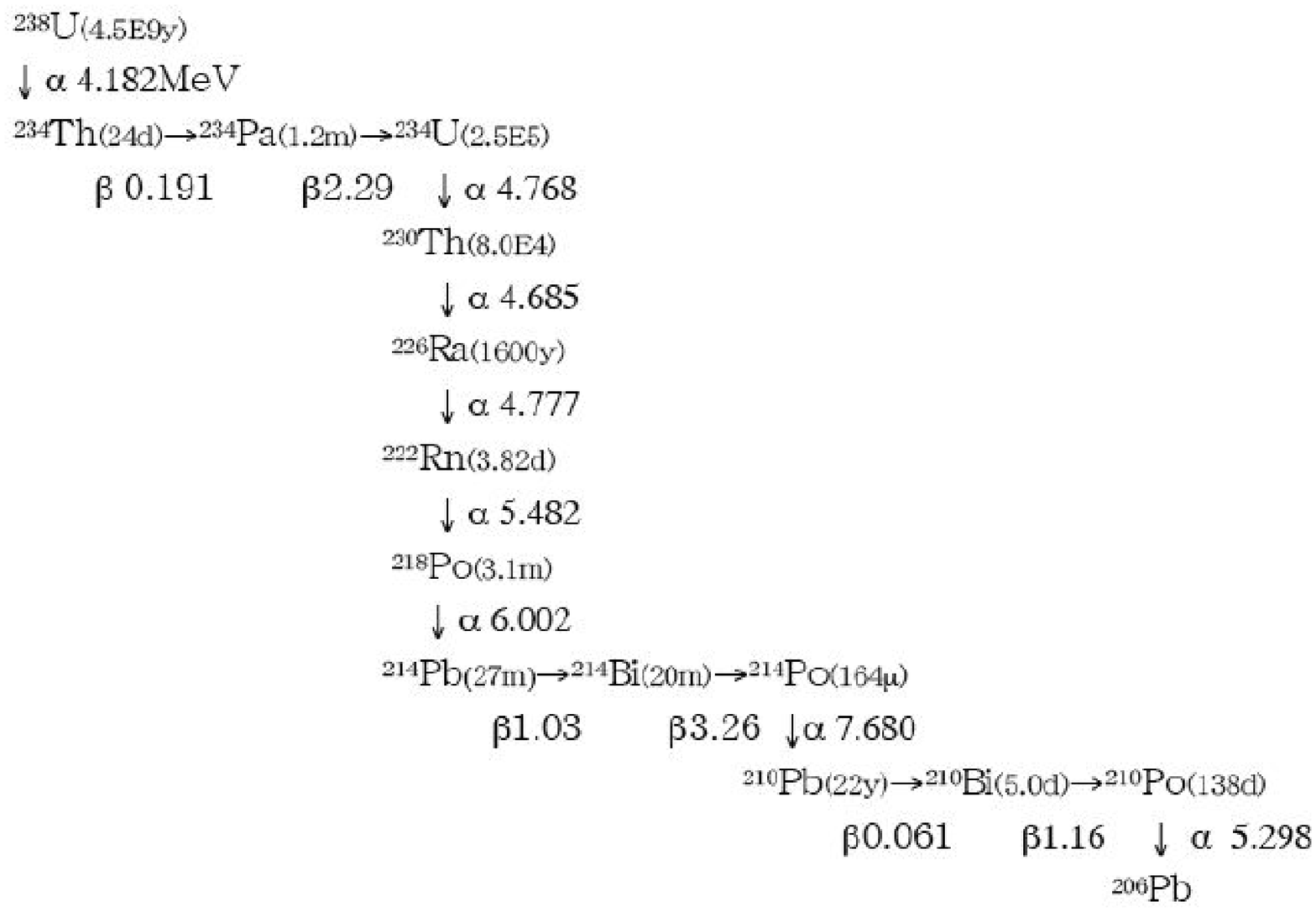}  
 \includegraphics[width=0.55\textwidth] {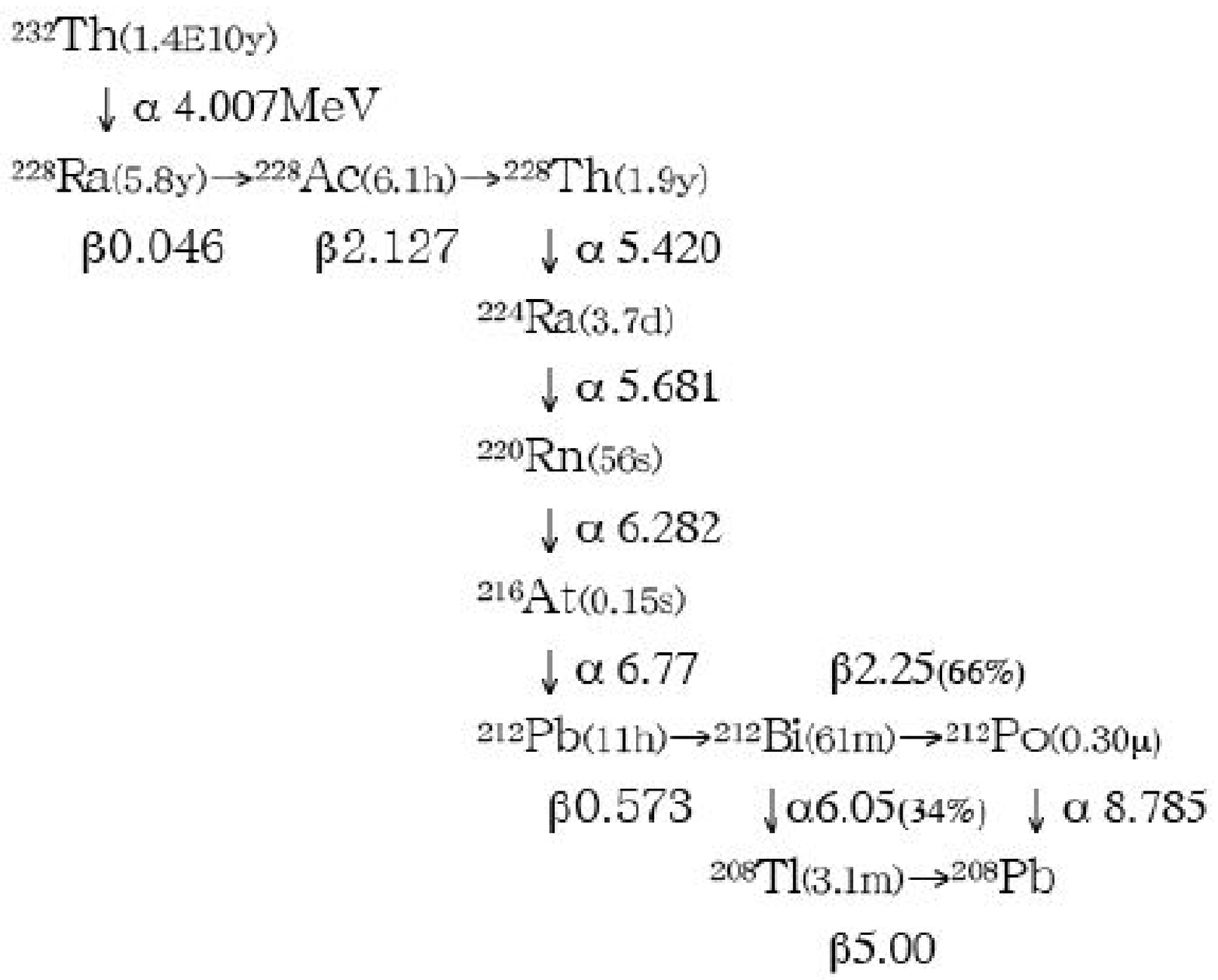} 
 \begin{verbatim}
  \end{verbatim}
 \includegraphics[width=0.4\textwidth] {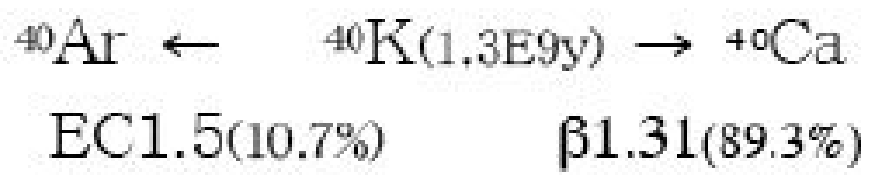} 
 \end{center}
\caption{Decay of $^{238}$U chain, $^{232}$Th chain and $^{40}$K.}
\label{fig:U_Th_K_decay}
\end{figure}

Background signals are caused by radioactive elements such as 
$^{238}$U$(\tau=4.5 \times 10^9y)$, $^{232}$Th$(\tau=1.4\times 10^{10}y)$, $^{40}$K$(\tau=1.3\times 10^9y)$ and $^{60}$Co$(\tau=7.6y)$. 
$^{238}$U, $^{232}$Th and $^{40}$K naturally exist in various materials due to their extremely long lifetime while
$^{60}$Co is artificially contaminated in steels in smelting process.\\
Fig.~\ref{fig:U_Th_K_decay} shows the decay chain of the $^{238}$U and $^{232}$Th. 
If radioactive elements decay within liquid scintillator, both $\beta$ and $\gamma$ rays  contribute as background (Internal background).
The $\alpha$ particles can not be a background because the visible energy of the  $\alpha$'s produced in natural radio isotopes are quenched to less than 0.5~MeV. 
 If these elements decays outside the liquid scintillator, only $\gamma$-rays contribute as background since the range of a  few MeV $\beta$ is only a few cm (External background). 
Table~\ref{tab:activity} shows  their background rates which satisfy the positron energy window. 
The highest decay energy of the $\beta$ decay is 5~MeV ($^{208}\textrm{Tl} \rightarrow ^{208}\textrm{Pb}$ in $^{232}$Th chain) and these radioactive elements do not satisfy the  neutron signal criterion of $\mbox{E}>5~\mbox{MeV}$.
The backgrounds for Gd signal are caused mainly by actual neutron absorption, where the neutrons are produced by cosmic-ray associated reaction with materials. 

\begin{table}[htbp]
 %\tbl{$\gamma and \beta$ production rates.}
 \begin{center}
\caption{U, Th, K, Co activities}
  {\footnotesize
   \begin{tabular}{|c|r|r|r|r|r|}
    \hline
    {} &{} &{} &{} &{} &{}\\[-1.5ex]
    Element & Bq/g & $\# \gamma$/decay & \# $\gamma$/g & \# ($\gamma + \beta$)/decay & 
    \# ($\gamma + \beta$)/g \\[1ex]
    {} &{} &(E$>$0.7MeV) &{} &(E$>$0.7MeV) &{}\\[-1.5ex]
    {} &{} &{} &{} &{} &{} \\[-1.5ex]
    \hline
    {} &{} &{} &{} &{} &{} \\[-1.5ex]
    $^{238}$U    &12,400   &2         &25,000 &3     &37,000 \\[1ex]
    $^{232}$Th  &4,100     &1.5     &6,200    &2      &8,200 \\[1ex]
    $^{40}$K      &255,000 &0.11  &28,000  &0.5  &130,000 \\[1ex]
    $^{60}$Co    & -               & 2        &         -      & 1      & -     \\[1ex]
    \hline
   \end{tabular} }
 \end{center}
 \vspace*{-13pt}
\label{tab:activity}
\end{table}

\begin{table}[htbp]

 %\tbl{$\gamma and \beta$ production rates.}
 \begin{center}
\caption{Internal background}
  {\footnotesize
   \begin{tabular}{|c|r|r|r|r|r|}
    \hline
    {}                      &{}            &{}                 &{}                 &{}               &{}                                    \\[-1.5ex]
    Component        & Quantity & $^{238}$U & $^{232}$Th& $^{40}$K  &  $\beta + \gamma $ rate  \\[1ex]
    {}                      &(kg)         &    (ppt)         & (ppt)             & (ppt)            & (Hz)                               \\[-1.5ex]
    {}                      &{}            &{}                 &{}                 & {}                &{}                                  \\[-1.5ex]
    \hline
    {}                      &{}            &{}                 &{}                 &{}                  &{}                                 \\[-1.5ex]
    Scintillator Oils & 25,000     & $<0.1$        & $<0.1$         & $<0.1$         & $<0.5$                          \\[1ex]
       \hline
    {}                      &{}             &{}                &{}                 &{}                  &{}                                  \\[-1.5ex]
    PP0                   & 150           & 4                 & $< 50 $        & 4.2                & $<0.17$                       \\[1ex]
         \hline
            {}              &{}              &{}               &{}                  &{}                  &{}                                  \\[-1.5ex]
       bis-MSB         & 2.5             & 48              & 60                 & 6.8                 & 0.008                           \\[1ex]
 \hline
    {}                       &{}             &{}                &{}                 &{}                  &{}                                \\[-1.5ex]
   Gd                        & 6.1           & 8.4              & 12                & $< 1200 $     & $<0.96$                     \\[1ex]   
  \hline
     {}                      &{}              &{}                &{}                 &{}                  &{}                             \\[-1.5ex]
    Acrylic               & 700           & 8                 & 50                &  8.2               & 1.2                            \\[1ex]
    \hline
     \hline
       {}                    &{}              &{}                 &{}                &{}                  &{}                         \\[-1.5ex]
    Internal Total      &                  &                     &                    &                     & $ <2.9 $ \\[1ex]
   \hline
    \end{tabular}
  }
  \end{center}
 \vspace*{-13pt}
 \label{tab:Internal_BKG}
\end{table}
    
%-------------------------------------------------
\subsubsection{Internal Background}
* The liquid scintillator\\
The amount of oil used for both LS-I and  LS-II is  25~tons.
The liquid scintillator will be formulated by dissolving PPO, bis-MSB and organic Gd into an oil which is made mainly of pseudocumene and paraffin oils. 
The U/Th/K contamination of these materials  has been measured by ICP-MS and flame  atomic absorption method. 
 The oils are produced from a number of steps of distillation and refinery processes and are intrinsically pure. 
 The U/Th/K contaminations are less than $10^{-13}$g/g and the total single rate from oil is less than 0.5~Hz.
The single rate from  PPO, bis-MSB and Gd samples were estimated to be  less than 1.2~Hz from its measured contaminants.\\

    * Acrylic vessels \\
    The weights of the inner and outer acrylic vessels are 240~kg and 900~kg, respectively. 
  Because half of the $\gamma$-rays from the outer acrylic vessel do not come in the scintillator region,
the effective mass of the outer vessel is 450~kg and the effective total acrylic weight is about 700~kg.
The single rate from acrylic tanks are estimated to be 1.2~Hz. \\
In summary, the internal background is expected to be $< 2.9$~Hz as shown in table~\ref{tab:Internal_BKG}.

\subsubsection{External Background}
For external background, only $\gamma$-rays can enter liquid scintillator.

   \begin{table}[htbp]
 %\tbl{$\gamma and \beta$ production rates.}
 \begin{center}
   \caption{external background~\cite{Furuta}}
  {\footnotesize
   \begin{tabular}{|c|r|r|r|r|r|r|r|}
    \hline
     {} &{} &{} &{} &{}&{}&{}\\[-1.5ex]
    Material & Quantity & $^{238}$U & $^{232}$Th& $^{40}$K & $^{60}$Co &  Hit rate  \\[1ex]
    {}                    &(ton)   &    (ppb)           & (ppb)              & (ppb)          & (mBq/kg)      & ($>$0.7MeV)(Hz)    \\[-1.5ex]
    {} &{} &{} &{} & {}&{}&{}\\[0.1ex]
    \hline
      {}&{}&{}  &{} &{} &{} &{}                        \\[-1.5ex]
      PMT Glass & 0.3   &46 &120 &10  & -  & 1.2 \\[1.5ex]
  \hline
    {}&{}&{}  &{} &{} &{} &{}                        \\[-1.5ex]
     PMT support & 2    & 1.9 & 4.7 & 0.08 & 9 &  0.2 \\[1.5ex]
\hline  
  {}&{}&{}  &{} &{} &{} &{}                        \\[-1.5ex]
     S. S. Tank & 32    & 1.9 & 4.7 & 0.08 & 9 &  0.05 \\[1.5ex]
      \hline
        {}&{}&{}  &{} &{} &{} &{}                        \\[-1.5ex]
     Iron Tank & 300    & 1.9 & 4.7 & 0.08 & 9 &  0.01 \\[1.5ex]
  \hline
    {}&{}&{}  &{} &{} &{} &{}                        \\[-1.5ex]
 Soil               & -   & 3,000 & 8,000 & 2,000  & - & 0.15\\[1.5ex]
  \hline
    {}&{}&{}  &{} &{} &{} &{}                        \\[-1.5ex]
Concrete        & 180   & 2,800 & 5,400 & 1,700  & - & 0.43\\[1.5ex]

           \hline
           \hline
             {}&{}&{}  &{} &{} &{} &{}                        \\[-1.5ex]
            Total   &  &  &  &   &  & 2.0 \\[1.5ex]

     \hline
    \end{tabular}
   }
  \end{center}
 \vspace*{-13pt}
       \label{tab:external_BKG}
\end{table}

* PMT Glass  \\
In KASKA, 300 10" PMT will be used. 
The total weight of the PMT glass is 300~kg.  
For Hamamatsu low background glass, 
U/Th/$^{40}$K contaminations are measured to be 46~ppb, 120~ppb and 10~ppb, respectively and 650/s of $\gamma$-rays which satisfy the positron energy cut are generated from the glass.
After penetrating 70~cm thick  buffer oil, the single rate in the LS becomes 1.2~Hz.

* $\gamma$-rays from the stainless steel and iron structure. \\
The mass of the stainless steel tank, PMT and steel tank  is 30~tons,  2~tons and  300~tons, respectively.   
Generally a steel contains $\sim 10m \mbox{Bq/kg}$ of $^{60}$Co as well as U/Th/K. 
$^{60}$Co emits two $\gamma$-rays whose energies are, 1.17~MeV and 1.33~MeV.
After passing through at least 90~cm of buffer oil, the single rate in the LS becomes 0.25~Hz. 

* gamma rays from soil \& concrete: \\
The radioactive contaminations in the soil at the near-B point are measured in a boring study to be, 
U/Th/K=3/8/2~ppm. 
A simulation shows the single rate of $\gamma$-ray hit from soils and concrete is only  0.6~Hz 

\begin{figure}[htbp]
 \begin{center}
  \includegraphics[width=0.7\textwidth] {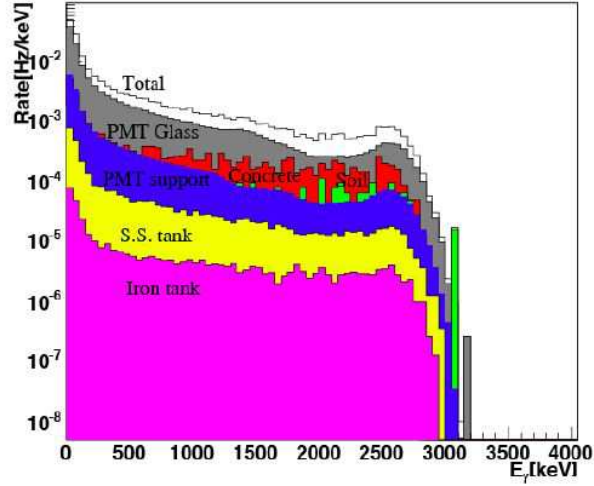}  
 \end{center}
\caption{External background spectrum~\cite{Furuta}}
\label{fig:externalBG}
\end{figure}

The total single rate which is originated from natural radio activities is expected to be $<5$~Hz. 
Fig.~\ref{fig:externalBG} shows the simulated external background spectrum. 

%===========================
\subsubsection{Cosmic Rays}

\begin{figure}[htbp]
 \begin{center}
  \includegraphics[width=0.7\textwidth] {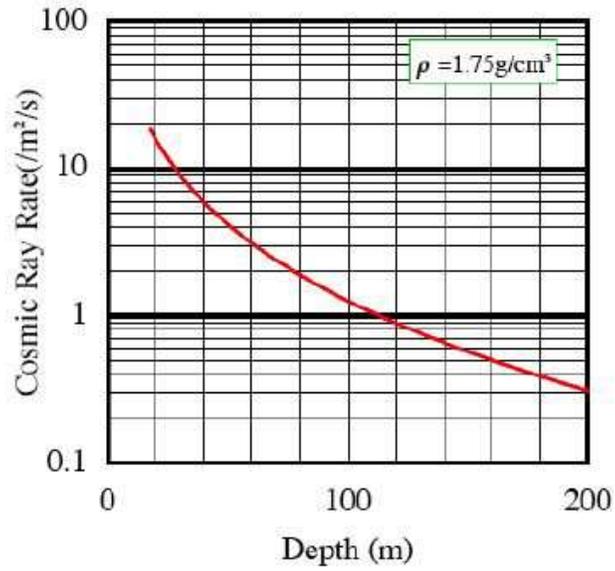}  
 \end{center}
\caption{Cosmic Ray Rate}
\label{fig:cosmic_rate}
\end{figure}

\begin{figure}[htbp]
 \begin{center}
 \includegraphics[width=1\textwidth] {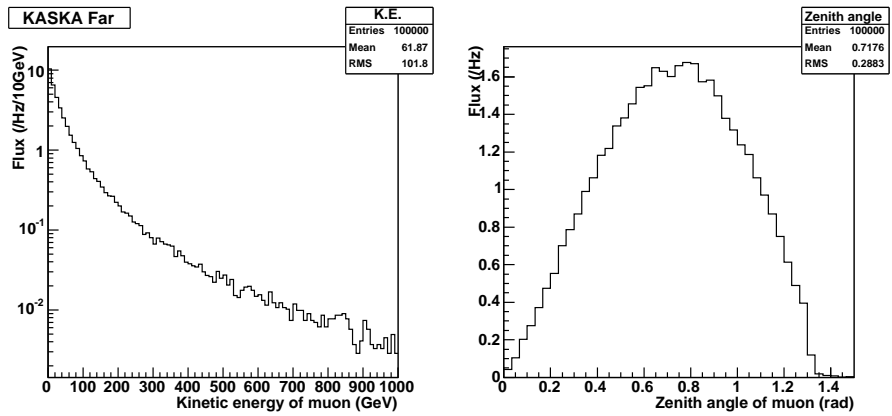} 
 \end{center}
\caption{Cosmi ray energy spectrum (left) and zenith angle distribution (right) at far detector depth~\cite{Kiyomi}.}
\label{fig:cosmicFar}
\end{figure}

In order to reduce cosmic-ray associated background, the detectors are placed at 150m depth underground for far detectors and at 50m depth for near detectors.  
 The mean density of the soil around Kashiwazaki-Kariwa Nuclear Power
 station is 1.75g/cm$^3$.  
30 weight \% is water component. 
The water equivalent overburden at 150~m underground is 260~meter-water-equivalent (mwe) and that at 50~m undergaround is 90~mwe.
Fig-\ref{fig:cosmic_rate} shows  the cosmic ray flux at various depth calculated by an empirical formula~\cite{nenpyo}; 
\begin{equation}
 I(h)=1.4 \times \frac{1,740,000}{h+400} \times 
 (h+11)^{-1.53}\exp(-7.0\times10^{-4}h) [/\mbox{m}^2/\mbox{s}]
\label{eq:mu_rate}
\end{equation}
where $h$ is the depth measured in  mwe  and the factor 1.4 is a normalization factor which is chosen to match
to our simulation result at 150m depth.
Fig.~\ref{fig:cosmicFar} shows the energy spectrum and zenith angle distribution at far detector depth~\cite{Kiyomi}. 
At 150m depth, the cosmic ray flux is $\sim 0.55/\mbox{m}^2/\mbox{s}$ and at 50m depth it is $\sim 6.6/\mbox{m}^2/\mbox{s}$.
The cosmic-ray hit rate at inner detector region (region-I, II, III) is $\sim$15~Hz for far detectors and $\sim$170~Hz for near detectors.  
The hit rate at outer detector region (region-IV, V) is $\sim$40~Hz for far detectors and $\sim$500~Hz for near detectors.
The mean energies of the cosmic rays are $\sim$70~GeV and $\sim$20~GeV, respectively. 

The cosmic-ray veto system consists of at least 1.6~m thick buffer oil in region-IV for vertical direction and 0.7~m thick buffer oil in region-IV and 0.5~m thick water in region-V in horizontal direction.
A minimum ionizing particle deposits at least 260~MeV of energy in region-IV at vertical direction,  110~MeV of energy in region-IV and 100~MeV of energy in region-V at horizontal direction. 
Together with the water cherenkov veto and cosmic-ray tracker, $>$99.5\% of cosmic-ray rejection  efficiency is expected. 
A fast neutron which collides with a proton in the region-IV and deposit visible energy of  few MeV is also expected to be vetoed by cosmic-ray anti counter.  
In order to avoid the instantaneous cosmic-ray associated backgrounds, 0.5~ms of dead time will be applied after each muon passage in the liquid scintillator region and  0.2~ms of dead time is set after the muon passage in non scintillating region. 
The livetime inefficiency introduced by those dead time is around 1.6~\% for far detectors and 20~\% for near detectors.
The $\bar{\nu}_e$/$\mu$ rate is higher for near detectors than far detectors and the mean muon energy at near detector depth is only 1/3 for far detector depth. 
Therefore the N/S of  cosmic-ray associated background in near detectors is a half of that of  far detectors. \\

\begin{figure}[htbp]
 \begin{center}
 \includegraphics[width=0.8\textwidth] {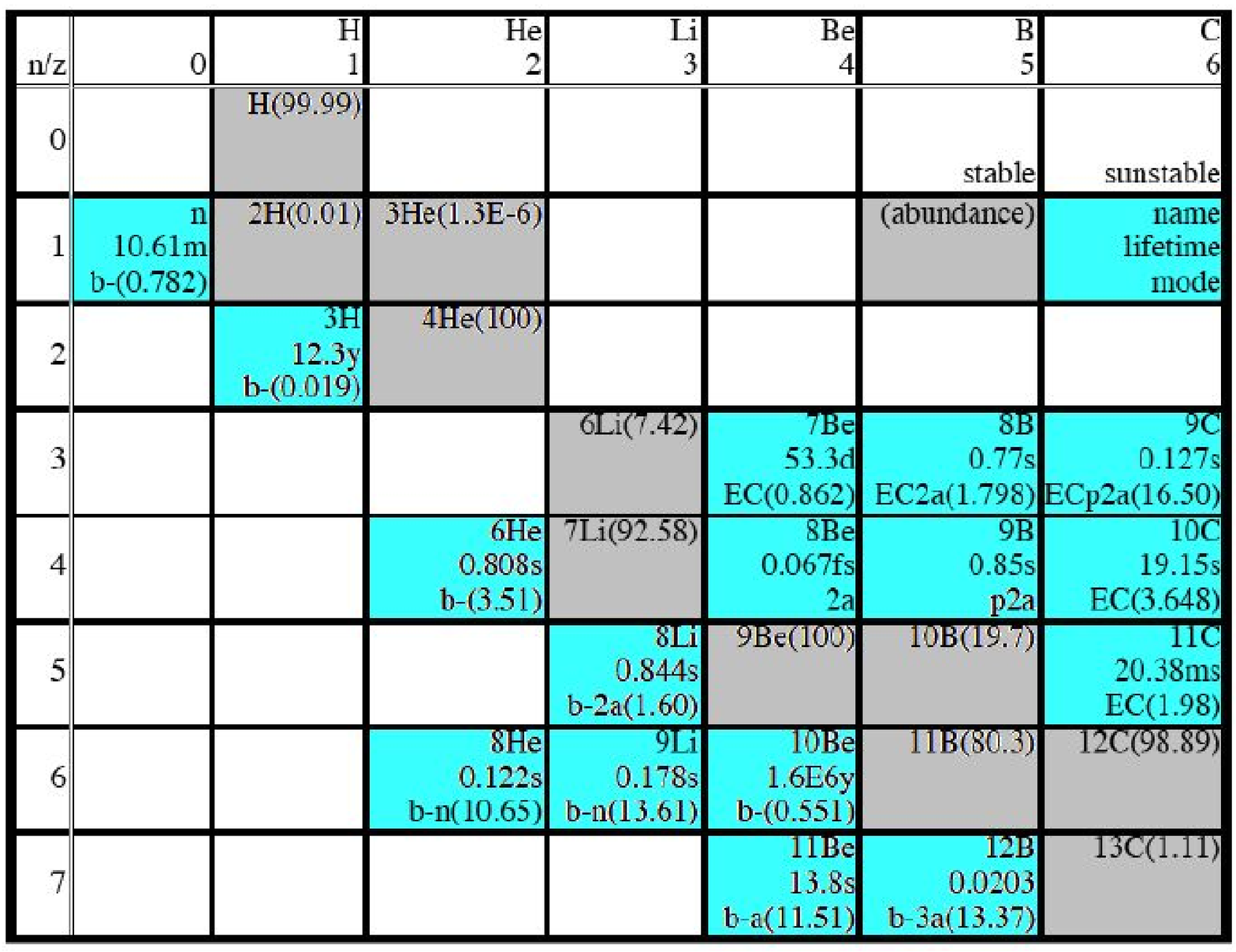} 
 \end{center}
\caption{Long life radio isotopes possibly produced by cosmic-ray spallation reaction of carbon~\cite{ToI96}.}
\label{fig:light_RI}
\end{figure}

%\subsubsection{Cosmic-ray associated background}
The liquid scintillators consist of hydrocarbonic oils and chemicals which consist of hydrogen, carbon, nitrogen and oxygen. 
When a cosmic ray hits such nucleus, it may break the nuclei producing radioactive nucleus. 
However, the atomic components of the LS's are roughly H:C:N:O=$1.8:1:O(10^{-3}):O(10^{-2})$
and carbon contributes mainly to such spallations.
An example of such reaction is,
\begin{equation}
\mu + ^{12}\mbox{C} \rightarrow \mu + ^{11}\mbox{C}^{*} +n
\label{eq:11C_spallation}
\end{equation}
The excitation of the produced nucleus decays instantly and its signal is killed by cosmic-ray veto. 
However, $^{11}$C decays with electron capture with half life of 20.4~ms, which is long enough to survive the cosmic-ray veto. 
The neutron produced will be absorbed by proton or Gd, which produces a delayed signal. 
Fig.~\ref{fig:light_RI} shows a list of long lived radioactive elements which can be produced by spallation of carbon nucleus. 
Table~\ref{tab:spallation_list} shows their production rate in region I and II, calculated by using the cross sections taken from \cite{Hagner} and assuming that the cross section is proportional to $\bar{E}_\mu(\mbox{GeV})^{0.75}$,
where  $\bar{E}_\mu$ is a mean energy of cosmic-rays.
\begin{table}[htbp]
\caption{Production rate (Region-I+II)}
  \begin{center}
  {\footnotesize
   \begin{tabular}{|c|r|r|r|}
    \hline
                                            & Far & Near   \\[1ex]
 \hline
    mean energy (GeV)         & 68  & 19     \\
    $\mu$ flux (/m$^2$/s)     & 0.55  & 6.6      \\
    \hline
     \hline
    {} & {} &{}  \\[-1.5ex]
  
 $^{11}$C(20.4ms)    & $2.6\times10^{-2}$    & $1.9\times 10^{-1}$   \\[1ex]
 $^{7}$Be(53.3d)    & $5.8\times10^{-3}$    & $4.8\times 10^{-2}$   \\[1ex]
 $^{11}$Be(13.8s )  &  $<5.5\times10^{-5}$    & $<5.0\times 10^{-4}$   \\[1ex]
 $^{10}$C(19.15s)    & $3.5\times10^{-3}$    & $2.4\times 10^{-2}$   \\[1ex]
 $^{8}$Li(0.844s )   & $1.4\times10^{-4}$    & $8.4\times 10^{-4}$   \\[1ex]
 $^{6}$He(0.808s)  & $4.6\times10^{-4}$    & $3.3\times 10^{-3}$   \\[1ex]
 $^{8}$B( 0.77s)  & $1.9\times10^{-4}$    & $1.5\times 10^{-3}$   \\[1ex]
 $^{9}$C(0.127s)   & $1.6\times10^{-4}$    & $7.5\times 10^{-4}$   \\[1ex]
 $^{9}$Li(0.178s)+$^{8}$He(0.122s)  & $6.0\times10^{-5}$    & $2.8\times 10^{-4}$   \\[1ex]
   \hline
     {} & {} &{}  \\[-1.5ex]
   sum &$3.6\times10^{-2}$ /s   & $2.7\times 10^{-1}$ /s  \\[1ex]
              \hline
   \end{tabular} }
 \end{center}
 \vspace*{-13pt}
\label{tab:spallation_list}
\end{table}
The total production rate within region-I and II is 0.036/s for far detector and 0.27/s for near detectors, both are much smaller than $\gamma$-ray rate from detector materials.
If the neutrons produced by the spallation reaction is absorbed by Gd,  they generate delayed signal. 
The rate of such neutron signal which survives the  cosmic-ray veto time is, 
 \begin{equation}
f_n <N_n f_I \left( \exp\left(-\frac{T_{\mbox{veto}}}{30\mu s}\right)+
\beta \exp\left(-\frac{T_{\mbox{veto}}}{200\mu s}\right) \right) ,
\end{equation}  
where $N_n$ is the number of neutrons produced in the spallation reaction,  $f_I$  is production rate of radioactive spallation nucleus in region-I and
$T_{\mbox{veto}}$ is the cosmic-ray veto time. 
$\beta$ is the fraction that a neutron produced in region-II 'spills-in' to the region-I and is absorbed by Gd, which is an order of 10\%. 
For the case of $^{11}$C and $T_{\mbox{veto}}=0.5$~ms,  $f_n <3.2$~/day for far detector. 
For other nuclei, it is much less than that and total neutron rate is less than 5~/day.

%If spallation products decay within 
%the delayed coincidence window with respect to the neutron signal, they become  semi-correlated backgrounds. 
%The rate of the semi-correlated background is 
%\begin{equation}
%f_{sc}\sim \frac{\Delta T_{coin}}{\tau_{sp}} f_n
%\end{equation}
%where $\Delta T_{coin}$ is the delayed coincidence window, $\tau_{sp}$ is the lifetime of the spallation nucleus.
%For $^{11}$C case, the semi-correlated background rate is 0.05~/day. 
%Other nuclei are also negligibly small. 

As described in the next section, there is a case that fast neutron produced by hard collision of $\mu$ and nucleus will be thermalized in the liquid scintillator and eventually absorbed by Gd. 
According to a MonteCarlo simulation, this kind of background takes place at a rate of 2.2/day in far detector assuming 
99.5\% efficiency for muon veto. 

\subsection{Accidental Backgrounds}

Accidental background is the events in which independent background signals accidentally happens within the delayed coincidence timing window and regarded as neutrino signals. 

The accidental background rate $f_{AB}$ is expressed as, 
\begin{equation}
f_{AB}=f_{e^{+}}f_n\omega
\end{equation}
where $f_{e^{+}}$ and $f_n$ are rates of $e^{+}$ like signals and neutron like signals. 
$\omega$ is the coincidence window. 
For KASKA experiment, the energy windows for positron and neutron signals are, 0.7~MeV to 9~MeV and 5~MeV to 11~MeV, respectively.
The timing window is 200~$\mu$s. 
 
Table~\ref{tab:single_rate} summarize the single rate and accidental background rate for far detector. 
Accidental background rate can be estimated precisely by opening coincidence window randomly. 
 
\begin{table}[htbp]
\caption{Single rate and accidental background summary.}
  \begin{center}
  {\footnotesize
   \begin{tabular}{|c|r|r|}
\hline  
Source    & $f_{e^+}(/s)$  & $f_n$ (/day)   \\[1ex]
 \hline
     \hline
    {}                            & {}                     &{}       \\[-1.5ex]
Internal                        & $<$2.9              & -            \\[1ex]
External                       & 2.0                & -             \\[1ex]
Cosmic ray spallation  &  0.036            & $<7.2$         \\[1ex]
   \hline
   sum                           &  $<5$         & $<7.2$ \\[1ex]
     \hline
   Accidental background &  \multicolumn{2}{|c|}{$<$0.01/day}  \\[1ex]
     \hline
   \end{tabular} }
 \end{center}
 \vspace*{-13pt}
     \label{tab:single_rate}
\end{table}

For near detectors, S/N is expected to be better than far detector because the mean energy of cosmic-ray is much smaller 
and the reaction cross section is smaller. 

%==================================
\subsection{Correlated Backgrounds}
Correlated background is a signal in which a single event produces two contiguous signals, each satisfies the prompt and delayed energy windows  for  $\bar{\nu}_e$ selection within coincidence time window.  
Fast neutron background and $\beta$-neutron emitting radio isotopes are categorized as correlated backgrounds.

%-----------------------------------------------------------
\subsubsection{ Fast Neutron Background }
When a fast neutron collides with  proton in a scintillator, there is a chance that the visible energy of the recoiled proton falls in the energy window for $\bar{\nu}_e$ prompt signals.  
Also there is a chance that the original neutron looses its energy scattered by protons around and eventually absorbed by Gd.  
In this case the fast neutron signals are misjudged as $\bar{\nu}_e$ events. 
Such fast neutrons are often produced by inelastic scattering of cosmic-ray muon,
\begin{equation}
\mu + A \rightarrow \mu + n + X
\end{equation}

\begin{figure}[htbp]
 \begin{center}
  \includegraphics[width=0.6\textwidth,clip] {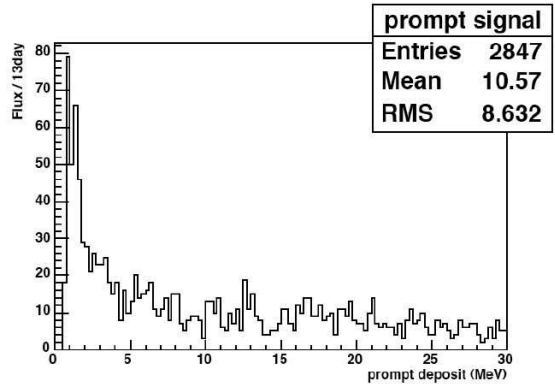} \\
 \end{center}
\caption{Visible energy distribution of delayed energy~\cite{Kiyomi}.}
\label{fig:neutron_prompt}
\end{figure}

According to the KASKA simulation, the fast neutron background (FNB) coming from the muons which go through outside of the veto system is negligibly small thanks to the thick buffer regions. 
Most of the FNB comes from the muons which come in the detector region but can not be detected due to the inefficiency of the veto system. 
For 99.5\% of veto efficiency,  0.2~/day after the event selection, or $\sim0.7\%$ of the $\bar{\nu}_e$ signal is expected from the simulation. 
The rate of FNB can be estimated by number of ways. 
Because most of the FNB  is associated with the veto inefficiency,  the rate can be estimated by applying veto inefficiency artificially when analyzing the data. 
It  can be  estimated from the energy distribution of the  prompt signal in the non-reactor 
neutrino energies or spallation background energies, such as beyond 15~MeV and less than 1~MeV. 
Fig.~\ref{fig:neutron_prompt} shows visible energy distributions calculated by a MonteCarlo simulation.
The figure shows the energy distribution is flat down to 4~MeV and there is an enhancement  at $E<4$~MeV. 
The enhancement can be estimated from the event rate below 1~MeV. 
It will be possible to understand the energy distribution with real data by tagging cosmic-rays. 
The radial position dependence of the events also gives information about fast neutron background since the neutron 
backgrounds tend to occur near the detector surface. 
The Palo-Verde type liquid scintillator and $\gamma$ catcher liquid scintillator have nentron/$\gamma$  pulse shape discrimination capability and this feature may be of help to separate the neutron and positron signals. 
Assuming the rate of fast neutron background can be estimated with $<$30~\% accuracy, the absolute error of this background estimation is $<$0.2~\%. 

%------------------------------------------------------------------------
\subsubsection{Spallation Background }
Some spallation products have decay mode to produce $\beta+n$.  
In such decays, the $\beta$ signal mimics the prompt signal of the reactor neutrino and  the neutron is absorbed by Gd producing exactly same signal as reactor neutrino delayed one.
Among these $\beta+n$ emitter,  $^9$Li and $^8$He are most serious background source because 
they can be produced from $^{12}$C and have long lifetimes and can not be rejected by the veto time applied after muon passage.  

KamLAND group showed that $^9$Li dominates and contribution from $^8$He was less than 15\%~\cite{Shimizu} in their data taken at 1000m deep underground and its amount is consistent with reference-\cite{Hagner}.
According to the reference, the production rate of $^9$Li+$^8$He in region-I  is 1.2~/day for far detectors. 
Because the branching ratio of the  $\beta - n$ decay mode is 50\% for $^9$Li and 16\% for $^8$He, the neutron associated background rate becomes at least half, which corresponds to $<$1.5~\% of the reactor $\bar{\nu}_e$ events.
The other $\beta - n$ emitters are assumed to be at least $O(10^{-2})$ times smaller because they have
to be produced from oxygen nuclei. 
The $^9$Li background rate can be estimated with following methods.\\
*Energy Spectrum: Because the Q value for neutron associated decay of $^9$Li is 11.2~MeV, 
this background can be estimated from the number of events between 9~MeV to 11~MeV.  
$\sim$5\% of $^9$Li signals falls in this non $\bar{\nu}_e$ energy window. \\
* Cosmic-ray track: 
Because decay mode of the $^9$Li is $\beta$ decay, the prompt like signal occurs within a few cm along the cosmic-ray track.\\
The production cross section and its extrapolation on mean muon energy was measured with accuracy 20\%~\cite{Hagner}. 
Assuming the background rate can be estimated with precision $<$30\%, the absolute error comes from this background is $<$0.5\%.  \\

%----------------------------------------
\subsubsection{Stopping $\mu$}
\begin{figure}[htbp]
 \begin{center}
  \includegraphics[width=0.6\textwidth] {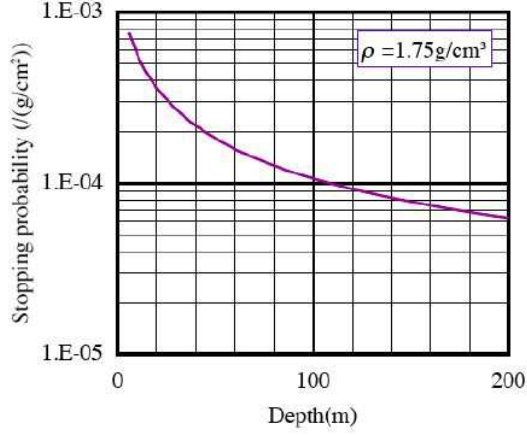}  
 \end{center}
\caption{Cosmic-ray stopping probability.}
\label{fig:stopping_probability}
\end{figure}

Fig.~\ref{fig:stopping_probability} shows the probability for cosmic-rays to stop after passing through 1~g/cm$^2$ of material.
In order for cosmic rays to satisfy  visible energy criteria of prompt energy selection, the muons have to stop at less than 40~cm from the PMT layer if weak scintillator is used for the buffer oil.
The probability to stop less than 40~cm in the region-III oil is 0.4\% and absolute rate is  $\sim$200/hr for far detectors. 
Assuming 99.5~\% of cosmic-ray veto efficiency, the survival rate becomes $\sim$1/hr. 
However, the stopping position is still in region-III and can easily be distinguished from $\bar{\nu}_e$ events which occur within region-I. 

Most of stopping muons decay as,
\begin{equation}
\mu \rightarrow e+\nu_{\mu}+\nu_e
\end{equation}
The maximum kinetic energy of electron or positron is 53~MeV which produce $\sim$5~MeV of visible energy in region-III.  Some of these events satisfy the delayed signal criteria. 
Assuming 10\% of delayed energy cut efficiency for the Michel electron, and timing cut efficiency of 
$e^{-1\mu s/2.1\mu s}$, the correlated event rate is 1.5~event/day.
For $\mu^{-}$ case, 7.9\% of stopped $\mu^{-}$ is captured on $^{12}$C and 80\% of which goes to unstable nucleus which typically emits neutron~\cite{white_paper},
\begin{equation}
\mu^-+^{12}\mbox{C} \rightarrow n+X+\nu_{\mu}
\end{equation}

If the stopping muon, which satisfies the prompt event selection, emits neutron, it can be a correlated background. 
Assuming 1~\% of neutron produced by such stopping muon is captured by Gd, the rate of this background is, $\sim$2.5 events/day. 

However,  these events can be easily distinguished from $\bar{\nu}_e$ events making use of the extreme position 
difference of the signals and reduced to a negligible level. \\

Table~\ref{tab:bkg_summary} summarizes the expected background rate and its estimation error. 

\begin{table}[htbp]
\caption{background summary}
 \begin{center}
 {\footnotesize
  \begin{tabular}{|l|c|c|}
   \hline
   {} &{} &{}\\[-1.5ex]
      source     & absolute & error    \\
   {} &{} &{}\\[-1.5ex]
   \hline
   {} &{} &{}\\[-1.5ex]
   Accidental             &   $<$0.02\%   & negligible    \\
   Fast Neutron         &   0.7\%  &  $<$0.2\%        \\
   $^9$Li+$^8$He   &   $<$1.5\%  & $<$0.5\%     \\
   Stopping $\mu$    &   nagligible & negligible               \\
    {} &{} &{}\\[-1.5ex]
       \hline
       {} &{} &{}\\[-1.5ex]
  Combined               &  $<$2.2\%   & $<$0.5\%   \\
   \hline
  \end{tabular} }
 \end{center}
       \label{tab:bkg_summary}
\end{table}

 For near detector case, the S/N is two times better than the far detector case and the absolute error after near/far comparison is 0.6\% or better.

%=================================
\section{Systematic Error}

\label{sec:systematics}

It is difficult to accurately estimate systematic error before experiment starts. 
In this section an upper limit of KASKA systematic error is estimated by comparing the recorded errors by CHOOZ experiment. 

%============================
\subsection{Near/Far cancellation}
\begin{figure}[htbp]
 \begin{center}
  \includegraphics[width=0.6\textwidth] {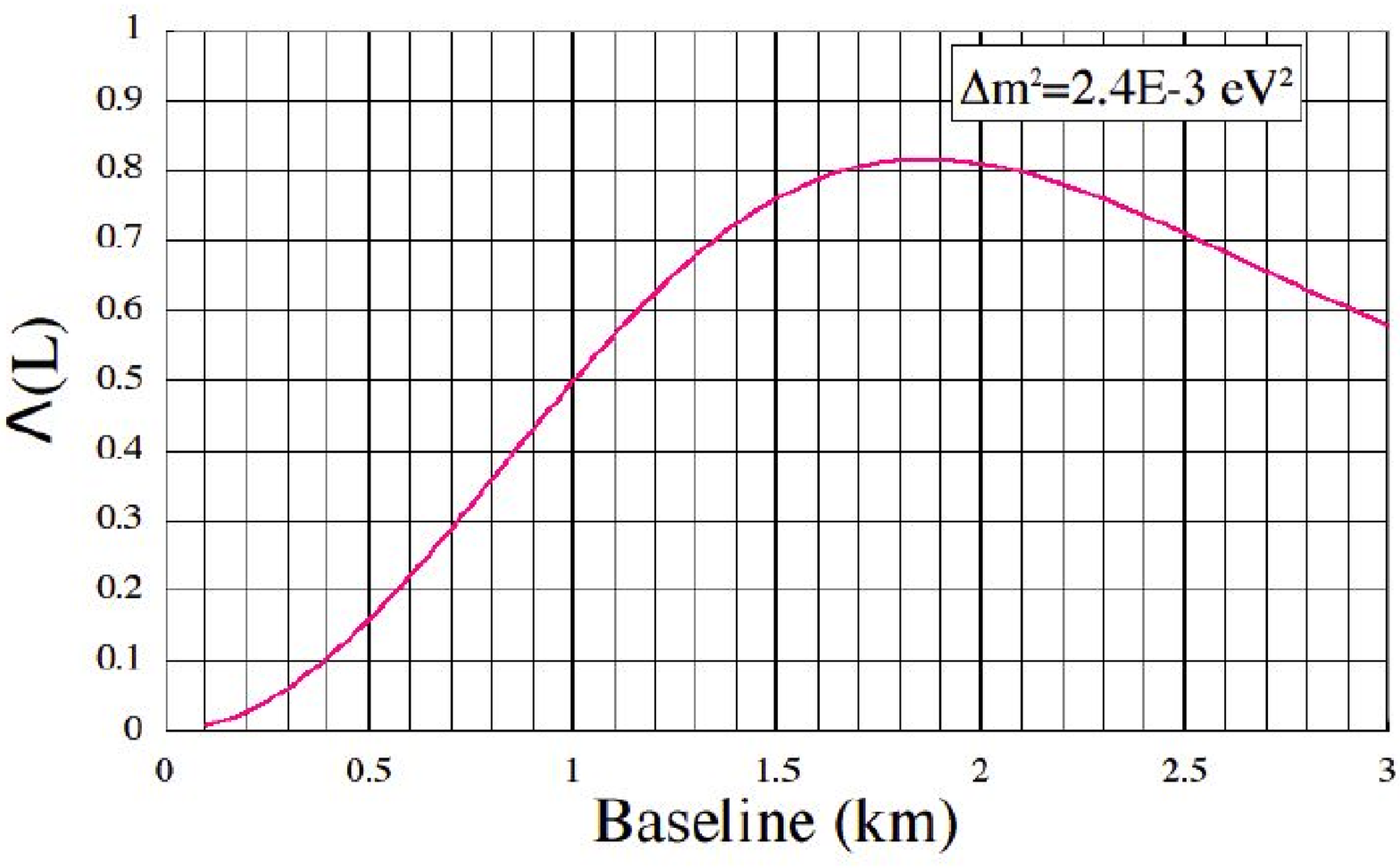}  
 \end{center}
\caption{Deficit probability ($\Lambda (L)$)}
\label{fig:Lambda}
\end{figure}

If neutrino oscillation takes place, the expected neutrino spectrum; $\mu_{\nu}(E)$ at distance $L$ from a reactor is calculated as 
\begin{equation}
 \mu_{\nu}(E,L)=\mu_{\nu}^0(E,L)(1-\sin^22\theta_{13} \sin^2\Phi_{13}(E,L))
\end{equation}
where $\mu_{\nu}^0(E,L)$ is the energy spectrum in case of no neutrino oscillation
calculated by the equation below. 
\begin{equation}
 \mu_{\nu}^0(E,L)=\frac{1}{4\pi L^2}N_p\sigma_{\nu p}(E) f_{\nu}(E)
\end{equation}
where $N_p$ is the number of target protons, $ f_{\nu}(E)$ is the number of generated neutrinos in reactors and $\sigma_{\nu p}(E) $ is the differential cross-section of the inverse neutron $\beta$ decay,
\begin{equation}
\bar{\nu}_e + p \rightarrow e^+ + n
\end{equation}
In the KASAK experiment, the oscillation will be searched for mainly by using the deficit of the total event rate. 
For such case, the expected number of events; $\mu_{\nu}$ is, 
\begin{equation}
 \mu_{\nu}=\mu_{\nu}^0(1-\sin^22\theta_{13}\Lambda(L) )
\end{equation}
where, 
\begin{equation}
 \mu_{\nu}^0=\int{\mu_{\nu}^0(E,L)dE}
\end{equation}
is the total number of events in case of no neutrino oscillation.
\begin{equation}
 \Lambda(L) =\frac{\int{\mu_{\nu}^0(E,L)\sin^2\Phi_{13}(E,L)dE}}
  {\int{\mu_{\nu}^0(E,L)dE}}
\end{equation}
is the deficit of event at distance $L$ from the reactor.
 Fig.~\ref{fig:Lambda} shows $\Lambda(L)$ for the case of $\Delta m^2=2.4\times 10^{-3}\mbox{eV}^2$.
 At the first oscillation maximum of $L\sim1.8$km,  only 20\% of $\bar{\nu}_e$ remains. 

For simplicity, we assume a case to measure $\sin^22\theta_{13}$ with a scheme of one reactor, and one far detector at a distance $L$, one near detector at a distance $L'$. 
If we observe $N_{obs}$ events in the far detector and $N_{obs}'$ events in the near detector,  we consider 
$\epsilon \mu_{\nu}=N_{obs}$ and $\epsilon' \mu_{\nu}'=N_{obs}'$, where $\epsilon$ and $\epsilon'$ represent detection efficiencies of far and near detectors, respectively.
Therefore, the most probable $\sin^22\theta_{13}$ is, 
\begin{equation}
 \sin^22\theta_{13}=\frac{1-r}{\Lambda(L)-r\Lambda(L')} 
 \label{eq:sin22q13}
\end{equation}
where
\begin{equation}
 r=\left(\frac{L}{L'}\right)^2\frac{\epsilon ' N_p'}{\epsilon  N_p}\frac{N_{obs}}{N'_{obs}}
\end{equation}
is expected to unity in case of no neutrino oscillation.
It is noted that $\sigma_{\nu p}$ and $f_{\nu}$ do not appear in Eq.(\ref{eq:sin22q13}) and 
the measurement accuracy does not directly depend on the errors on such parameters.
If value of Eq.(\ref{eq:sin22q13}) is significantly larger than 0, it indicates that  non-0 $\sin^22\theta_{13}$ is observed.

 Since $\sin^22\theta_{13}$ is known to be small, the measurement error is,
\begin{equation} 
 \delta \sin^22\theta_{13}
  = \frac{1}{\Delta\Lambda} 
  \left(\sqrt{\frac{1}{N_{obs}}+\frac{1}{N_{obs}'}} 
 + \frac{ \delta_{uncorr}(\epsilon N_p)}{\epsilon N_p}
  +2\frac{\delta (L/L')}{(L/L')}\right)
 \end{equation}
 where $\Delta \Lambda = \Lambda (L)-r\Lambda (L') \approx 0.7$ for KASKA base lines. 
  In order to obtain a small error, having a large $\Delta\Lambda$ is essential. 
  $\delta_{uncorr}$ means uncorrelated error for both detectors.
  By taking the ratio, the common uncertainty components cancel out and only uncorrelated components remain. 
 The term of square root is the statistic error. 
 
In the KASKA experiment, the main event selections are, 
\begin{center}
   (1) $ 0.7MeV<E_{prompt}<9MeV $ \\
   (2) $ 5MeV<E_{delay}<11MeV $ \\
   (3) $ 1\mu s < \Delta t < 200 \mu s $ \\
\end{center}
The detection efficiency is multiplication of each event selection.
\begin{equation}
\epsilon= \epsilon_{e^+} \epsilon_n  \epsilon_{\Delta t} 
\end{equation}
Errors come in from the relative uncertainties of each parameter. 
  
  %==========================================
\subsection{Detector  Associated Error; $\delta_{uncorr}(\epsilon N_p)/\epsilon N_p$}

%----------------------------------------
\subsubsection{Number of target protons; $N_p$}
The number of target proton is calculated by,  
\begin{equation}
 N_p=M_{GdLS} \rho_p
\end{equation}
where $\rho_p$ is the proton density per unit LS mass and $M_{GdLS}$ is the mass of the Gd-LS  within the fiducial region. 
The Gd-LS for all the detectors will be stored in a storage tank before introducing in the detectors and sent to each detector from the tank. 
So that in principle, the component of the LS is same for all the detectors. \\
KASKA does not apply fiducial cut because it introduces a large uncertainty. 
Instead, the reactor neutrino signal is identified by existence of the neutron-captured Gd signal, regardless where the prompt signal occurs. 
Because region-II scintillator does not contain Gd, there is a natural cut-off of the fiducial region. 
The main error of the neutrino detection efficiency comes from the
relative difference of the fiducial mass of the Gd-LS between the near/far detectors.  
The mass will not be defined by the volume of the container because it
possibly deforms under the pressure of the liquids.  
Instead, it will be defined by total mass of the LS which is measured precisely
when filled in the detector. 
The Gd-LS will be delivered by 200~liter oil drums and the weight difference before and after the 
filling is measured with 0.3\% accuracy weighing machine. 
The three weighing machines used in the far and near detectors are cross calibrated occasionally
during the filling process. 
The diameter of the chimney is 30~cm and 0.1\% of volume error corresponds to 10~cm of head level difference,
which can easily be detectable by a level monitor. 

\begin{figure}[htbp]
 \begin{center}
  \includegraphics[width=0.6\textwidth] {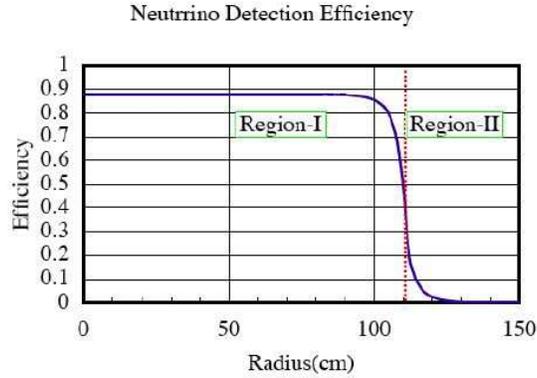}  
 \end{center}
\caption{Radial dependence of detection efficiency. 
Average radius of 1.1~m is used as a typical size. 
Diffusion length of neutron is assumed to be 6~cm. }
\label{fig:spill_in_out}
\end{figure}

If the neutron produced in a $\bar{\nu}_e$ event goes out from region-I, it introduces an inefficiency (spill-out effect). 
Contrarily even if a $\bar{\nu}_e$ event occurs in region-II, it is considered to be $\bar{\nu}_e$  signal if the produced neutron comes in the region-I and is absorbed by Gd (spill-in effect). 
The spill-in effect largely cancels the spill-out inefficiency. 
Fig.-\ref{fig:spill_in_out} depicts such effects. 
The error introduced due to the edge effects is
\begin{equation}
\frac{\Delta N_p}{N_p} \sim \frac{3(\Delta \sigma_{I}-(\rho_{II}/\rho_I)\Delta \sigma_{II})}{2R},
\end{equation}
where $\sigma_I, \sigma_{II}$ and $\rho_I, \rho_{II}$ are diffusion length of neutrons and proton density in region-I and II, respectively. 
$R$ is average radius of region-I.
The main error component comes from uncorrelated uncertainty of the diffusion length. 
Because diffusion length is insensitive to liquid scintillator components, the error is assumed to be negligibly small. 

In summary the relative uncertainty of the fiducial mass is safely claimed to be better than 0.5\%.

%-------------------------------------------
\subsubsection{ $e^+$ Energy Cut Efficiency; $\epsilon_{e^+}$  }

\begin{figure}[htbp]
 \begin{center}
  \includegraphics[width=0.5\textwidth] {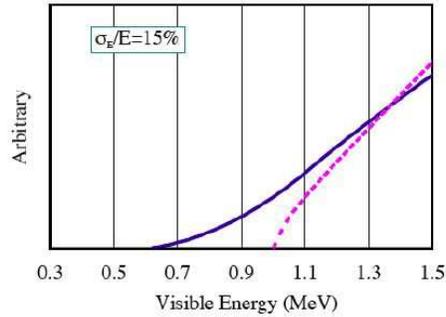} \\
 \end{center}
\caption{Energy spectra near the threshold. 
The dashed line shows original energy spectrum and the solid like shows the spectrum smeared by finite energy resolution of $\sigma_E/E=15\%$.
}
\label{fig:threshold}
\end{figure}

The threshold for the positron energy is 0.7~MeV which is 3$\sigma$ below the positron annihilation energy of 1.022~MeV. 
Fig.~\ref{fig:threshold} shows the reactor neutrino spectrum near the energy threshold. 
Even if the energy resolution is $\sigma_E/E=15\%$ the rate below 0.7~MeV is less than 0.03\% and even if 0.1~MeV of
detector-relative systematic bias takes place, the relative difference in the number of event is less than 0.1\%.  

%----------------------------------------------------------
\subsubsection{ Neutron Capture Efficiency; $\epsilon_n$  }
The efficiency for neutron signal cut is a multiply of neutron absorption efficiency on Gd; $\epsilon_{abs}$ and the detection efficiency of Gd $\gamma$ signal; $\epsilon_{Gd\gamma}$ ,

\begin{equation}
 \epsilon_n=\epsilon_{abs}\epsilon_{Gd\gamma}.
\end{equation}

The neutron capture efficiency for Gd is expressed as,

\begin{equation}
\epsilon_{Gd}
 =\frac{1}{1+\frac{\rho_H}{\rho_{Gd}}\frac{\sigma^n_H}{\sigma^n_{Gd}}} 
\end{equation}

where $\rho$ is the number density of the corresponding element. $\sigma^n$ is the neutron absorption cross section as shown in the table~\ref{tab:n_abs}. \\
\begin{table}[htbp]
\caption{neutron absorption cross section}
 \begin{center}
 {\footnotesize
  \begin{tabular}{|c|r|r|r|}
   \hline
   {} &{} &{} &{}\\[-1.5ex]
         Element      & Q value (MeV)    & Abundance(\%)  &  cross section (b) \\
   {} &{} &{} &{}\\[-1.5ex]
   \hline
   {} &{} &{} &{}\\[-1.5ex]
   $^1H $                       & 2.2  & 99.985           & $0.3326 \pm 0.0007   $ \\
 \hline
   $^{155}Gd$             & 8      & $14.80 \pm 0.05$ & $60900  \pm 500 $ \\
   $^{157}Gd$             & 8      & $15.65 \pm 0.03$ & $254000 \pm 800 $ \\
   \hline
   $\it{Gd}$  Average &      &                  & $48800  \pm 400 $ \\
   \hline
  \end{tabular} }
 \end{center}
 \label{tab:n_abs}
\end{table}
The ratio of the absorption cross section is $\sigma^p_{abs} /\sigma^{Gd}_{abs}=6.82 \times 10^{-6}$. 

The atomic component of the KASKA liquid scintillator is  

\begin{center}
	\#H:\#C:\#O:\#Gd=1.85:1:0.018:0.00009
\end{center}
and $\rho^p_{abs} /\rho^{Gd}_{abs}=2.01 \times 10^4$.
Therefore, the neutron absorption cross section on Gd is, $ \epsilon_{Gd}=0.88 $.
The capture efficiency error associated with the Gd concentration error is,

\begin{equation}
\frac{\delta \epsilon_{Gd} }{\epsilon_{Gd}}
 =(1-\epsilon_{Gd}) \frac{\Delta \rho_{Gd}}{\rho_{Gd}}=0.12 \frac{\Delta \rho_{Gd}}{\rho_{Gd}}
\end{equation}
Therefore the capture efficiency is insensitive to the change of the component of the liquid scintillator. 
Because the Gd-LS for all detectors is taken from the same storage tank, the variation of the
component is supposed to be small and the error on $\epsilon_{Gd}$ is estimated to be better than 0.5\%. 

%--------------------------------------------------------------
\subsubsection{ n Energy Cut Efficiency; $\epsilon_n$  }
\begin{figure}[htbp]
 \begin{center}
  \includegraphics[width=0.4\textwidth,clip] {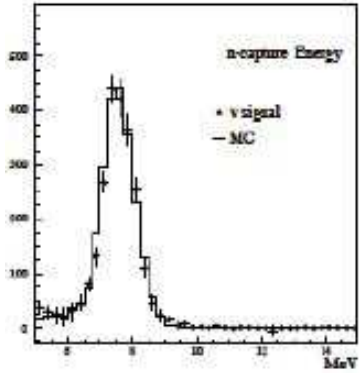} \\
 \end{center}
\caption{Gd-captured neutron signal observed 
by the CHOOZ experiment~\cite{CHOOZ} }
\label{fig:chooz_Gd_signal}
\end{figure}

 The total energy of the gamma rays from neutron captured Gd is 8~MeV. 
The energy threshold for the neutron signal is 5~MeV, which is low enough and the
uncertainty of the threshold does not cause error of the event selection. 
Fig.~\ref{fig:chooz_Gd_signal} shows the Gd-captured neutron signal observed by the CHOOZ group~\cite{CHOOZ}. 

CHOOZ group assigned 0.4\% as absolute error with energy threshold 6~MeV. 
Because the energy threshold can be set lower than CHOOZ case, thanks to a new buffer layer in front of PMT array, 
and energy containment is better, 
and because  error of KASKA is relative error, it is expected to be much better than 0.4\%. 
The threshold energy can be calibrated by $n+^{12}C\rightarrow ^{13}C +\gamma (5MeV)$ events.
%----------------------------------------------------
\subsubsection{ Efficiency for the timing cut; $\epsilon_{\Delta t}$  }

The efficiency for the timing cut is,
\begin{equation}
\epsilon_{\Delta t}=\frac{1}{\tau}\int_{t_1}^{t_2}e^{-\frac{t}{\tau}}dt
= e^{-\frac{t_1}{\tau}}- e^{-\frac{t_2}{\tau}}
\end{equation}
and its error is,
\begin{equation}
\frac{\delta\epsilon}{\epsilon}
=\frac{1}{e^{-\frac{t_1}{\tau}}-e^{-\frac{t_2}{\tau}}}
(\frac{t_1e^{-\frac{t_1}{\tau}}-t_2e^{-\frac{t_2}{\tau}}}{\tau}\frac{\Delta \tau}{\tau}
-e^{-\frac{t_1}{\tau}}\frac{\Delta t_1}{\tau}
+e^{-\frac{t_2}{\tau}}\frac{\Delta t_2}{\tau})
\end{equation}

For KASKA, $\tau \approx 30\mu s, t_1=1\mu s$ and $t_2=200\mu sec$. 
So that, 

\begin{equation}
\epsilon_{\Delta_t}=0.966
\end{equation}

and

\begin{equation}
\frac{\delta\epsilon}{\epsilon}
=0.058\frac{\Delta \tau}{\tau}-1.0\frac{\Delta t_1}{\tau}+0.0014\frac{\Delta t_2}{\tau}
\end{equation}

The absolute timing will be quantized in unit of system clock frequency such as $O(10\mbox{ns})$ and 
the timing of scintillation signal is $O(10\mbox{ns})$.  
Therefore the error induced by absolute timing is negligibly small, $O(0.03\%)$.
Because the components of the Gd-LS is same for all the  detectors, essentially $\Delta \tau$ is small.
The absorption time $\tau$ can be measured  $in-situ$ with real data with precision of a few \% and the $\Delta \tau$ term is order of 0.1\%. 

\subsection{Error from Reactor Neutrino Flux; $f_{\nu}$}

The reactor neutrino flux; $f_{\nu}$ is known to  accuracy of 3.4\%.
For one reactor and 2 detectors case,  the uncertainty completely cancels.  
However, for KASKA case, there are 7 reactors and three detector locations whose near/far baseline ratios are different.
In this case the cancellation becomes imperfect and some ambiguity remains. 
This effect is estimated analytically in ref.~\cite{Sugiyama:04}. 
The following is an intuitive description. 
The neutrino flux is calculated as, 
\begin{equation}
f_{\nu} \sim P_{th}  \times n_f \times \eta_{\nu}
\end{equation}
where $P_{th}$ is the reactor power, $n_f$ is the fission rate per thermal power and $\eta_{\nu}$ is the neutrino flux per fission. 
Their uncertainties are, 2\%, 1\%, and 2.5\% respectively. 
However,  $\eta_{\nu}$ is common to all reactors and  cancel out by near/far comparison. 
All the reactors in Kashiwazaki-Kariwa nuclear power station will be more than 10 years old and their fuel components have became equilibrium . 
Thus the ratio of the fission rate of each  fissile element are similar when averaged over a few years of run period. 
Although the original error is $n_f\sim$1\%, it will be canceled out to the level of negligibly small. 
The ambiguity from the baseline difference comes from the error of thermal power measurement, multiplied by the  standard deviation of the square of the ratio of far/near baselines,
\begin{equation}
\frac{\delta \mu}{\mu} \sim \frac{\delta P}{P} 
\sqrt{\frac{1}{7}\sum_r{\left(1-\frac{R_r}{\hat R}\right)^2}}
\label{eq:baseline_error}
\end{equation}
where $R_r$ is the square of the ratio of far and near baselines, $R_r=(L_{r}/L'_{r})^2$ and $\hat R$ is the average of $R_r$;
\begin{equation}
\hat R = \frac{1}{7} \sum_r R_r
\end{equation}

For KASKA case, the value of the square root in Eq.(\ref{eq:baseline_error}) $\sim 0.07$ and the error  introduced from the neutrino flux uncertainty is estimated to be  0.14\%.

%=============================
\subsection{systematics summary}
Since the current upper limit of $\sin^22\theta_{13}$  was measured by CHOOZ experiment and  the types of systematic uncertainties  of KASKA experiment are similar to those of CHOOZ, it is appropriate to compare systematic uncertainties of both experiment to see the improvement of the sensitivity. 
Table~\ref{tab:sys_summary} and \ref{tab:sys_summary_II} compares  the systematic errores so far discussed.  
The total systematic error can be controlled to be better  than 1\% and possibly be as low as 0.5\%. 
The $\sin^22\theta_{13}$  sensitivity shown in   Fig.-\ref{fig:sensitivity} are calculated with these estimations. 

\begin{table}[htbp]
\caption{Detector Associated Systematics Errors.}
 \begin{center}
 {\footnotesize
  \begin{tabular}{|l|c|c|}
   \hline
   {} &{} &{}\\[-1.5ex]
      source     & CHOOZ & KASKA    \\
   {} &{} &{}\\[-1.5ex]
   \hline
   {} &{} &{}\\[-1.5ex]
   $e^+$ energy ($\epsilon_{e^+}$)        &   0.8\%   & $<$0.1\%     \\
    $e^+$ position                                      &   0.1\%  &  --                 \\
    $n$ capture ($\epsilon_{abs}$)            &   1.0\%  & $<$0.5\%     \\
    $n$ energy ($\epsilon_{Gd\gamma}$) &   0.4\%  & $<$0.4\%     \\
    $n$ position                                          &   0.4\%  & --                  \\
    $n$ delay ($\epsilon_{\Delta t}$)         &   0.4\%  & $<$0.1\%     \\
    $e^+-n$ distance                                   &   0.3\%  & --                  \\
    $n$ multiplicity                                     &  0.5\%   & --                  \\
    $N_p$                                                   &  0.8\%   & $<$0.5\%     \\
    {} &{} &{}\\[-1.5ex]
       \hline
       {} &{} &{}\\[-1.5ex]
  Combined                                              &  1.76\%   & $<$0.8\%   \\
   \hline
  \end{tabular} }
 \end{center}
       \label{tab:sys_summary}
\end{table}

\begin{table}[htbp]  
\caption{Systematics Summary}  
 \begin{center}
 {\footnotesize
  \begin{tabular}{|l|c|c|}
   \hline
   {} &{} &{}\\[-1.5ex]
      source                                        & CHOOZ        & KASKA      \\
   {} &{} &{}\\[-1.5ex]
   \hline
   {} &{} &{}\\[-1.5ex]
   $\sigma_{\nu p}$                          &   1.9\%          & negligible              \\
    Detection efficiency ($\epsilon$)  &   1.76\%        &  $<$0.8\%    \\
   Reactor Power ($P_{th}$)            &   0.7\%          & 0.2\%         \\
   Energy released per fission            &   0.6\%          & negligible                  \\
   Background                                   &   0\%             & $<$0.6\%     \\
    {} &{} &{}\\[-1.5ex]
       \hline
       {} &{} &{}\\[-1.5ex]
  Combined                                      &  2.7\%           & $<$1.0\%     \\
   \hline
  \end{tabular} }
 \end{center}
        \label{tab:sys_summary_II} 
\end{table}

%=================================
\section{Conclusions}

The KASKA project enjoys world's most intense low energy $\bar{\nu}_e$ flux from Kashiwazaki-Kariwa nuclear power station and has opportunity to pin down several neutrino oscillation parameters, such as 
$\theta_{13}$,  precise $\theta_{12}$ and $\Delta m_{13}^2$. 
Moreover, by combining with accelerator-based neutrino oscillation measurements, there are possibilities to find non-0 $\sin \delta$ and  to solve $\theta_{23}$ degeneracy.  

The first phase of the project is to measure $\sin^2 \theta_{13}$ with sensitivity ten times better than the current limit. 
The detector R\&D has progressed well and it will be possible to start data taking 3 years after start of  construction. 

\appendix

\section{Acknowledgement}
KASKA R\&D activities have been supported by Grants-in-Aid for Scientific Research of Japanese Ministry of Education, Culture, Sports, Science and Technology and  Collaborative Development Research Program of KEK. 
The mechanical engineering center of KEK helps development of detector structure. 
The KASKA group greatly appreciates them.

%========================================

%\include{tex_files/cover}
\end{document}